\documentclass[%
 reprint,
 amsmath,amssymb,
 aps,
]{revtex4-2}

\usepackage{graphicx}% Include figure files
\usepackage{dcolumn}% Align table columns on decimal point
\usepackage{bm}% bold math
\usepackage[utf8]{inputenc}
\usepackage{physics}
\usepackage{bbm}
\usepackage[dvipsnames]{xcolor}
\UseRawInputEncoding %fixes an error related to bib file
\usepackage[french,english]{babel}
\usepackage{soul}
\usepackage{tabularray}

\usepackage[hidelinks]{hyperref}
\usepackage{cleveref}
\usepackage[T1]{fontenc}
\hypersetup{colorlinks=true,linkcolor=blue}

\begin{document}

\preprint{APS/123-QED}

\title{Suppressing spin qubit decoherence during shuttling via confinement modulation}

\author{Daniel Q. L. Nguyen}
\email{q.l.d.nguyen@tudelft.nl}
\affiliation{QuTech and Kavli Institute of Nanoscience, Delft University of Technology, Delft, Netherlands}
\author{Maximilian Rimbach-Russ}
\affiliation{QuTech and Kavli Institute of Nanoscience, Delft University of Technology, Delft, Netherlands}
\author{Stefano Bosco}
\affiliation{QuTech and Kavli Institute of Nanoscience, Delft University of Technology, Delft, Netherlands}

\date{\today}

\begin{abstract}
{Reliable long-range qubit shuttling is a powerful tool for scalable quantum computing architectures. We investigate strategies to improve the coherence of moving spin qubits by performing continuous dynamical decoupling by modulating their confinement potential. Specifically, we introduce temporal and spatial breathing shuttling protocols that leverage spin-orbit interactions in hole-spin systems to electrically drive the qubit while moving. This enables efficient dressed-state shuttling, where the spin is continuously rotated during transport, suppressing the effect of low-frequency noise. Using the filter function formalism, we identify driving regimes that efficiently mitigate both global and local magnetic and electric noise sources. We find that confinement-modulated shuttling can significantly enhance coherence during transport, while revealing distinct limitations depending on the correlation length of the noise. Applying our framework to germanium hole-spin qubits, we show that these protocols provide a practical route toward noise-resilient long-range coherent quantum links.}
\end{abstract}

\maketitle

\section{Introduction}
\begin{figure}[t]
    \centering
    \includegraphics[width=1\linewidth]{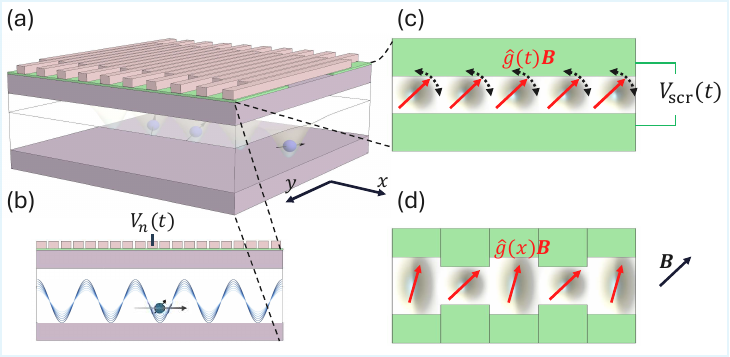}
    \caption{\textbf{Breathing shuttling protocols.} (a) Sketch of a conveyor-mode shuttling architecture. (b)-(c) Temporal breathing.  Modulating the amplitude of the potential applied to the top gates (b), or  the screening gates (c), enables to modulate the confinement potential of the spin qubit during shuttling. (d) Spatial breathing. A periodic change in the design of the screening gates modulates of the confinement in space, leading to a time-dependent driving of the spin even at constant shuttling velocities.}
    \label{fig:model}
\end{figure}
Semiconductor spin qubits~\cite{Burkard2023-ex} are a promising platform to host large-scale quantum computers, owing to their compatibility with established semiconductor fabrication techniques~\cite{zwerverQubitsMadeAdvanced2022,steinackerIndustrycompatibleSiliconSpinqubit2025,george12SpinQubitArraysFabricated2025,nicklEightQubitOperation3002025,vorreiterPrecisionHighspeedQuantum2025}. Large-scale architectures benefit from long-range coherent quantum links that enable the coherent transfer of quantum information over long distances~\cite{Taylor2005,Vandersypen2017,Knne2024,494s-jd8h}, extending the connectivity beyond nearest neighbors and providing greater flexibility in device layout and wiring constraints. Recently, conveyor-mode shuttling~\cite{Taylor2005,langrockBlueprintScalableSpin2023}, where the spin qubit is coherently moved over several microns by smoothly displacing its electric potential, gained increasing attention~\cite{Seidler2022,Struck2024,DeSmet2025,xueSiSiGeQuBus2024,ademi2025distributingentanglementdistantsemiconductor,krzywda2026coherenceprotectionmobilespin,matsumoto2025twoqubitlogicteleportationmobile}. It has been shown that conveyor-mode shuttling can be performed using shared-control shuttling lanes that require only a few distinct electrical signals~\cite{xueSiSiGeQuBus2024,ademi2025distributingentanglementdistantsemiconductor}, making conveyor-mode shuttling a promising tool for semiconductor-based large-scale quantum architectures.

The coherence time of a spin qubit is typically limited by low-frequency noise that may originate from ensembles of nuclear spins or two-level fluctuators coupling to the spin via spin-orbit interaction or magnetic gradient fields~\cite{Chekhovich2013,Schriefl2006,struckLowfrequencySpinQubit2020,paqueletwuetzReducingChargeNoise2023,Kpa2023,shehataModelingSemiconductorSpin2023,wangDephasingPlanarGe2025,wangModelingPlanarGermanium2024,wangUltrafastCoherentControl2022}. Through motional narrowing, mobile spins traversing distances large compared to the spatial correlation length of the noise experience enhanced coherence due to spatial averaging, providing additional protection compared to stationary spins~\cite{krzywda2026coherenceprotectionmobilespin,langrockBlueprintScalableSpin2023}. To further extend the lifetime of a shuttled qubit beyond motional narrowing, non-trivial qubit dynamics, such as driving the qubit while shuttling, can greatly enhance the shuttling fidelity by effectively dressing the qubit~\cite{Bosco2024,Cywinski2008-aj}. This dressing modifies the qubit's response to low-frequency noise and suppresses decoherence during transport, in close analogy to continuous dynamical decoupling~\cite{Laucht2016,tsoukalas2025dressedsinglettripletqubitgermanium}. As a result, coherent control during shuttling not only preserves the quantum state but also enables high-fidelity transport over much longer distances.

In this work, we analyze how the coherence times of a moving qubit in an inhomogeneous effective magnetic field can be enhanced using continuous dynamical decoupling. Furthermore, we present a set of protocols that enable such decoupling sequences during shuttling~\cite{Bosco2024,krzywda2026coherenceprotectionmobilespin} focusing on hole-spin qubits in germanium-based platforms~\cite{Scappucci2020,jirovecSinglettripletHoleSpin2021,hendrickxFourqubitGermaniumQuantum2021,hendrickxSweetspotOperationGermanium2024,wangOperatingSemiconductorQuantum2024,saez-mollejoExchangeAnisotropiesMicrowavedriven2025,johnRobustLocalisedControl2025,ivlevOperatingSemiconductorQubits2025} as an example. Our approach relies on globally modulating the gate-voltage signals of existing shuttling architectures, without introducing additional control lines. These time-dependent gate voltages dynamically reshape the confinement potential along the transport path and, because of the strong spin-orbit interaction in hole systems, induce a controlled modulation of the effective $g$-tensor~\cite{Bosco_2021,Martinez_2022,Sarkar_2023,Liles_2021}. In this dressed-state regime, the moving qubit experiences continuous protection against low-frequency noise, leading to enhanced coherence times without  the precisely timed control pulses required in  pulsed dynamical decoupling schemes.

In realistic devices, electrostatic disorder and strain gradients along the shuttling lane cause spatial variations in both the Larmor frequency and the orientation of the Larmor vector~\cite{ademi2025distributingentanglementdistantsemiconductor,strainmapping}. We show that, even in the presence of such variations, resonant driving that continuously tracks the local Larmor frequency retains the benefits of dynamical decoupling and enables high-fidelity shuttling.

This paper is organized as follows. In Sec.~\ref{sec:continuous_dynamical_decoupling}, we discuss the qubit's susceptibility to noise at different frequencies. We use a simple driven spin Hamiltonian and filter function formalism to analyze the  optimal driving regime to  enhance coherence times in the presence of low-frequency noise~\cite{Green_2013,Cywinski2008-aj,Bosco2024}. We also present results for a general time-dependent Larmor frequency. Section~\ref{sec:model} introduces the model Hamiltonian for planar germanium hole-spin qubits enabling electrical control of the spin degree of freedom via confinement modulation, which constitutes a specific realization of a time-dependent Larmor frequency. Possible confinement-modulation schemes compatible with existing gate architectures are shown in Fig.~\ref{fig:model}, enabling control of both the shuttling and transverse in-plane confinement. Such modulation may be achieved using time-dependent gate voltages or through appropriately shaped screening gates, without modifying conventional conveyor-mode shuttling signals. Building on this, in Sec.~\ref{sec:time-modulated_breathing_shuttle} we derive a time-dependent effective Hamiltonian, evaluate the corresponding modified filter functions for electrically and magnetically induced noise - local and global - and present numerical simulations quantifying the resulting enhancement of coherence factors. Lastly, we repeat the analysis for spatially dependent confinements in Sec.~\ref{sec:spatially-modulated_breathing_shuttle}. Our findings show that temporal and spatial breathing protocols enable dynamical decoupling during shuttling, with temporal modulation providing robust suppression across noise types, while spatial modulation offers a simpler implementation but reduced efficiency for localized noise.

\section{Continuous dynamical decoupling}\label{sec:continuous_dynamical_decoupling}
\begin{figure*}[t]
    \centering
    \includegraphics[width=1\linewidth]{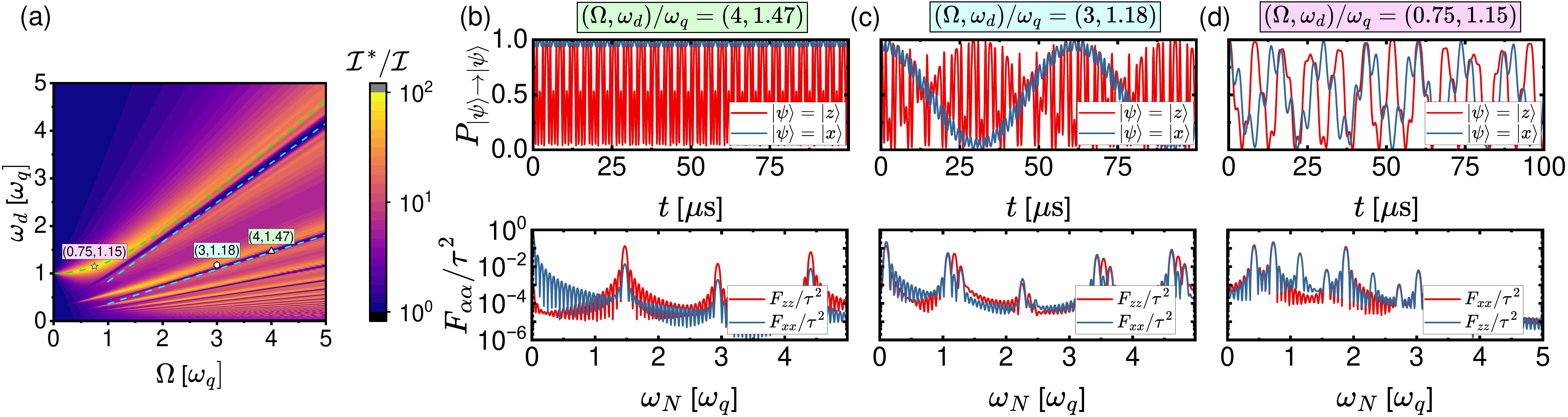}
    \caption{\textbf{Noise decoupling of driven two-level system}. (a) Ratio $\mathcal{I}^\ast/\mathcal{I}$ of the infidelity without driving ($\mathcal{I}^\ast$) ($\Omega=0$) to that under driving ($\mathcal{I}$) as a function of $\omega_d$ and $\Omega$. Larger values in yellow indicate improved coherence. The green dashed line denotes the resonance condition including the Bloch-Siegert shift [Eq.~\eqref{eq:general_resonance_frequency_condition}], while the cyan dashed lines indicate the first two solutions of Eq.~\eqref{eq:Condition_large_coupling_dark_branch_condition}. (b)-(d) Transition probabilities $P_{\ket{\uparrow}\rightarrow\ket{\uparrow}}(t)$ and $P_{\ket{+}\rightarrow\ket{+}}(t)$ together with the relevant filter functions $F_{xx}$ and $F_{zz}$ for $(\omega_d,\Omega)=(0.75,1.15)$, $(4,1.47)$, and $(3,1.18)$, respectively. Panel (b) corresponds to the dark-branch condition, where the $\sigma_x$ eigenstate is approximately eigenstate of the driven system, yielding a peak in $F_{xx}$ at $\omega=0$. Panel (c) is slightly detuned from this condition, leading to oscillations in both bases and suppression of the zero-frequency components. Panel (d) corresponds to resonant driving in the strong-coupling regime ($\Omega\sim\omega_q$), where the oscillations are no longer fully coherent due to effects beyond the rotating wave approximation. The total evolution time is $10\,\mu\text{s}$ with $\omega_q=100\,\text{rad/}\mu\text{s}$.}
    \label{fig:toymodel_plots}
\end{figure*}

To understand the dynamics of a spin shuttled in an inhomogeneous and modulated Zeeman field, we begin by investigating a model Hamiltonian describing a driven two level system 
\begin{align}\label{eq:driven_two_level_system}
    H_c=\frac{\hbar\omega_q(t)}{2}\sigma_z+\hbar\Omega(t)\sin\Phi(t)\sigma_x.
\end{align}
Here, $\omega_q(t)$ denotes the (possibly time-dependent) Larmor frequency, $\Omega(t)$ is the driving amplitude, and $\Phi(t)$ denotes the driving phase.

To illustrate the basic mechanism of continuous dynamical decoupling, we first analyze the simplified case of a constant Larmor frequency and then extend the discussion to time-dependent Larmor frequencies $\omega_q(t)$. We remark that such time-dependencies can arise when the spin moves in a spatially inhomogeneous Zeeman field.
\subsection{Free-induction decay}
We first consider a constant Larmor frequency, $\omega_q(t)\equiv\omega_q$, and no driving, $\Omega(t)=0$. In this case, the qubit undergoes free precession and the time-evolution operator of the control Hamiltonian (i.e., the system Hamiltonian without noise) reduces to $U_c(t)=\exp(-i\omega_q t\sigma_z/2)$. Using the filter function formalism~\cite{Green_2013,Bosco2024}, one can express the average gate infidelity in the presence of noise as
\begin{align}\label{eq:average_gate_infidelity_FF_formalism}
    \mathcal I\simeq\frac{1}{2\pi\hbar^2}\int_{-\infty}^\infty d\omega S(\omega) \text{Tr}(F(\omega,\tau)),
\end{align}
This equation is valid if noise is small, wide sense stationary, isotropic, and its components mutually independent in spin-space, see App.~\ref{app:Filter_function_formalism}. The power spectral density function of the noise source  $S(\omega)$ is assumed to be proportional to $1/\omega$ with $1/\omega$ denoting the time-scale describing the characteristic dynamics of the noise source. Additionally, we take the trace of the matrix of filter functions defined as
\begin{align}\label{eq:filterfunction_Global_Noise}
    F(\omega,\tau)=\int_0^\tau\int_0^\tau dt_1dt_2\hat R_c(t_1)\hat R_c^T(t_2)e^{-i\omega(t_1-t_2)}.
\end{align}

The filter function provides the susceptibility of the control dynamics to noise components acting along different spin directions at frequency $\omega$. The $3\times3$ rotation matrix satisfies $R_c(t)\boldsymbol\sigma=U^\dagger_c(t)\boldsymbol\sigma U_c(t)$, with $U_c(t)$ being the time-evolution of the noiseless control Hamiltonian, and is in case of the constant Larmor frequency without drive given as a time-dependent rotation about the axis $z$, that is, $R_c(t)=R_z(\omega_qt)$. For spin qubits low frequency noise dominates decoherence and an efficient dynamical decoupling protocol aims to suppress the filter function at frequencies close to 0, such that the weight in the integral in Eq.~\eqref{eq:average_gate_infidelity_FF_formalism} is suppressed where $S(\omega)$ is large.

The suppression of the filter function at a given frequency can be expressed directly in terms of transition probabilities $P_{\psi\rightarrow\phi}(t)=|\bra{\phi}U_c(t)\ket{\psi}|^2$. In particular, we find that  
\begin{align}\label{eq:Iff_Condition_For_filterfunction_Suppression}
    F(0,\tau)=0^{3\times3}\Leftrightarrow
    \frac{1}{\tau}\int_0^\tau dt\,P_{\psi\rightarrow\phi}(t)=\frac{1}{2}
    \quad \forall \psi,\phi \ ,
\end{align}
which provides a necessary and sufficient criterion for suppressing the filter function at frequency $\omega=0$ with $0^{3\times3}$ denoting the $3\times3$ zero matrix. For general $\omega$ and the derivation of this equation we refer the reader to App.~\ref{app:Filter_function_formalism}. Specifically, suppression of the zero-frequency component requires that the time-averaged transition probability between any two single qubit states equals $1/2$. For free precession, the $z$-eigenstates are eigenstates of the Hamiltonian with no driving. Thus, $P_{\ket{z}\rightarrow\ket{z}}(t)=1$ which implies that the matrix of filter function for zero frequency noise cannot be $0^{3\times3}$, see Eq.~\eqref{eq:Iff_Condition_For_filterfunction_Suppression}. In fact one finds the $z$ filter function to be maximal at zero frequency $F_{zz}(0,\tau)=\tau^2$.

\subsection{Rabi driving}
\subsubsection{Weak driving amplitude}
We now include a weak driving with constant amplitude ($\Omega(t)\equiv\Omega\ll\omega_q$), and consider resonant driving $\Phi(t)=\omega_d t$ with $\omega_d=\omega_q$. The time-evolution of Eq.~\eqref{eq:driven_two_level_system} is  approximated by $U_c(t)\approx\exp(-i\omega_qt\sigma_z/2)\exp(-i\Omega t\sigma_y/2)$ in the rotating wave approximation (RWA). The corresponding rotation matrix in the filter function [Eq.~\eqref{eq:filterfunction_Global_Noise}] is therefore
\begin{align}\label{eq:control_matrix_weak_coupling_resonant_driving_RWA}
    R_c(t)\approx R_z(\omega_qt)R_y(\Omega t)
\end{align}
where $\Omega$ is the Rabi frequency. In contrast to  free precession, the qubit now undergoes a precession around the  $z$ axis, complemented by an additional precession around $y$. Importantly, no single initial state retains a static projection along any axis. This can lead to the suppression of the zero-frequency filter function $F(0,\tau)$ as implied by Eq.~\eqref{eq:Iff_Condition_For_filterfunction_Suppression}.

Using the approximation in Eq.~\eqref{eq:control_matrix_weak_coupling_resonant_driving_RWA} the zero-frequency filter function $F(0,\tau)$ is suppressed at finite $\tau,\omega_q$, and $\Omega$ such that $\omega_q\tau=2l\pi$ and $\Omega \tau=2k\pi$ ($l,k\in\mathbb N$). For long times $\tau$, the zero frequency filter function is suppressed also at arbitrary, finite $\Omega$ and $\omega_q$ in the RWA. With zero Rabi amplitude ($\Omega=0$ or $|\omega_d-\omega_q|\gg\Omega$) we recover the free-induction decay  leading to $F_{zz}(0,\tau)=\tau^2$ being maximal.

Beyond weak driving, when $\Omega\sim \omega_q$, we can still decouple the qubit efficiently from low frequency noise when the driving frequency satisfies the general resonance condition~\cite{Yan2015-pm}
\begin{align}\label{eq:general_resonance_frequency_condition}
    \omega_d=\omega_q+\frac{\omega_d \Omega^2}{(\omega_d+\omega_q)^2}+\frac{(2\omega_q-\omega_d)^4}{64(\omega_d+\omega_q)^4}.
\end{align}
Here, the relevant states still undergo complete oscillations [see Fig.~\ref{fig:toymodel_plots}(d)]. We further find that the real and positive solution to this equation matches well the driving region with enhanced coherence as shown in Fig.~\ref{fig:toymodel_plots}(a). When the Rabi frequency $\Omega$ becomes comparable to or exceeds the Larmor frequency, counter-rotating contributions become significant and redistribute the weight of the filter function toward higher-frequency peaks. Nevertheless, the transition probability between the two $\sigma_z$ eigenstates remains oscillatory with an approximately $\tfrac{1}{2}$ mean value, as required by Eq.~\eqref{eq:Iff_Condition_For_filterfunction_Suppression} and shown in Fig.~\ref{fig:toymodel_plots}(d).

\subsubsection{Strong driving amplitude}\label{sec:toymodel_Rabi_driving_strong_driving_amplitude}
In the strong coupling regime $\Omega\gg \omega_q$ the transition probability for the $z$-eigenstates is well approximated by~\cite{PhysRevLett.98.013601} 
\begin{align}
P_{\ket{\uparrow}\rightarrow\ket{\uparrow}}(t)\approx\cos^2\left(\frac{\Omega}{\omega_d}\sin\omega_dt\right).
\end{align}
Using Eq.~\eqref{eq:Iff_Condition_For_filterfunction_Suppression} with the explicit states $\ket{\psi},\ket{\phi}=\ket{\uparrow}$ and the Jacobi-Anger expansion, we find that the filter function at zero frequency is suppressed when
\begin{align}
J_0(\Omega/2\omega_d)+\sum_{m=1}^{\infty}J_{2m}(\Omega/2\omega_d)\frac{\sin(2m\omega_d\tau)}{2m\omega_d\tau}\stackrel{!}{=}0 \ .
\end{align}

For sufficiently long times $\omega_d\tau\gg1$, or under the condition $2\omega_d\tau=k\pi$ ($k\in\mathbb N$), the oscillatory terms can be neglected and the condition reduces to
\begin{align}\label{eq:Condition_large_coupling_dark_branch_condition}
J_0(\Omega/2\omega_d)=0 ,
\end{align}
or equivalently $\omega_d=\Omega/2j_k$, where $j_k$ denotes the $k$-th root of the zeroth Bessel function of the first kind. At these driving frequencies the filter function component $F_{zz}(0,\tau)$ is suppressed. However, we observe that this condition corresponds to the darker branches or cyan dashed lines in Fig.~\ref{fig:toymodel_plots}(a), indicating a shorter decoherence time. This is because at these points the states $\ket{\pm x}=(\ket{\uparrow}\pm\downarrow)/\sqrt{2}$ become approximate eigenstates of the time evolution operator, such that $P_{\ket{+x}\rightarrow \ket{+x}}(t)$ remains close to unity, as illustrated in Fig.~\ref{fig:toymodel_plots}(b). Consequently, at large couplings and driving frequencies satisfying Eq.~\eqref{eq:Condition_large_coupling_dark_branch_condition} the filter function $F_{xx}(0)$ becomes the limiting factor to the qubits coherence. By detuning the frequency  away from this condition the $x$-states are no longer eigenstates and coherent oscillations between $\ket{\pm x}$ are recovered as illustrated in Fig.~\ref{fig:toymodel_plots}(c), which enhances the suppression of low-frequency noise and leads to the neighboring bright branches with increased coherence times.

\subsection{Varying Larmor frequency}\label{sec:Varying_Larmor_frequqency}
\begin{figure}
    \centering
    \includegraphics[width=1\linewidth]{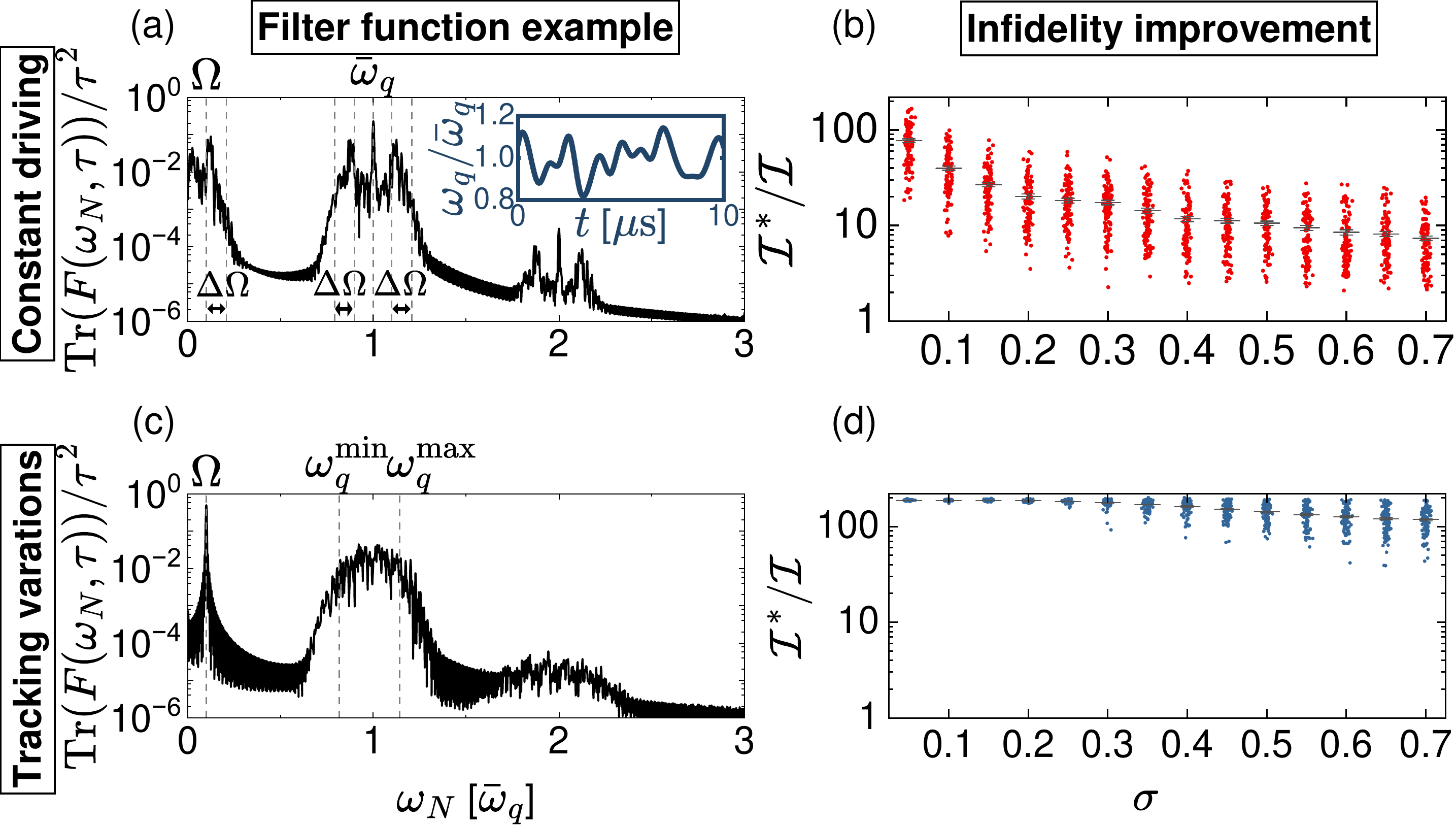}
    \caption{\textbf{Driven two-level system with varying Larmor frequency.} Filter functions of an exemplary Larmor frequency landscape using (a) constant driving with the mean Larmor frequency and (c) the tracking protocol defined in Eq.~\eqref{eq:instantaneous_driving}. The corresponding frequency landscape is shown in the inset of (a). Panels (b) and (d) show the corresponding average infidelity enhancement taking the ratio of infidelity without driving ($\mathcal{I}^\ast$) to that under driving ($\mathcal{I}$) for different variances $\sigma$ in the Larmor frequency. For each chosen variance $\sigma$ we plot the results from $100$ sampled Larmor frequency landscapes for the constant-driving and tracking protocols, respectively. The total evolution time is $10\,\mu\text{s}$  with $\omega_q=100\,\text{rad/}\mu\text{s}$.}
    \label{fig:varying_larmor}
\end{figure}
To benchmark driving protocols where the Larmor frequency varies in time, we consider a non-trivial time-dependence of the Larmor frequency $\omega_q(t)$ in Eq.~\eqref{eq:driven_two_level_system}. Starting from a chosen mean value $\bar{\omega}_q$, we generate discrete values $\omega_q(t_j)$ at times $t_j = j\Delta t$ by sampling from a Gaussian distribution with mean $\bar{\omega}_q$ and standard deviation $\sigma_{\bar{\omega}_q}$. An example of $\omega_q(t)$ landscape is shown in the inset of Fig.~\ref{fig:varying_larmor}(a). We  consider again constant driving amplitudes $\Omega(t)=\text{const.}$ and discuss varying driving amplitudes in App.~\ref{app:Varying_Rabi_frequency}.
\subsubsection{Driving with the mean Larmor frequency}
At a given position- or time-dependent Larmor frequency one can drive the spin at the mean Larmor frequency $\Phi(t)=\bar\omega_qt$ assuming weak driving amplitudes. At times where the spin is driven at a detuned frequency the Rabi frequency increases leading to the broadening of the peaks in the filter function in Fig.~\ref{fig:varying_larmor}(a) of width
\begin{align}\Delta\Omega=\sqrt{(\delta\omega_q^\text{max})^2+\Omega^2}-\Omega.
\end{align}
If the variations of the Larmor frequency are smooth on the scale of the Larmor frequency, $\Delta\Omega$ is well approximated by $\delta\omega_q^\text{max}=\max\limits_{t\in[0,\tau]}|\omega_q(t)-\bar\omega_q|$. Importantly, for sufficiently large detuning, the $z$-eigenstates no longer undergo full rotations, which reduces the mixing between spin states and thus weakens the suppression of the filter function at zero frequency. Since the drive remains fixed at $\bar{\omega}_q$, it continues to match only the average Larmor frequency, resulting in a peak in the filter function at $\omega=\bar{\omega}_q$ of Fig.~\ref{fig:varying_larmor}(a).

To quantify this effect, we generate multiple realizations of such Larmor frequency landscapes and evaluate the corresponding fidelity for each case under this driving, as shown in Fig.~\ref{fig:varying_larmor}(b). While the fidelity improvement heavily depends on the specific Larmor frequency landscape, increasing the variation strength $\sigma$ generally leads to a worse average performance due to the reduced efficiency of the Rabi driving at larger detuning.

\subsubsection{Tracking Larmor frequency variations}\label{sec:tracking_larmor_frequency_variations}
Considering a fully characterized Larmor frequency landscape, one can adapt the driving field to be continuously resonant. Using
\begin{align}\label{eq:instantaneous_driving}
    \Phi(t)=\int_0^t dt'\omega_q(t')
\end{align}
and in the RWA where $\omega_q(t)$ needs to remain large and does not vary rapidly compared with $\Omega$, we recover the conventional Rabi driving along the entire shuttling lane yielding the control matrix
\begin{align}
    R_c(t)\approx R_z(\Phi(t))R_y(\Omega).
\end{align}
The peak associated with the driving frequency therefore broadens and ranges between frequencies $\omega_q^\text{min}=\min\limits_{t\in[0,\tau]}\omega_q(t)$ and $\omega_q^\text{max}=\max\limits_{t\in[0,\tau]}\omega_q(t)$ as shown for an explicit example in Fig.~\ref{fig:toymodel_plots}(c). 

Importantly, this driving continues to suppress the zero-frequency filter function. As a consequence, instantaneous tracking of the Larmor frequency performs significantly better than driving at the mean Larmor frequency as can be seen in Figs.~\ref{fig:varying_larmor}(b) and~\ref{fig:varying_larmor}(d) where one can find an order of magnitude improvement to the fidelity between the two driving regimes at variations as low as $10\%$. This comes at the cost of more stringent hardware and architectural constraints.

Although less detrimental than for constant driving, large variations in the Larmor frequency can also reduce the efficiency of the tracking protocol. In particular, when $\omega_q(t)$ becomes small, the associated peak in the filter function broadens towards low frequencies, where the noise power spectral density is usually the largest. Moreover, when $\omega_q(t)\sim\Omega$, or varies too rapidly in time, beyond-RWA effects emerge~\cite{Zeuch2020,RimbachRuss2023}, leading to imperfect Rabi rotations and thus reduced decoupling performance.
\section{Shuttling of hole-spin qubits}\label{sec:model}
After having discussed continuous dynamical decoupling in a general driven two-level system, we now turn to a more realistic description of the shuttling of gate-defined quantum dot spin qubits. The Hamiltonian of a particle with spin 1/2 moving in a conyevor-mode shuttling potential can be modeled as~\cite{langrockBlueprintScalableSpin2023,Bosco2024}
\begin{align}\label{eq:moving_particle_Hamiltonian}
\begin{aligned}
    H=&\frac{p_x^2}{2m}+\frac{p_y^2}{2m}+V(x-x_0(t),y)
    +\tfrac{1}{2}\mu_B\boldsymbol \sigma\cdot \hat g \boldsymbol B.
\end{aligned}
\end{align}
Here, the first two terms are the kinetic energy with effective mass $m$, $V(x,y)$ is the confining potential along in-plane directions, and the last term is the Zeeman field, where we have introduced the effective $\hat g$-tensor. The displacement of the quantum dot by the conveyor potential is captured by the time-dependent position $x_0(t)=v t$, with velocity $v$. This Hamiltonian can be used to model moving electrons or holes. We focus here on holes and consider a $\hat g$-tensor that depends on the confinement because of strong spin-orbit interaction (SOI). Such strong SOI is usually present in hole spin systems, e.g. planar germanium quantum dots, Ge/Si core/shell nanowires or silicon field-effect transistors~\cite{kloeffelProspectsSpinBasedQuantum2013,Scappucci2020,PhysRevLett.129.247701,PhysRevResearch.3.013081,PRXQuantum.2.010348,Bassi2025,Yu2023}. Furthermore, due to SOI the effective $\hat g$-tensor in hole-spin based platforms  contains terms proportional to quadratic momenta~\cite{spindependentmass,PhysRevB.106.235408,PhysRevLett.129.247701}. Considering in-plane harmonic confinement and expanding the Zeeman contribution, we can write the Hamiltonian as
\begin{align}\label{eq:Moving_QD_Hamiltonian_with_SOC}
\begin{aligned}
H=&\frac{p_x^2}{2m}\left(\mathbbm1_2+B\boldsymbol\nu\cdot\boldsymbol\sigma\right)+\frac{m\omega_x^2}{2}(x-vt)^2\\
&+\frac{p_y^2}{2m}\left(\mathbbm1_2+B\boldsymbol\mu\cdot\boldsymbol\sigma\right)+\frac{m\omega_y^2}{2}y^2+H_z^{(0)}(x)\\
&+\frac{p_xp_y}{2m}B\boldsymbol\xi\cdot\boldsymbol\sigma.
\end{aligned}
\end{align}
 The vectors $\boldsymbol{\nu}$, $\boldsymbol{\mu}$, and $\boldsymbol{\xi}$ encode SOI originating from the $\hat g$-tensor and depend on the magnetic-field orientation, the shuttling direction relative to the crystallographic axes, strain, and out-of-plane electric fields. Explicit expressions for these vectors can be found in App.~\ref{app:Explicit formula of spin-orbit coefficients} focusing on planar germanium. The term $H_z^{(0)}(x)$ collects Zeeman contributions that are independent of the momentum operators. In particular, it incorporates position-dependent variations of the strain components~\cite{strainmapping}, which give rise to spatial variations of the Larmor frequency along the shuttling lane. All terms can be derived from the Luttinger-Kohn-Bir-Pikus Hamiltonian, which provides an accurate description of strained, gate-defined hole-spin qubits~\cite{Terrazos_2021}, see App.~\ref{app:Explicit formula of spin-orbit coefficients} for explicit expressions of $H_z^{(0)}(x)$.

In what follows, we investigate time-dependent periodic modulation of the orbital frequency $(\omega_x,\omega_y)\rightarrow(\omega_x(t),\omega_y(t))$ in Sec.~\ref{sec:time-modulated_breathing_shuttle} as well as a protocol with spatial periodicity of the $y$-confinement while fixing the $x$-confinement $(\omega_x,\omega_y)\rightarrow(\omega_x,\omega_y(x))$ in Sec.~\ref{sec:spatially-modulated_breathing_shuttle} connecting the shuttling protocols to a driven 2-level Hamiltonian. We further restrict ourselves to a homogeneous Zeeman field $H_z^{(0)}(x)\equiv H_z^{(0)}$ for the numerical simulations. For more general cases, we refer to Sec.~\ref{sec:Varying_Larmor_frequqency}, where we quantify the efficiency of continuous dynamical decoupling in presence of time-varying Larmor frequencies.

\section{Temporal breathing potential}\label{sec:time-modulated_breathing_shuttle}
\subsection{Effective spin Hamiltonian}\label{sec:time_modulated_breathing_potential_effective_spin_hamiltonian}
We now derive an effective spin Hamiltonian from Eq.~\eqref{eq:Moving_QD_Hamiltonian_with_SOC} under the assumption of time-modulated orbital frequencies. Going first into the co-moving frame using the translation operator $T=\exp(-ivtp_x/\hbar)$ , we transform the position operator to $x\rightarrow x+vt$ and introduce a dynamic term $i\hbar T^\dagger T=-vp$. The frames $R_\alpha(t)=\exp\left(-\frac{i}{2\hbar}\ln\sqrt{\frac{\omega_{\alpha,0}}{\omega_\alpha(t)}}\{\alpha,p_\alpha\}\right)$ squeeze and lengthen the $\alpha=x$ and $\alpha=y$ coordinates in time, matching at each time the squeezing and lengthening of the lateral confinement in the original frame. Details of the unitary transformations are provided in App.~\ref{app:Unitary_transformations}. Applying both frames in either order, since $[R_x(t),R_y(t)]=0$, yields the Hamiltonian
\begin{align}\label{eq:Moving_QD_Hamiltonian_with_SOC_after_frames_temporal}
    \tilde H=&\sum_{\alpha=x,y}\frac{\omega_\alpha(t)}{\omega_{\alpha,0}}\left(\frac{p_\alpha^2}{2m}+\frac{m\omega_{\alpha,0}^2}{2}\alpha^2\right)+\frac{\dot\omega_{\alpha}(t)}{4\omega_\alpha(t)}\{\alpha,p_\alpha\}\nonumber\\
    &+\frac{\omega_x(t)}{\omega_{x,0}}\frac{p_x^2}{2m}B\boldsymbol\nu\boldsymbol\sigma+\frac{\omega_y(t)}{\omega_{y,0}}\frac{p_y^2}{2m}B\boldsymbol\mu\boldsymbol\sigma\\
    &+\sqrt{\frac{\omega_x(t)\omega_y(t)}{\omega_{x,0}\omega_{y,0}}}\frac{p_xp_y}{2m}B\boldsymbol\xi\cdot\boldsymbol\sigma+H_z^{(0)}-\sqrt{\frac{\omega_x(t)}{\omega_{x,0}}}vp_x.\nonumber
\end{align}
Switching into these frames allows us to work in the basis of time-independent harmonic oscillators. Therefore, in the two-dimensional orbital harmonic product basis $\ket{n_x,n_y}$ with frequencies $\omega_{x,0}$ and $\omega_{y,0}$ the matrix elements of $\{\alpha,p_\alpha\}$ follow a simple selection rule, changing the orbital quantum number by 2. In particular between the ground orbital state and any other orbital state the matrix elements of $\{\alpha,p_\alpha\}$ are non-zero only for $\bra{0_x,0_y}\{\alpha,p_\alpha\}\ket{2_x,0_y}\neq0$ if $\alpha=x$ and $\bra{0_x,0_y}\{\alpha,p_\alpha\}\ket{0_x,2_y}\neq0$ if $\alpha=y$. 

Assuming $\dot\omega_\alpha(t)/\omega_\alpha(t)\ll1$ and $B|\boldsymbol \nu|,B|\boldsymbol \mu|\ll1$, we find the Hamiltonian is approximately block diagonal in the orbital basis, leading to the effective spin Hamiltonian in the low-energy subspace
\begin{align}\label{eq:Effective_spin_Hamiltonian_breathing_protocol}
    \tilde H_\text{eff}=\frac{\hbar\omega_x(t)}{4}B\boldsymbol\nu\cdot\boldsymbol\sigma+\frac{\hbar\omega_y(t)}{4}B\boldsymbol\mu\cdot\boldsymbol\sigma+\frac{\hbar\tilde\omega_q}{2}\tilde{\boldsymbol n}\cdot\boldsymbol\sigma.
\end{align}
The term with $\sim p_xp_y\boldsymbol\xi\cdot\boldsymbol\sigma$ in Eq.~\eqref{eq:Moving_QD_Hamiltonian_with_SOC_after_frames_temporal} does not contribute to the effective Hamiltonian because of the assumption of separable confinements in $x$ and $y$ direction. Moreover, the condition $\dot\omega_\alpha(t)/\omega_\alpha(t)\ll1$ corresponds to an orbital adiabatic limit and corresponds to negligible diabatic transitions between the instantaneous orbital eigenstates. The derivative is proportional to the driving $\dot \omega_\alpha\propto \omega_d$ that is on the order of the Zeeman splitting in recent experiments and our proposed protocol. Since the Zeeman splitting is much smaller than the orbital splittings for spin qubits, i.e., $\omega_d\ll \omega_\alpha(t)$, the adiabatic limit is typically met in current experiments and we can safely neglect diabatic corrections. We note that Eq.~\eqref{eq:Effective_spin_Hamiltonian_breathing_protocol} corresponds to the zeroth-order expression in the time-dependent Schrieffer-Wolff approximation. While first-order correction vanish, non-trivial correction can appear in second  or higher orders. 
Interestingly, if we include spatial variations in strain $H_z^{(0)}(x)$ or a constant linear or cubic spin-orbit coupling, i.e. a coupling of the form $p_\alpha \sigma$ or $p_\alpha^3 \sigma$~\cite{Terrazos_2021}, we obtain non-trivial first-order corrections. These additional terms can modify the time-dependent Zeeman field $\tilde{\omega}_q$ in Eq.~\eqref{eq:Effective_spin_Hamiltonian_breathing_protocol}.

\subsection{Noise models}
So far we limited our analysis to an ideal moving spin. We now focus on the impact of different noise environments on the dynamics of the shuttled spin. We remark that depending on the type of noise, the unitary transformations performed in the previous section to derive the spin evolution, may not commute with the noise Hamiltonian leading to additional dynamics acting on the moving spin.

\subsubsection{Global magnetic noise sources}\label{sec:temporal_long_range_magnetic_noise_sources}
For a global fluctuating magnetic field, the noise Hamiltonian is 
\begin{align}\label{eq:Effective_Noise_Hamiltonian_Global_magnetic_noise_temporal_breathing}
    H_{N}^\text{MF}=\boldsymbol\beta(t)\boldsymbol\cdot\sigma \ .
\end{align}
This Hamiltonian  commutes with the orbital transformations applied in the previous section and the free-induction decay filter function in Eq.~$\eqref{eq:filterfunction_Global_Noise}$ describes well the qubit's response to such noise sources.

\subsubsection{Local magnetic noise sources}\label{sec:temporal_local_magnetic_noise_sources}
If the  magnetic noise sources are local, e.g. the noise is caused by localized nuclear spins, the dynamics of the shuttled particle influences the noise experienced by the spin. 
To show this, we consider a local magnetic noise Hamiltonian in the lab frame of our system with atomic density $n_0$
\begin{align}\label{eq:Untransformed_Noise_Hamiltonian_Local_Magnetic_noise_}
    H_N=&\frac{1}{2n_0}\sum_kG(x-x_k,\ell_c)G(y-y_k,\ell_c)\boldsymbol \beta_k(t)\cdot\boldsymbol\sigma,\\
    G(x,\ell_c)=&\exp(-x^2/\ell_c^2)/\sqrt{2\pi}\ell_c.
\end{align}
Each noise source is modeled by a Gaussian envelope function centered at coordinates $(x_k,y_k)$. All  envelopes are assumed to have the same spatial correlation length $\ell_c$ for simplicity, and are weighted by individual stochastic, time-dependent variables $\boldsymbol\beta_k(t)$. Accounting for the unitary transformations introduced in Sec.~\ref{sec:time_modulated_breathing_potential_effective_spin_hamiltonian},
i.e., $T$, $R_x$, and $R_y$ (in this order), and correspond to translations and rescalings of the orbital coordinates, the noise Hamiltonian $H_N$ transforms into $\tilde H_N$ by the substitutions $x \rightarrow {\ell_x(t)\,x/}{\ell_{x,0}} + x_0(t)$ and $y \rightarrow {\ell_y(t)}\,y/{\ell_{y,0}}$. Projecting the noise Hamiltonian in the effective low-energy spin-basis leads to
\begin{align}\label{eq:temporal_effective_noise_hamiltonian_local_magnetic_noise}
\bra{0_x,0_y}\tilde H_N\ket{0_x,0_y}\equiv \sum_k\mathcal A(x_k,y_k,t)\boldsymbol \beta_k(t)\cdot\boldsymbol\sigma,
\end{align}
where the renormalized noise amplitude
\begin{subequations}
\begin{align}
    \mathcal A(x_k,y_k,t)=&\frac{1}{2\pi n_0D_x(t)D_y(t)}e^{-\frac{(x_k-x_0(t))^2}{D_x^2(t)}-\frac{y_k^2}{D_y^2(t)}}\\
    D_{x,y}(t)=&\sqrt{2\ell_c^2+\ell_{x,y}^2(t)}
\end{align}
\end{subequations}
emerges from the projection of the orbital harmonic ground state $\langle x,y|0_x,0_y\rangle=e^{-{x^2}/{2\ell_{x,0}^2}-{y^2}/{2\ell_{y,0}^2}}/{\sqrt{\pi\ell_{x,0}\ell_{y,0}}}$. The term $\mathcal A(x_k,y_k,t)$ describes the spatial overlap between a noise source with Gaussian spatial correlation profile and the Gaussian instantaneous orbital ground-state wavefunction of width $\ell_x(t)$ and $\ell_y(t)$. The effect of noise sources far from the coordinates $x=x_0(t),y=0$ is suppressed as a Gaussian function.

In contrast to global magnetic noise, the two-point noise correlator involves products of the projected noise amplitudes $\mathcal A$ evaluated at times $t_1$ and $t_2$. Assuming for simplicity that the noise sources are isotropic in spin space and mutually uncorrelated, that is $\langle\boldsymbol\beta_k^{(i)}(t_1)\boldsymbol\beta_{k'}^{(j)}(t_2)\rangle=\delta_{kk'}\delta_{ij}\int d\omega\, e^{-i\omega(t_2-t_1)}S(\omega)/2\pi$, allows us to formulate a general decoupling framework that depends only on the amplitude of the noise. This is a worst case scenario, in practice, however, noise is often anisotropic. For example, in planar hole-spin systems in Ge the hyperfine coupling to nuclear spins is characterized by a dominant out-of-plane contribution~\cite{Bosco2021,Fischer2008}.

Taking the continuum limit $\sum_kf(x_k,y_k)\rightarrow \nu n_0\int\int dxdyf(x,y)$ where $\nu$ denotes the average percentage of defects, the filter function for local magnetic noise can be expressed as~\cite{Bosco2024}
\begin{widetext}
\begin{subequations}
\begin{align}\label{eq:Matrix_of_filter_function_local_noise}
    F_L(\omega,\tau)=&\int_0^\tau\int_0^\tau dt_1dt_2K(t_1,t_2)\hat R_c(t_1)\hat R_c^T(t_2)e^{-i\omega(t_1-t_2)},\\
    K(t_1,t_2)=&\nu n_0\int\int dxdy \mathcal A(x,y,t_1)\mathcal A(x,y,t_2)  =\frac{\nu\exp\left(-\frac{(x_0(t_1)-x_0(t_2))^2}{D_x^2(t_1)+D_x^2(t_2)}\right)}{4n_0\pi\sqrt{(D_x^2(t_1)+D_x^2(t_2))(D_y^2(t_1)+D_y^2(t_2))}}.\label{eq:Kernel_temporal_breathing_local_magnetic_noise} 
\end{align}
\end{subequations}
\end{widetext}
In comparison to the filter function $F$ in Eq.~\eqref{eq:filterfunction_Global_Noise}, valid for global magnetic noise, $F_L$ in Eq.~\eqref{eq:Matrix_of_filter_function_local_noise} acquires a kernel $K_M(t_1,t_2)$ accounting for the spatial dependent renormalization of individual noise source. For noise correlation lengths $\ell_c$ much longer than the total shuttled distance, the kernel is approximately time-independent and reduces, as expected, to $F$ in Eq.~\eqref{eq:filterfunction_Global_Noise}. Furthermore, the time dependence of the confinement lengths $\ell_{x,y}(t)$ causes the spatial overlap with the noise sources to vary in time, and introduces oscillatory components in the kernel $K_M(t_1,t_2)$ at the  frequencies modulating the confinement. Consequently, even static noise in the lab frame is perceived as a time-dependent noise by the spin. In Fourier space, the kernel carries significant weight at these frequencies, which overlap with the peaks of the unmodified filter function $F$ computed as in Eq.~\eqref{eq:filterfunction_Global_Noise}. Their interplay redistributes spectral weights, leading to a small but finite contribution near $\omega \approx 0$ in the filter function $F_L$. Thus, strong driving amplitudes in orbital frequencies relative to their dc component may limit the efficiency of the decoupling schemes against local magnetic noise.

\subsubsection{Global electric noise sources}\label{sec:temporal_long_range_electric_noise_sources}
A global charge noise $H_N=xE_x(t)+yE_y(t)$, with stochastic random variables $E_x(t)$ and $E_y(t)$, requires SOI to  couple to the spin. Under the assumption of weak spatial dependence of the effective Zeeman field $\boldsymbol b(x,y)=\hat g(x,y)\boldsymbol B(x,y)$, we can linearize around the instantaneous quantum dot position as
\begin{align}\label{eq:temporal_linearized_effective_magnetic_field}
\boldsymbol b(x+vt,y)\approx \boldsymbol b(vt,0)+\sum_{\alpha=x,y}\sqrt{\frac{\omega_{\alpha,0}}{\omega_\alpha(t)}}\alpha\partial_\alpha \boldsymbol b(vt,0),
\end{align}
after the unitary transformations used in Sec.~\ref{sec:time_modulated_breathing_potential_effective_spin_hamiltonian}. Importantly, within a first order Schrieffer-Wolff transformation the terms linear in $x$ and $y$ yield a non-trivial contribution in the effective noise Hamiltonian which reads
\begin{align}\label{eq:temporal_effective_hamiltonian_long_range_electric_noise}
    \tilde H_{N,\text{eff}}(t)=\sum_{\alpha=x,y}k_\alpha(t)\boldsymbol \beta_\alpha(t)\boldsymbol\sigma.
\end{align}
with $k_\alpha(t)=\omega_{\alpha,0}^2/\omega^2_\alpha(t)$ and $\boldsymbol\beta_\alpha(t)={E_\alpha(t)}\boldsymbol b(vt,0)/{m\omega_{\alpha,0}}$; $\omega_{\alpha,0}$ is defined as the dc component of the oscillating orbital frequency in $\alpha=x,y$, respectively. The noise vectors $\boldsymbol\beta_\alpha(t)$ thereby incorporate both the stochastic electric fields $E_\alpha(t)$ and SOI induced by the field gradients. A similar effective Hamiltonian can be obtained for intrinsic SOI linear in the momenta $p_\alpha$, which requires a second-order Schrieffer-Wolff transformation. In both cases, the coupling retains the same scaling $k_\alpha(t)\propto 1/\omega_\alpha^2(t)$.

In analogy to the undriven case where $\omega_\alpha(t)=\omega_{\alpha,0}=\text{const.}$,  we assume that the error vector $\boldsymbol \beta_\alpha(t)$ is wide-sense stationary $\langle\boldsymbol\beta_\alpha(t_1)\boldsymbol\beta_\alpha(t_2)\rangle=\langle\boldsymbol\beta_\alpha(t_1-t_2)\boldsymbol\beta_\alpha(0)\rangle$ and has a power spectral density $S_\alpha(\omega)\propto 1/\omega$. We remark that the coherent oscillations of the spin state at different times are not a function of the time-difference alone and can therefore not be incorporated in the power spectral density $S$. However, their effects can be accounted for in the filter function~\cite{Bosco2024} by adding a kernel $k(t)=k_x(t)+k_y(t)$ to Eq.~\eqref{eq:filterfunction_Global_Noise}. If the noise vectors $\boldsymbol\beta_{x}$ and $\boldsymbol\beta_{y}$ are uncorrelated, the modified filter function reads
\begin{align}\label{eq:temporal_filter_function_global_electric_noise}
    F_\text{El}(\omega)=\int_0^\tau\int_0^\tau dt_1dt_2k(t_1)\hat R_c(t_1)k(t_2)\hat R_c^T(t_2)e^{-i\omega(t_1-t_2)}.
\end{align}
Under the realistic assumption that the quantum dot is more strongly confined in the $y$-direction, we can approximate $k(t)\approx k_x(t)$, as the $y$-orbital contribution is suppressed by the larger gap $\hbar\omega_y(t)$ (i.e., $k_y\ll k_x$). If, in addition, the modulation of the confinement occurs only along the $y$-direction and the confinement remains elliptically shaped, then $k_x(t)$ remains constant, and we obtain $k(t)\approx k_x(t)=\text{const.}$, thereby recovering the filter function in Eq.~\eqref{eq:filterfunction_Global_Noise} found for global magnetic noise.

We note that noise produced by a global electric field along the out-of-plane direction lead to a contribution $\boldsymbol\beta_z(t)$, which may also acquire a time-dependent prefactor if the confinement in the growth direction is modulated. In practice, this contribution is typically negligible due to the stronger confinement and larger orbital gap in the out-of-plane direction. In this work, we therefore restrict ourselves to in-plane confinement modulation. Furthermore, we assume that the noise sources $\beta_\alpha(t)$  are not altered by the drive. The additional kernel $k(t)$ only reflects the time-dependent susceptibility of the effective spin Hamiltonian to electric noise arising from the modulation of the orbital energy spacing $\omega_\alpha(t)$.

Lastly, if the qubit is dressed~\cite{Laucht2016,tsoukalas2025dressedsinglettripletqubitgermanium} near resonance, the kernel $k(t)$ acquires significant spectral weight at the driving frequency, which overlaps with the dominant frequencies of the filter function. In analogy to local magnetic noise, this overlap redistributes spectral weight towards low frequencies, leading to a finite contribution near $\omega\approx 0$ and thus reducing the efficiency of continuous dynamical decoupling if the drive of the confinement is too strong.

\subsubsection{Local electric noise sources}\label{sec:temporal_local_electric_noise_sources}
To investigate the impact of local electric noise sources, we consider the following noise Hamiltonian
\begin{align}\label{eq:Untransformed_Noise_Hamiltonian_Local_Electric_noise_}
    H_N^\text{EL,L}=\frac{1}{2n_0}\sum_kG_{\ell_c}(x-x_k)G_{\ell_c}(y-y_k) \beta_k(t).
\end{align}
The kernel for local electric noise can again be derived with a Schrieffer-Wolff approximation to find an effective non-trivial noise Hamiltonian in spin space. The calculation requires assuming a specific form of the SOI. For instance, a small gradient in the effective Zeeman field $\boldsymbol b(x,y)=\hat g(x,y)\boldsymbol B(x,y)$ can be linearized around the instantaneous quantum dot position $(vt,0)$ as  in Eq.~\eqref{eq:temporal_linearized_effective_magnetic_field}. With this assumption, we find a spin-dependent effective noise Hamiltonian with renormalized noise modulation $\mathcal A$ and stochastic noise amplitudes $\boldsymbol\beta_{\alpha,k}\sim\partial_\alpha\boldsymbol b(vt,0)\beta_k(t)$. Including this renormalization, we find that Eq.~\eqref{eq:Kernel_temporal_breathing_local_magnetic_noise} is modified by the process-dependent kernels 
\begin{align}\label{eq:Kernel_temporal_breathing_local_electric_noise}
\begin{aligned}
K_{E,\alpha}&(t_1,t_2)=\,K_M(t_1,t_2)
\frac{\omega_{\alpha,0}^4}
{\omega_\alpha^2(t_1)\omega_\alpha^2(t_2)}\\
&\times
\begin{cases}
\dfrac{D_x^2(t_1)+D_x^2(t_2)-2(vt_1-vt_2)^2}
{(D_x^2(t_1)+D_x^2(t_2))^2}, & \alpha=x,\\[6pt]
\dfrac{1}{D_y^2(t_1)+D_y^2(t_2)}, & \alpha=y .
\end{cases}
\end{aligned}
\end{align}
Again, we used $\omega_{\alpha,0}$ for the dc component of the oscillating orbital frequency for $\alpha=x,y$, respectively.

The kernels $K_{E,x}$ and $K_{E,y}$ describe noise sources that couple to the spin via orbital states along the shuttling direction and the in-plane perpendicular direction, respectively. When spatial correlations are long, $\ell_c\gg \ell_x(t),\ell_y(t),L_s$, where $L_s$ denotes the total shuttled distance, the kernel scales as $\propto 1/\ell_c^4\omega_\alpha^2(t_1)\omega_\alpha^2(t_2)$ with $\alpha=x,y$. 
In this case, the filter function approaches our results for global electric noise (see Sec.~\ref{sec:temporal_long_range_electric_noise_sources}), because the Gaussian envelopes in Eq.~\eqref{eq:Untransformed_Noise_Hamiltonian_Local_Electric_noise_} vary  across the device and can be approximated by linear functions, matching Eq~\eqref{eq:temporal_effective_hamiltonian_long_range_electric_noise}.

The kernel $K_{E,x}$ exhibits a distinct behavior. If we neglect the explicit time dependence of $\ell_x(t)$ and $\ell_y(t)$, $K_{E,x}$ depends only on the time difference, $K_{E,x}(t_1,t_2)\equiv K_{E,x}(t_1-t_2)$. In this case, the filter function can be written as a convolution of $\tilde K_{E,x}(\omega)$ with $F(\omega,\tau)$, where $\tilde K_{E,x}(\omega)$ is the Fourier transform of $K_{E,x}$ with respect to $t_1-t_2$ (see App.~\ref{app:Filter_function_formalism}). The function $\tilde K_{E,x}(\omega)$ vanishes at $\omega=0$ and exhibits a  double peak at $\omega=\pm\sqrt{2}\,v/D_x$. Consequently, the filter function inherits this structure and produces pairs of peaks around the frequencies where $F(\omega,\tau)$ is maximal. While this derivation neglected the time-dependence of the confinement lengths, we find by numerical simulations that  the filter function  inherits  double peak features even when the confinement is  time-modulated.

The filter function peak splitting has a simple physical origin that can be understood by considering a quasi-static local electric noise source in a region with a linear magnetic-field gradient. Depending on the sign of the electric field of the defect, the quantum dot can be pulled toward or away from the noise source. Because of SOI this displacement modifies the effective Zeeman field experienced by the spin. As the quantum dot passes the noise source, the displacement reverses and the induced field shift changes sign. For slowly varying noise these two contributions compensate each other. The fluctuations add constructively only if the noise changes sign while the wavefunction traverses the noisy region, which  happens at frequencies $\sim\sqrt{2}\,v/D_x$. This provides an intuitive explanation for the double-peak  of the kernel $\tilde K_{E,x}(\omega)$ at $\omega=\pm\sqrt{2}\,v/D_x$.

When the Larmor vector is constant, the free-induction decay filter function $F(\omega,\tau)$ without kernel [Eq.~\eqref{eq:filterfunction_Global_Noise}] peaks at $\omega=0$ and the kernel $\tilde K_{E,x}(0)=0$ suppresses the central peak and shifts the dominant weight to frequencies $\sim\sqrt{2}\,v/D_x$. 

For resonant driving, if the Rabi frequency satisfies $\Omega\approx\sqrt{2}\,v/D_x$, a peak reappears at zero frequency in the modified filter function, reducing the efficiency of dynamical decoupling. Therefore, a more efficient suppression of this noise is achieved at Rabi frequencies $\Omega>\sqrt{2}\,v/D_x$. We find that this effect is a general feature of local electric noise sources that couple to the spin via SOI and may become relevant in other architectures where the confinement of the shuttled spin is elongated along the transport direction, such that the contribution of $K_{E,x}$ dominates over that of $K_{E,y}$ as $\omega_y\gg\omega_x$.

For quantum dots elongated along the shuttling direction, local noise hotspots can lead to fast, spatially localized dephasing and thus reduced $T_2^\ast$. In these cases, applying a $\pi$ pulse as the quantum dot traverses a hotspot can modify the accumulated phase. Depending on whether the noise is of magnetic or electric origin, this can either suppress or enhance dephasing, leading to an increase or decrease of the coherence time $T_2$, respectively. Thus, such an experiment may allow one to identify the dominant coupling nature of local hotspots.
\subsection{Numerical simulations}
\begin{figure*}
    \centering
    \includegraphics[width=1\linewidth]{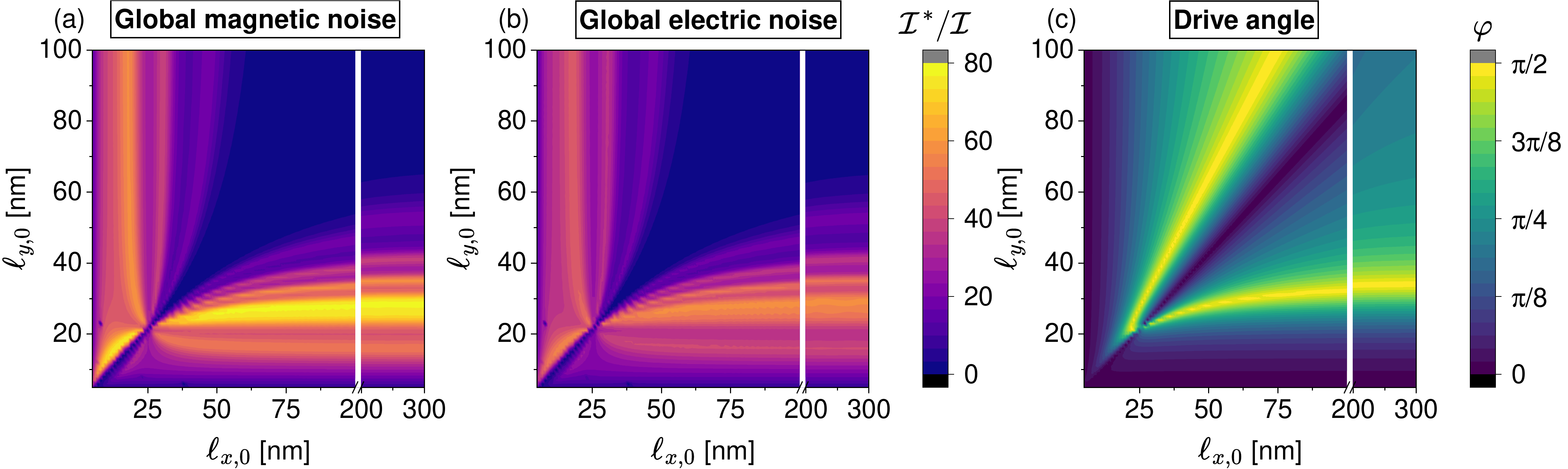}
    \caption{\textbf{Temporal breathing shuttle with uniform driving.} (a)-(b) Ratio $\mathcal I^\ast/\mathcal I$ of the infidelity without time-dependent confinement modulation ($\mathcal I^\ast$) to that under driving $\omega_\alpha(t)=\omega_{\alpha,0}(1+\gamma\sin\omega_dt)$ ($\mathcal I$) at a fixed relative driving amplitude $\gamma=0.05$. We consider a total evolution time $1\,\mu\text{s}$ while sweeping different confinement lengths $\ell_{\alpha,0}=\sqrt{\hbar/m\omega_{\alpha,0}}$ ($\alpha=x,y$) and assume (a) global magnetic noise and (b) global electric noise. (c) Relative angle $\varphi$ between the Larmor vector and the drive for different confinements $\ell_{\alpha,0}$. Small angles $\varphi$ correspond to driving fields aligned to the static Zeeman field, leading to inefficient driving and decoupling from quasi-static noise. We choose a total evolution time $1\,\mu\text{s}$. Asymmetries arise from the chosen in-plane magnetic field $|\boldsymbol B|=0.1\,\text{T}$ direction and shuttling trajectory enclosing different angles with the crystallographic in-plane axes, see App.~\ref{app:Explicit formula of spin-orbit coefficients}.}
    \label{fig:temp_uniform}
\end{figure*}
We now consider constant shuttling velocities $x_0(t)=vt$ and numerically compute the average gate infidelity of different driving protocols using Eq.~\eqref{eq:average_gate_infidelity_FF_formalism}. By modulating the amplitudes of all gate voltages applied to all top gates along the shuttling lane (see Fig.~\ref{fig:model}), the orbital frequencies can be driven uniformly as
\begin{align}
    \omega_\alpha(t)=\omega_{\alpha,0}(1+\gamma\sin\omega_d t)
\end{align}
with a dimensionless driving amplitude $\gamma<1$. In practice $\gamma$ can depend on position because of local differences in the lever arm caused by static charges. Here, we neglect this effect for simplicity and we refer to App.~\ref{app:Varying_Rabi_frequency} for more details on the effect of varying Rabi frequencies.

To find the optimal driving frequency $\omega_d$, we use the positive and real-valued solution of Eq.~\eqref{eq:general_resonance_frequency_condition} that accounts for the Bloch-Siegert shift. As the figure of merit for the decoupling effectiveness of our protocols we use the ratio of the average gate infidelity without drive $\mathcal I^\ast$ and the average gate infidelity with drive $\mathcal I$ using Eq.~\eqref{eq:average_gate_infidelity_FF_formalism} and formulas of the modified filter functions for the respective kind of noise. If the driving direction is parallel to the Larmor vector, the decoupling protocol is not effective as the spin does not precess. This is illustrated in Fig.~\ref{fig:temp_uniform}, in which we consider uniform driving at different combinations for $\ell_{\alpha,0}$ which relates to the time-averaged orbital frequencies $\ell_{\alpha,0}=\sqrt{\hbar/m\omega_{\alpha,0}}$ with $\alpha=x,y$. In particular, Figs.~\ref{fig:temp_uniform}(a) and~\ref{fig:temp_uniform}(b) show no enhancement of the coherence time when the static Larmor frequency is aligned to the driving field. We examine this further in Fig.~\ref{fig:temp_uniform}(c)  and introduce the angle $\varphi$  between the static Zeeman and driving field; at $\varphi=0$, the drive is parallel to the Larmor vector and does not induce spin rotations. This angle depends on system and device parameters such as the orientation of the shuttling direction relative to the crystallographic axes $\theta_s$, strain anisotropy $\varepsilon_{xx}-\varepsilon_{yy}$, magnetic-field direction $\boldsymbol B$, and the confinement lengths $\ell_{x,0}$ and $\ell_{y,0}$; the explicit expression of $\varphi$ is given in Eq.~\eqref{eq:angle_uniform_driving} in App.~\ref{app:Explicit formula of spin-orbit coefficients}.

Furthermore, increasing the driving strength improves noise suppression, provided the condition $\Omega\lesssim\omega_d$ holds. This is because a larger Rabi frequency shifts the first peak of the filter function to higher frequencies, thereby reducing its overlap with low-frequency noise. Since the Rabi frequency $\Omega$ scales with $\gamma\omega_{x,0}$ and $\gamma\omega_{y,0}$, increasing the confinement strengths $\omega_{x,0}$ and $\omega_{y,0}$, i.e., considering a tighter dot, enhances the driving amplitude even at fixed relative modulation $\gamma$.

For local noise sources, the peaks of  the filter functions have widths $\sim v/\sqrt{4\ell_c^2+2\ell_{x,0}^2}$~\cite{Bosco2024}. To efficiently suppress low-frequency noise, the lowest-frequency peak of the filter function must be shifted beyond the characteristic bandwidth. This requires $\Omega\gtrsim v/\sqrt{4\ell_c^2+2\ell_{x,0}^2}$. In other words, the qubit performs spin rotations on a timescale shorter than the time it spends within a given local noise region. As a result, the coherent drive averages out the impact of individual noise sources, providing additional decoupling beyond motional narrowing that arises from spatial averaging during shuttling.

Fig.~\ref{fig:temp_lateral} shows the enhanced fidelity as a function of $\gamma$ and $\ell_{y,0}$ using lateral driving, where only one orbital direction is modulated,
\begin{align}
    \omega_y(t)=\omega_{y,0}(1+\gamma\sin\omega_d t),
\end{align}
while $\omega_x(t)=\omega_x$ remains constant. This protocol may be implemented, for instance, by applying a sinusoidal signal to the screening gates as suggested in Fig.~\ref{fig:model}. We consider local magnetic [Figs.~\ref{fig:temp_lateral}(a) and (c)] and electric noise [Figs.~\ref{fig:temp_lateral}(b) and (d)] of varying spatial correlation. For comparison, we also plot the corresponding global noise case. In particular, increasing the driving amplitude $\gamma$ is generally beneficial when the qubit is mostly subjected to global magnetic noise. For both global electric noise and local (electric or magnetic) noise, increasing $\gamma$ beyond a threshold reduces the coherence. This occurs because the noise kernels develop a peak at the driving frequency $\omega_d$, where the bare filter function also peaks as discussed in Secs.~\ref{sec:temporal_local_magnetic_noise_sources} and~\ref{sec:temporal_local_electric_noise_sources}. Since the effective filter function is given by a convolution of the kernel with the bare filter function, this overlap increases the susceptibility of the system to previously quasi-static noise sources. Consequently, the interplay of faster driving and higher susceptibility to quasi-static noise at increasing $\gamma$ leads to a non-trivial dependence of the enhanced coherence time on $\gamma$.

\section{Spatial breathing shuttle}\label{sec:spatially-modulated_breathing_shuttle}
\subsection{Effective spin Hamiltonian}\label{sec:spatial_effective_spin_Hamiltonian}
Starting from the Hamiltonian in Eq.~\eqref{eq:Moving_QD_Hamiltonian_with_SOC} where we replace $\omega_y\rightarrow\omega_y(x)$ to account for the spatial dependence in the $y$-confinement, we proceed similarly as in Sec.~\ref{sec:time_modulated_breathing_potential_effective_spin_hamiltonian}. First, we apply the translation operator $T$ which introduces time dependence in the angular frequency along the $y$-direction $\omega_y(x)\rightarrow\omega_y(x+vt)$. Using the squeezing operator $R=\exp\left(-\frac{i}{2\hbar}\ln\sqrt{\frac{\omega_{y,0}}{\omega_y(x+vt)}}\{y,p_y\}\right)$ and projecting to the orbital ground state yields (see App.~\ref{app:Unitary_transformations})
\begin{align}\label{eq:effective_Hamiltonian_spatial_breathing}
    \tilde H_\text{eff}=\frac{\hbar}{16}\langle\omega_y(\partial_x\ell_y)^2\rangle_xB\boldsymbol\nu\cdot\boldsymbol\sigma+\frac{\hbar\langle\omega_y\rangle_x}{4}B\boldsymbol\mu\cdot\boldsymbol\sigma+H_z^{(0)}.
\end{align}
To improve readability, we omitted the explicit spatial and temporal dependence and write $\omega_y\equiv\omega_y(x+vt)$. We further use $\langle \cdot \rangle_x$ to denote the expectation value in the ground state of the harmonic oscillator along the $x$-direction with angular frequency $\omega_x$. 
By taking the expectation value we integrate out the $x$-dependence of $\omega_y(x+vt)$ and reduce the expression into an effective spin Hamiltonian.

Assuming spatial periodicity $\omega_y(x+L)=\omega_y(x)$ [see Fig.~\ref{fig:model}(d)], both the first term $\langle\left(\partial_x\omega_y(x+vt)/\omega_y(x+vt)\right)^2\rangle_x$ and the second term $\langle\omega_y(x+vt)\rangle_x$ in Eq.~\eqref{eq:effective_Hamiltonian_spatial_breathing} are periodic in time with  period $2\pi L/v$. In this protocol, the driving frequency is directly related to the shuttling velocity via $\omega_d=v/L$. Because of the derivative taken for the first term in Eq.~\eqref{eq:effective_Hamiltonian_spatial_breathing}, the first two terms  generally oscillate out of phase. Consequently, when $\boldsymbol \nu\not\perp\boldsymbol\mu$ the effective Hamiltonian corresponds to a driven two-level system, with a  time-dependent direction of the drive in spin space. However, in the parameter regime used for the numerical simulations, the contribution of the first term in Eq.~\eqref{eq:effective_Hamiltonian_spatial_breathing} is much smaller than the second term. Therefore, the Hamiltonian approximately takes the form of a static field with a single driving tone.
\begin{figure}
    \centering
    \includegraphics[width=1\linewidth]{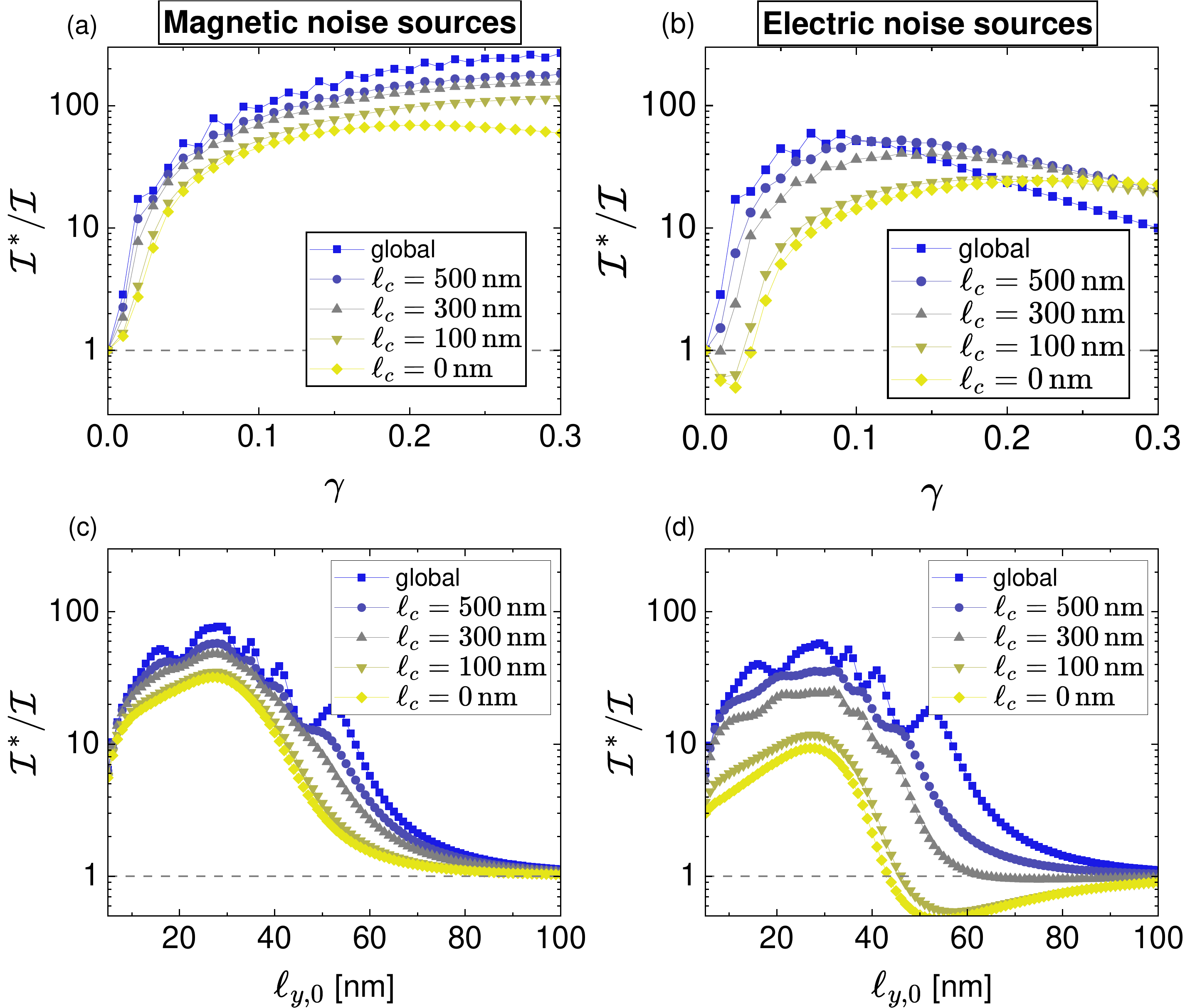}
    \caption{\textbf{Temporal breathing shuttle with lateral driving.} Ratio $\mathcal I^\ast/\mathcal I$ of the infidelity without time-dependent confinement modulation  ($\mathcal I^\ast$) to that under lateral driving $\omega_y(t)=\omega_{y,0}(1+\gamma\sin\omega_dt)$  ($\mathcal I$). Panel (a) [(b)] shows the dependence on the relative driving amplitude $\gamma$, while panel (c) [(d)] shows the dependence on the confinement length $\ell_{y,0}=\sqrt{\hbar/m\omega_{y,0}}$, assuming many local magnetic [electric] noise sources with spatial correlations $\ell_c=0\,\text{nm},100\,\text{nm},300\,\text{nm},500\,\text{nm}$. Values $\mathcal I^\ast/\mathcal I>1$ (above the gray dashed line) indicate improved coherence. For reference the infidelities obtained assuming global magnetic [electric] noise are also shown. We fix the confinement along shuttling direction $\ell_x=250\,\text{nm}$ to be constant, a total evolution time $1\,\mu\text{s}$, shuttling velocity $1\,\text{m}/\text{s}$, $\ell_{y,0}=\sqrt{\hbar/m\omega_{y,0}}=36\,\text{nm}$ for panels (a)-(b), and $\gamma=0.05$ for panels (c)-(d). Other parameters of this simulation are detailed in App.~\ref{app:Explicit formula of spin-orbit coefficients}.}
    \label{fig:temp_lateral}
\end{figure}

\subsection{Noise models}
If the system is dominated by global magnetic noise, the filter function is not modified by a kernel and simply reduces to the free-induction decay filter function in Eq.~\eqref{eq:filterfunction_Global_Noise}.
\subsubsection{Local magnetic noise sources}
Local magnetic noise sources as given in Eq.~\eqref{eq:Untransformed_Noise_Hamiltonian_Local_Magnetic_noise_} yield a non-trivial kernel in the filter function. To derive the kernel we use the unitaries provided in the previous section that transform the spatial coordinates as $(x,y)\rightarrow\left(x+v t,{\ell_y(x+v t)}y/{\ell_{y,0}}\right)$ on the bare noise Hamiltonian for local magnetic noise sources. Proceeding as in Sec.~\ref{sec:temporal_local_magnetic_noise_sources}, and assuming $\ell_c\rightarrow0$ (i.e., replacing the Gaussian envelopes by delta functions) together with $\omega_y(x)=\omega_{y,0}(1+\gamma\sin x/L)$, we expand to linear order in $\gamma$. This yields the following closed-form expression for the kernel
\begin{align}
    K_M(t_1,t_2)=&\frac{\nu\exp\left(-\frac{v^2(t_1-t_2)^2}{2\ell_x^2}\right)}{8n_0\pi\ell_x\ell_{y,0}}\left[1\right.\nonumber\\
    &+\left.\frac{\gamma}{2} e^{-\ell_x^2/8L^2}\sin\left(\frac{v(t_1+t_2)}{2L}\right)\right]+O(\gamma^2).
\end{align}

Again, because of the Gaussian component of the kernel $K_M(t_1,t_2)$, the resulting modified filter function  exhibits peaks of widths $\sim v/\ell_x$ even for long shuttling times. As previously mentioned, the locality of noise leads to natural improvements in coherence times due to motional narrowing. Further significant enhancements to the coherence are achieved by dressing the state with sufficiently large Rabi frequencies $\Omega\gg v/\ell_x$. 

We further find that the modulation induced by the spatial dependence of $\omega_y(x)$ introduces a driving frequency $v/L$, where $L$ denotes the spatial periodicity of the confinement modulation. Efficient dressing requires that the driving frequency, up to Bloch-Siegert corrections, matches the qubit splitting, i.e., $v/L\simeq\omega_q$. However, achieving sizable couplings in Eq.~\eqref{eq:effective_Hamiltonian_spatial_breathing} requires that the modulation is not averaged out over the spatial extent of the wavefunction. 

To see this, consider $\omega_y(x+vt)=\omega_{y,0}\left(1+\gamma\sin[(x+vt)/L]\right)$. Projecting onto the orbital ground state along $x$ as required in Eq.~\eqref{eq:effective_Hamiltonian_spatial_breathing} yields $\langle \omega_y \rangle_x = \omega_{y,0}\left(1+\gamma e^{-\ell_x^2/4L^2}\sin(vt/L)\right)$, such that the modulation amplitude is suppressed by a Gaussian factor for $L\gg\ell_x$. In this limit, the confinement varies slowly across the wavefunction and the effective driving becomes weak.

Combining these conditions, efficient coupling requires $L\sim \ell_x$, such that the resonance condition $v/L\simeq\omega_q$ implies $v/\ell_x\sim\omega_q$. Together with the requirement $\Omega\gg v/\ell_x$ for decoupling from local noise, this leads to $\Omega\gg\omega_q$, which is generally incompatible with near-resonant driving.

Finally, the reduced efficiency of this protocol in suppressing local noise is largely independent of the specific noise mechanism, as we will show in Sec.~\ref{sec:spatial_local_electric_noise_sources}. From a practical perspective, this approach can still be experimentally appealing, as it requires only simple control signals. However, its performance is contingent on high material quality, such that decoherence is dominated by global magnetic and electric field fluctuations rather than localized noise sources.

\subsubsection{Global electric noise sources}
For global electric noise sources, the effective noise Hamiltonian depends on the spin coupling mechanism. Electric noise sources in $x$-direction do not acquire additional time-dependencies and, thus, the effective Hamiltonian is analogous to global magnetic noise sources as given in Eq.~\eqref{eq:Effective_Noise_Hamiltonian_Global_magnetic_noise_temporal_breathing} for which the filter function formula in Eq.~\eqref{eq:filterfunction_Global_Noise} can be used. In contrast, electric noise sources coupling to $y$-orbitals, can be expressed by
\begin{align}
    H^\text{El}_N&=\int dx\frac{\exp[-x^2/\ell_x^2]}{\sqrt{\pi}\ell_x}\frac{\omega_{y,0}^2}{\omega_y^2(x+vt)}\boldsymbol{\beta}(t)\cdot\boldsymbol\sigma
\end{align}
The noise Hamiltonian acquires an additional time dependence that originates from the constant motion of the wavefunction together with the spatially dependent orbital frequency. Moreover, the effective coupling is obtained as a spatial average over the ground-state wavefunction in the $x$-direction, as reflected by the integral.

\subsubsection{Local electric noise sources}\label{sec:spatial_local_electric_noise_sources}
In analogy to local magnetic noise sources, strongly localized electric noise sources cannot be efficiently suppressed with this protocol beyond motional narrowing. The respective kernel exhibits filter function peaks of width $\sim v/\ell_x$. This broadening implies that efficient dynamical decoupling would require Rabi frequencies $\Omega\gg v/\ell_x$. At the same time, the driving must remain close to resonance with the qubit splitting, which is again usually incompatible with the condition $\Omega\gg v/\ell_x$.

In App.~\ref{app:Filter_function_formalism}, we report the kernel for noise coupling mediated by the $x$- and $y$-orbitals in the limit of small $\gamma$ and very local noise sources $\ell_c=0$. For noise sources that indirectly couple to the spin via $x$-orbitals we again expect the emergence of separate double peaks as found in Sec.~\ref{sec:temporal_local_electric_noise_sources} which is an effect irrespective of confinement modulation.

\subsection{Numerical simulations}
\begin{figure}
    \centering
    \includegraphics[width=1\linewidth]{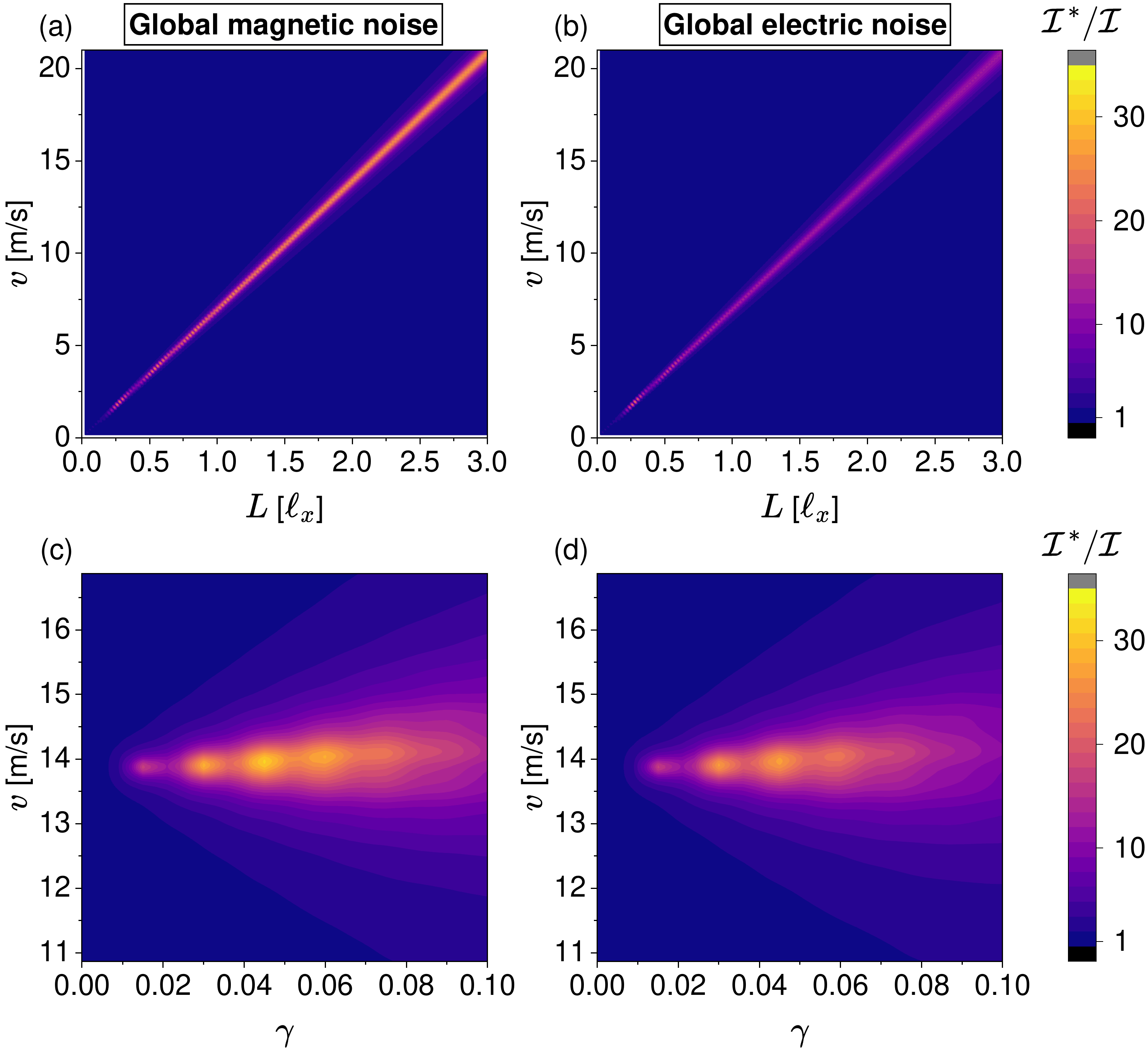}
    \caption{\textbf{Spatial breathing shuttle with lateral confinement modulation}. Ratio $\mathcal I^\ast/\mathcal I$ of the infidelity without spatially-dependent confinement  ($\mathcal I^\ast$) to that with spatially dependent $y$ confinement $\omega_y(x)=\omega_{y,0}(1+\gamma\sin x/L)$  ($\mathcal I$) against the shuttling velocity $v$ along the $y$ axis, using the protocol illustrated in Fig.~\ref{fig:model}(d). The $x$-axis in panel (a) [(b)] sweep the spatial periodicity $L$ of the screening gates, while panel (c) [(d)] sweep the relative driving amplitude $\gamma$ along the $x$-axis, assuming global magnetic [electric] noise. The infidelity ratio $\mathcal I^\ast/\mathcal I>1$ is greatly enhanced when matching $v/L$ to the Larmor frequency at which the system performs continuous dynamical decoupling. We fix the confinement in along shuttling direction $\ell_x=50\,\text{nm}$ to be constant, a total evolution time $1\,\mu\text{s}$, shuttling velocity $1\,\text{m}/\text{s}$, $\ell_{y,0}=\sqrt{\hbar/m\omega_{y,0}}=30\,\text{nm}$, $\gamma=0.03$ for panels (a)-(b), and $L=100\,\text{nm}$ for panels (c)-(d). Other parameters required for this simulation are detailed in App.~\ref{app:Explicit formula of spin-orbit coefficients}}
    \label{fig:spatial_plots}
\end{figure}
We  restrict ourselves to the simple example $\omega_y(x)=\omega_{y,0}(1+\gamma\sin x/\lambda)$. The relevant time-dependent term in Eq.~\eqref{eq:effective_Hamiltonian_spatial_breathing} is $\langle\omega_y(x+vt)\rangle_x=\omega_{y,0}[1+\gamma e^{-{\ell_x^2}/{4L^2}}\sin vt/L]$, where we observe a strong suppression of the amplitude of the driven component if  $L\ll\ell_x$. Furthermore, the term $\sim\boldsymbol\nu$ in Eq.~\eqref{eq:effective_Hamiltonian_spatial_breathing} is small in all of our numerical simulations relative to the other terms arising from being quadratic in $\gamma$ in the lowest non-trivial order.

Figs.~\ref{fig:spatial_plots}(a) and~\ref{fig:spatial_plots}(b) plot the infidelity $\mathcal I^\ast$ with constant confinement $\omega_y(x)=\omega_{y,0}$ divided by the infidelity $\mathcal I$ found with spatially modulated confinement $\omega_y(x)=\omega_{y,0}(1+\gamma\sin x/\lambda)$ against the shuttling velocity $v$ and spatial periodicity $L$ for global magnetic and electric noise, respectively. The infidelities were computed using Eq.~\eqref{eq:average_gate_infidelity_FF_formalism} as well as the filter functions of the respective noise sources. The line of improved fidelity follows the resonance condition $\omega_d=v/L\simeq\omega_L$, where the modulation-induced driving matches the Larmor frequency. For small spatial periodicity, $L\ll\ell_x$, the driving amplitude becomes exponentially suppressed, making the protocol ineffective for dressing the qubit.

Figs.~\ref{fig:spatial_plots}(c) and~\ref{fig:spatial_plots}(d) show the ratio of infidelities as a function of velocity and relative driving amplitude $\gamma$ for global magnetic and electric noises. Again, only  certain velocities enable decoupling from low-frequency noise and enhance the shuttling fidelity. Around $\gamma\sim 0.06$, the coupling frequency 
$\Omega=\omega_{y,0}\gamma e^{-\ell_x^2/4L^2}B|\boldsymbol{\mu}|/8$ 
becomes comparable to the Larmor frequency. In this regime, the qubit no longer undergoes complete rotations, reducing the suppression of low-frequency noise, as discussed in Sec.~\ref{sec:toymodel_Rabi_driving_strong_driving_amplitude}. Consequently, optimal decoupling is achieved for $\gamma\lesssim0.06$ in this simulation.

\section{Conclusion}
\begin{table*}[t]\centering\label{tab:summary}
\caption{Conditions to decouple from low-frequency noise under various noise sources using one of the two shuttling protocols assuming that the Rabi frequency $\Omega$ is always smaller than the Larmor frequency $\omega_q$ of the shuttled spin.}
\centering
\begin{tblr}{
  cells = {l,m},cell{1}{1} = {r=2}{},cell{6}{1} = {r=2}{},vline{2-3}={-}{}
}
\textbf{Noise type} 
& \textbf{Temporal breathing shuttle}& \textbf{Spatial breathing shuttle}\\&
$\omega_{\alpha}(t)=\omega_{\alpha,0}(1+\gamma\sin\omega_d t)$, $\alpha=x,y$&
$\omega_{y}(x)=\omega_{y,0}(1+\gamma\sin x/L)$

\\\hline\hline
Global magnetic noise &Best against this noise source&Best against this noise source\\\hline
Global electric noise &$\gamma\ll1$&$\gamma\ll1$, $v=\omega_q L$\\ \hline
Local magnetic noise  &$\Omega\gg v/D_x,\gamma\ll1$&Generally not possible unless spin-orbit coupling\\\cline{1-2}
Local electric noise  &$\Omega\gg2v/D_x,\gamma\ll1$& is sizable allowing $\Omega\gg2v/D_x,\gamma\ll1$ even\\
&$\Omega\sim v/D_x$ may reduce shuttling fidelity& at spatial a periodicity 
$L<\ell_x$
\end{tblr}
\end{table*}
We have investigated continuous dynamical decoupling for driven two-level systems with time-dependent Larmor frequency, drawing an analogy with a shuttled spin qubit moving through a spatially inhomogeneous environment. Within the filter-function formalism, we identified driving regimes in which coherent control suppresses the qubit's susceptibility to low-frequency noise and thereby enhances shuttling fidelity. In particular, driving protocols that continuously adapt to variations in the Larmor frequency enable robust coherence enhancement, even in the presence of strong inhomogeneities.

Focusing on planar germanium hole-spin qubits, we derive effective spin Hamiltonians for confinement-modulated shuttling protocols and analyze the corresponding filter functions in the presence of magnetic and electric noise sources with different spatial correlations. The resulting conditions for suppressing low-frequency noise are summarized in Tab.~\ref{tab:summary}.

We have considered two possible implementations. In the temporal breathing protocol, the confinement is modulated in time during shuttling and can be realized using additional time-dependent signals on existing gate electrodes~\cite{ademi2025distributingentanglementdistantsemiconductor}. In contrast, the spatial breathing protocol relies on a spatially periodic modulation of the confinement, inducing an effective drive through motion under a resonance condition set by the velocity and spatial periodicity.

For local noise, efficient decoupling requires the Rabi frequency to exceed the inverse time set by the velocity and the effective spatial correlation length, $\Omega \gg v/D_x$ (see Tab.~\ref{tab:summary}), with a stricter condition $\Omega \gg 2v/D_x$ for electric noise. In spatial breathing, this can only be achieved if SOI is sufficiently strong to generate sizable Rabi frequencies even when the spatial periodicity is smaller than the confinement length along the shuttling direction, $L<\ell_x$. These conditions act on top of motional narrowing, where spatial averaging already suppresses noise, and enable additional decoupling from individual local noise sources encountered during motion. Furthermore, the confinement modulation should remain moderate, as strong driving enhances coupling to electric and local noise (see Tab.~\ref{tab:summary}).

Overall, temporal modulation provides a robust route to continuous dynamical decoupling, while spatial modulation achieves similar protection under suitable conditions, with the best performance achieved in both protocols when the dominant noise originates from fluctuations of the global magnetic field.

While this work focused on continuous dynamical decoupling, confinement modulation during shuttling more generally enables coherent control of mobile spin qubits. In particular, the same mechanism can be used to implement pulsed dynamical decoupling sequences or single qubit gate operations directly during the shuttling process. Thus, confinement-modulated shuttling provides a route to combine coherent control and noise protection within transport protocols for semiconductor spin-qubit architectures.
%\newpage
\begin{acknowledgments}
We thank members of the Bosco, Rimbach-Russ, Veldhorst, and Vandersypen groups for valuable discussions. 
This research  was supported by the EU through the H2024 QLSI2 project, by the Army Research Office under Award Number: W911NF-23-1-0110, and by NCCR Spin (grant number 225153). M.R.-R. acknowledges support from the Dutch Research Council (NWO) under Award Number Vidi TTW 22204. The views and conclusions contained in this document are those of the authors and should not be interpreted as representing the official policies, either expressed or implied, of the Army Research Office or the U.S. Government. The U.S. Government is authorized to reproduce and distribute reprints for Government purposes notwithstanding any copyright notation herein.

\end{acknowledgments}

\appendix
\section{Filter function formalism}\label{app:Filter_function_formalism}
For a single qubit, the Hamiltonian~\cite{Green_2013}
\begin{align}
    H(t)=H_c(t)+H_N(t)
\end{align}
models the effect of noise, captured by the general noise Hamiltonian $H_N(t)=\boldsymbol\beta(t)\cdot\boldsymbol\sigma$. The control Hamiltonian $H_c(t)$ corresponds to the noise-free control Hamiltonian for which the corresponding solution of the Schr{\"o}dinger equation is $U_c(t)$ executing a target operation $U(T)$ over a time interval $[0,T]$. In the interaction frame
\begin{align}
    \tilde H_N(t)=U_c^\dagger(t)H_N(t)U_c(t)
\end{align}
we define $\tilde U(t)$ as the solution of
\begin{align}
    i\hbar\frac{\text{d}}{\text{d}t}\tilde U(t)=\tilde H_N(t)\tilde U(t)
\end{align}
such that the time evolution of $H(t)$ is then given by $U(t)=U_c(t)\tilde U(t)$. The goal of dynamical decoupling is to find a control Hamiltonian that yields the target operation and mitigates the impact of noise, i.e., reduces $\tilde U(t)$ to the identity $\mathbbm 1$.

The time-evolution operator $\tilde U(t)$ can be approximated within the Magnus expansion 
\begin{align}
\tilde U(t)\approx\exp\left(-i\int_0^t d\tau\boldsymbol\beta(\tau)\hat R_c(\tau)\cdot\boldsymbol\sigma/\hbar\right)\equiv e^{-i\boldsymbol \phi_N(t)\cdot\boldsymbol\sigma/2}, 
\end{align}
if noise is small $\int_0^\tau dt\|\boldsymbol\beta(t)\|/\hbar\ll \pi$. Here $(\hat R_c(t))_{ij}=\text{Tr}(U_c^\dagger(t)\sigma_iU_c(t)\sigma_j)/2$ defines the $3\times3$ rotation matrix of the control Hamiltonian satisfying $\hat R_c(t)\boldsymbol\sigma=U_c^\dagger(t)\boldsymbol\sigma U_c(t)$, and $\boldsymbol{\phi}_N$ is the error vector. Following this, one can find the fidelity
$\mathcal{F_\text{av}}(t)=\tfrac{1}{4}\langle|\text{Tr}(\tilde U(t))|^2\rangle=\tfrac{1}{2}[\langle\cos2\phi_N(t)\rangle+1]$, taking the ensemble average over all realizations of the noise Hamiltonians denoted by the angular brackets. By considering the lowest order expansion and neglecting higher order cumulants of the noise, the cosine can be related to the infidelity $\mathcal I=1-\mathcal F\approx \langle\phi_N^2(\tau)\rangle$ as well as to the two-point cross-correlation~\cite{Cywinski2008-aj,Green_2013}
\begin{align}
\mathcal I=\sum_{i,j,k=1}^3\int\int dt_1dt_2\langle\beta_i(t_1)\beta_j(t_2)\rangle \hat R_{ik}(t_1)\hat R_{jk}(t_2)/\hbar^2.
\end{align}
We further assumes that noise is wide sense stationary and the cross-correlation depends only on the time difference $\langle\beta_i(t_1)\beta_j(t_2)\rangle=\tfrac{1}{2\pi}\int_{-\infty}^{\infty}d\omega S_{ij}(\omega)e^{i\omega(t_2-t_1)}$, introducing the cross-power spectral density $S_{ij}(\omega)$. Under the assumption that noise is isotropic and its components mutually independent in spin-space, i.e., $S_{ij}(\omega)=\delta_{ij} S(\omega)$ one can derive Eq.~\eqref{eq:average_gate_infidelity_FF_formalism} using the definition of the filter function Eq~\eqref{eq:filterfunction_Global_Noise}.

For the derivation of Eq.~\eqref{eq:Iff_Condition_For_filterfunction_Suppression} we note that the filter function Eq.~\eqref{eq:filterfunction_Global_Noise} can be expressed as
\begin{subequations}
\begin{align}
 F(\omega,\tau)=&Y(\omega,t)Y^\dagger(\omega,t)\\
Y(\omega,\tau)=&\int_0^\tau dt e^{i\omega t}R(t)
\end{align}
\end{subequations}
Furthermore, using $|\bra{\phi}U(t)\ket{\psi}|^2=\frac{1}{2}(1+\boldsymbol n_\phi^T\cdot R(t)\boldsymbol n_\psi)$ with $\boldsymbol n_\psi, \boldsymbol n_\phi$ denoting the corresponding Bloch vectors of $\ket{\psi},\ket{\phi}$ it follows
\begin{align}
&\frac{1}{\tau}\int_0^\tau dtP_{{\psi}\rightarrow{\phi}}(t)e^{i\omega t}=\frac{1}{\tau}\int_0^\tau dt\frac{1}{2}e^{i\omega t}(1+\boldsymbol n_\phi^T\cdot R(t)\boldsymbol n_\psi)\\
=&\frac{e^{i\omega\tau}-1}{2i\omega\tau}+\frac{1}{2}\boldsymbol n_\phi^T\cdot Y(\omega,\tau)\boldsymbol n_\psi/\tau)
\end{align}
Taking the limit $\omega\rightarrow0$ then yields Eq.~\eqref{eq:Iff_Condition_For_filterfunction_Suppression}. For general $\omega$ we find that the filter function $F(\omega,\tau)$ is zero iff $\frac{1}{\tau}\int_0^\tau dtP_{{\psi}\rightarrow{\phi}}(t)e^{i\omega t}=\frac{e^{i\omega\tau}-1}{2i\omega\tau}$ for any two arbitrary states $\ket{\psi},\ket{\phi}$.

For the numerical integration in Eq.~\eqref{eq:average_gate_infidelity_FF_formalism} we use a lower frequency cutoff of around $63\,\text{Hz}$ and a higher frequency cutoff of $1\,\text{GHz}$. With the chosen upper frequency cutoff the frequency interval includes several peaks coming from Rabi, driving and Larmor frequencies and we find that further increasing the upper frequency does not visually affect the results shown in the figures.\\

In the main text, we derive the modified filter functions in the case of coherent modulations of the noise source $\beta(t)\rightarrow k(t)\beta(t)$. With and without coherent modulation, $\beta(t)$ is assumed to be a stochastic variable that constitutes $1/f$ noise. $k(t)$ is a coherent modulation of the noise amplitude, which is independent of the ensemble averaging. As such we can use $\langle k(t_1)\beta_i(t_1)k(t_2)\beta_j(t_2)\rangle=k(t_1)k(t_2)\langle\beta_i(t_1)\beta_j(t_2)\rangle$. Eq.~\eqref{eq:average_gate_infidelity_FF_formalism} can therefore be rederived using a modified filter function introduced for global electric noise sources in Eq.~\eqref{eq:temporal_filter_function_global_electric_noise}.\\

For local noise sources the procedure is as follows. With the noise Hamiltonian being as reported in Eq.~\eqref{eq:temporal_effective_noise_hamiltonian_local_magnetic_noise} and using the assumptions of independent noise and identical power spectral density functions $\langle\boldsymbol\beta_k^{(i)}(t_1)\boldsymbol\beta_{k'}^{(j)}(t_2)\rangle=\delta_{kk'}\delta_{ij}\int d\omega\, e^{-i\omega(t_2-t_1)}S(\omega)/2\pi$ we can write
\begin{align}
\begin{aligned}
&\left\langle \Big(\sum_k \mathcal A_k(t_1)\beta_k^{(i)}(t_1)\Big)
\Big(\sum_{k'} \mathcal A_{k'}(t_2)\beta_{k'}^{(j)}(t_2)\Big) \right\rangle
\\&= \sum_k \mathcal A_k(t_1)\mathcal A_k(t_2)\langle \beta^{(i)}_k(t_1)\beta^{(j)}_k(t_2)\rangle,
\end{aligned}
\end{align}
where we used $\mathcal A_k(t)=\mathcal A(x_k,y_k,t)$. Again, Eq.~\eqref{eq:average_gate_infidelity_FF_formalism} can be derived using a modified filter function and kernel $K(t_1,t_2)$ as given in Eqs.~\eqref{eq:Matrix_of_filter_function_local_noise} and~\eqref{eq:Kernel_temporal_breathing_local_magnetic_noise}.

If the kernel depends only on the time difference $K(t_1,t_2)\equiv K(t_1-t_2)$, then we can define its 1D Fourier transform as 
$\tilde K(\omega)=\tfrac{1}{2\pi}\int d\Delta t\, K(\Delta t)e^{i\omega \Delta t}$. 
Writing $K(\Delta t)=\int d\omega'\,\tilde K(\nu)e^{-i\omega'\Delta t}$ and inserting this representation into Eq.~\eqref{eq:Matrix_of_filter_function_local_noise} for $F_L$, the kernel introduces an additional phase factor $e^{-i\omega'(t_1-t_2)}$. Combining this with the existing phase $e^{-i\omega(t_1-t_2)}$ yields $e^{-i(\omega+\omega')(t_1-t_2)}$, such that the remaining double integral is identical to the global filter function evaluated at the shifted frequency $\omega+\omega'$. One therefore obtains
\begin{align}
F_L(\omega,\tau)=\int d\omega'\,\tilde K(\omega'-\omega)\,F(\omega',\tau),
\end{align}
after a change of variables, showing that $F_L$ is given by a convolution of the Fourier transform of the kernel with the global filter function $F_G$ if the kernel depends only on the time difference.

In the main text, the kernel for electric noise sources coupling along the $x$-orbitals depends only on the time difference if confinements are not modulated in time. In which case, the corresponding Fourier transform yields
\begin{align}
    \tilde K_{E,x}(\omega)=\frac{\nu\ell_x^2\omega^2 e^{-\frac{D_x^2\omega^2}{2v^2}}}{32\pi n_0\ell_c\ell_yv^3\hbar^2\omega_x^2},
\end{align}
with $D_x=\sqrt{\ell_c^2+\ell_x^2}$. Consequently, $\tilde K_{E,x}$ is minimal at $\tilde K_{E,x}(0)=0$ and maximal at frequencies $\omega=\pm \frac{\sqrt2v}{\sqrt{2\ell_c^2+\ell_x^2}}$

Here, we report the kernels required for local electric noise sources in the spatial breathing shuttle protocol Sec.~\ref{sec:spatial_local_electric_noise_sources} approximated to linear order in $\gamma$ and assuming $\ell_c=0$:
\begin{widetext}
\begin{subequations}
\begin{align}
    K_x(t_1,t_2)=& e^{-\frac{v^2(t_1-t_2)^2}{2\ell_{x}^2}}\frac{\nu}{16\pi n_0\ell_{x,0}\ell_{y,0}}\left(1-\frac{(t_1-t_2)^2v^2}{\ell_{x,0}}+\gamma f_x(t_1,t_2)\right),\\
    f_x(t_1,t_2)=&e^{-\frac{\ell_{x,0}}{8L^2}}\frac{4L^2(\ell_{x,0}^2-(t_1-t_2)^2v^2)-\ell_{x,0}^4}{8\ell_{x,0}^2L^2} \sin[(t_1+t_2)v/2L],\\
    K_y(t_1,t_2)=& e^{-\frac{v^2(t_1-t_2)^2}{2\ell_{x}^2}}\frac{\nu}{16\pi n_0\ell_{x,0}\ell_{y,0}}\left(1-\frac{5}{2}e^{-\ell_{x,0}^2/8L^2}\gamma\sin[(t_1+t_2)v/2L]\right).
\end{align}
\end{subequations}
\end{widetext}
\section{Varying Rabi frequencies}\label{app:Varying_Rabi_frequency}
We consider the driven two-level system Hamiltonian in Eq.~\eqref{eq:driven_two_level_system} and allow for variations in the coupling amplitude
\begin{align}
    H(t)=\frac{\hbar\omega_q(t)}{2}\sigma_z+\hbar\Omega(t)\sin\Phi(t)\sigma_x.
\end{align}
Effective decoupling occurs for resonant driving for variations in the Larmor frequency $\omega_q(t)$ we can use the tracking protocol introduced in Sec.~\ref{sec:tracking_larmor_frequency_variations}. Going into the toggling frame $R=\exp(-i\int_0^tdt'\omega_q(t')\sigma_z/2\hbar)$ yields the transformed Hamiltonian
\begin{align}
    \tilde H(t)=\frac{\hbar\Omega(t)}{2}[\sin2\Phi(t)\sigma_x+\cos2\Phi(t)-\sigma_y].
\end{align}
Dropping the terms that oscillate with $2\Phi(t)$ within a regime $\omega_q(t)\gg\Omega(t),\dot \Omega(t)/\omega_q(t)$ yields the lab-frame unitary evolution $U(t)\approx e^{-i\Phi(t)\sigma_z/2}e^{-i\Theta(t)\sigma_y/2}$ which in terms of rotation matrices is
\begin{align}
    R(t)\approx R_z(\Phi(t))R_y(\Theta(t)).
\end{align}
Importantly, all eigenstates of the Pauli operators undergo rotations. As a result, over many rotations the corresponding transition probabilities average to $1/2$. According to Eq.~\eqref{eq:Iff_Condition_For_filterfunction_Suppression}, this constitutes the condition for suppressing the filter function at low frequencies. Hence, provided the accumulated rotation $\Theta(t)$ is sufficiently large, variations in the coupling amplitude $\Omega(t)$ do not significantly affect the decoupling efficiency, as long as the dynamics ensures continuous mixing between all qubit states.

\section{Spin-orbit coefficients in planar Germanium hole systems}\label{app:Explicit formula of spin-orbit coefficients}
\begin{table*}[]
\centering
\label{tab:appendix_parameters_for_g_tensor}
\caption{Parameters used for the simulations in Figs.~\ref{fig:temp_uniform},~\ref{fig:temp_lateral} and~\ref{fig:spatial_plots}. Choices are motivated by Refs~\cite{urielstrainholedriving,Martinez_2022,Terrazos_2021,mvtj-zhrl}.}
\begin{tabular}{l|l|l|l|l|l|l|l|l|l|l|l|l|l|l}
   $q$  & $\kappa$    & $\tilde\kappa$     & $b_v$      &  $d_v$    &   $\varepsilon_{xx}$     &$\varepsilon_{yy}$         &  $\varepsilon_{xz}$      & $\varepsilon_{yz}$       & $\varepsilon_{xy}$  &   $\Delta_\text{HL}$    & $\lambda$    &$\lambda'$       &  $\tilde\lambda$      & $\gamma_h$    \\ \hline
0.06 & 3.41 & 3.41 & -2.16$\,e\text{V}$ & -6.06$\,e\text{V}$ & -0.0061 & -0.0061 & 0.0001 & 0.0001 & 0 & 0.046$\,e\text{V}$ & 63.2 & 43.443 & 116.201 & 2.62 \\ 
\end{tabular}
\end{table*}
For planar germanium hole systems, the $\hat g$-tensor required in Eq.~\eqref{eq:moving_particle_Hamiltonian} follows from the Luttinger-Kohn-Bir-Pikus Hamiltonian via a Schrieffer-Wolff transformation to the predominantly heavy-hole subspace~\cite{Terrazos_2021}. Assuming a separable confinement $V(x,y,z)=V(x,y)V(z)$, the $g$-tensor components can be approximated by~\cite{Martinez_2022}
\begin{subequations}
\begin{align}
    g_{xx}=&3q-\frac{6\tilde\kappa b_v(\varepsilon_{xx}-\varepsilon_{yy})}{\Delta_\text{HL}}-\frac{6\lambda \bar p_x^2}{m_e\Delta_\text{HL}}+\frac{6\lambda' \bar p_y^2}{m_e\Delta_\text{HL}},\\
    g_{yy}=&-3q-\frac{6\tilde\kappa b_v(\varepsilon_{xx}-\varepsilon_{yy})}{\Delta_\text{HL}}-\frac{6\lambda' \bar p_x^2}{m_e\Delta_\text{HL}}+\frac{6\lambda \bar p_y^2}{m_e\Delta_\text{HL}},\\
    g_{zz}=&6\kappa+\frac{27}{2}q-2\gamma_h,\\
    g_{xy,yx}=&\pm\frac{4\sqrt{3}\kappa d_v\varepsilon_{xy}}{\Delta_\text{HL}}\mp\frac{12\tilde \lambda \bar p_x\bar p_y}{m_e\Delta_\text{HL}},\\
    g_{xz,yz}=&\frac{4\sqrt{3}\kappa d_v\varepsilon_{xz,yz}}{\Delta_\text{HL}},
\end{align}
\end{subequations}
where $\kappa$ and $q$ are linear and cubic Zeeman parameters, $\Delta_\text{HL}$ is the heavy-hole-light-hole splitting set by confinement and strain, and $m_e$ is the bare electron mass. The scaling parameters $\lambda,\lambda',\kappa'$ depend on the Luttinger parameters $\gamma_{1,2,3}$, deformation potentials $a_v$, $b_v$, $d_v$, and interband couplings $\eta_h$, $\tilde{\eta}_h$. Refs.~\cite{urielstrainholedriving,Martinez_2022,Terrazos_2021} and in the supplementary of Ref.~\cite{mvtj-zhrl} provide details and values for the parameters which were also used for our numerical simulations of Figs.~\ref{fig:temp_uniform},~\ref{fig:temp_lateral}, and~\ref{fig:spatial_plots} in the main text.

The momenta $(\bar p_x,\bar p_y)$ are in the coordinate system $(\bar x,\bar y)$ aligned with the in-plane crystallographic axis, i.e. $\bar x\parallel [100]$ and $\bar y\parallel [010]$. They are related to the coordinate system $(x,y)$ we used in Sec.~\ref{sec:model} where $x$ is aligned with the shuttling direction via a constant 2-D rotation $R_z(\theta_s)=e^{-i\theta_s(xp_y-yp_x)/\hbar}$ with angle $\theta_s$
\begin{subequations}\label{eq:2D_coordinate_rotation_on_position_and_momenta}
\begin{align}
    \bar x&=R_z^\dagger(\theta_s) xR_z(\theta_s)=x\cos\theta_s - y\sin\theta_s,\\
    \bar y&=R_z^\dagger(\theta_s) yR_z(\theta_s)=x\sin\theta_s + y\cos\theta,\\
    \bar p_x&=R_z^\dagger(\theta_s) p_xR_z(\theta_s)=p_x\cos\theta_s - p_y\sin\theta_s,\\
    \bar p_y&=R_z^\dagger(\theta_s) p_yR_z(\theta_s)=p_x\sin\theta_s + p_y\cos\theta_s,
\end{align}
\end{subequations}
Physically, the angle $\theta_s$ depends on the shuttling direction $x$ and encloses the angle between the trajectory and the $[100]$ crystallographic axis ($\bar x$). For the potential $V(x, y)$ used in Eq.~\eqref{eq:moving_particle_Hamiltonian} we assume that it is elliptically shaped along the shuttling direction $\parallel x$ given by
\begin{align}
    V(x,y)=\frac{m\omega_x^2}{2}(x-x_0(t))^2+\frac{m\omega_y^2}{2} y^2.
\end{align}
Consequently, the Hamiltonian can be written as
\begin{align}
\begin{aligned}
    H=&\frac{ p_x^2}{2m}+\frac{ p_y^2}{2m}+\frac{m\omega_x^2}{2}({x}-x_0(t))^2+\frac{m\omega_y^2}{2} y^2\\
    &+\tfrac{1}{2}\mu_B \sigma\cdot \hat g \boldsymbol B.
\end{aligned}
\end{align}
Which in turn can be expressed as the Hamiltonian in Eq.~\eqref{eq:Moving_QD_Hamiltonian_with_SOC}. As mentioned in the main body of this manuscript, the term $H_z^{(0)}$ can be position dependent through strain gradients leading to variations in Rabi and Larmor frequency while shuttling, but is assumed to be constant for Figs.~\ref{fig:temp_uniform},~\ref{fig:temp_lateral}, and~\ref{fig:spatial_plots}. The vectors $\boldsymbol\nu$, $\boldsymbol\mu$, and $\boldsymbol\xi$ are given as
\begin{widetext}
\begin{subequations}
\begin{align}
    \boldsymbol{\nu}&=\begin{pmatrix}
        \frac{6m_\parallel\mu_B\cos\theta_B(-\lambda\cos^2\theta_s\cos\phi_B+\lambda'\cos\phi_B\sin^2\theta_s+\tilde\lambda\sin2\theta_s\sin\phi_B)}{m_e\Delta_\text{HL}}\\
        -\frac{6m_\parallel\mu_B\cos\theta_B(\tilde\lambda\cos\phi_B\sin2\theta_s+(\lambda'\cos^2\theta_s-\lambda\sin^2\theta_s)\sin\phi_B)}{m_e\Delta_\text{HL}}\\
        0
    \end{pmatrix},\\
    \boldsymbol{\mu}&=\begin{pmatrix}
        \frac{6m_\parallel\mu_B\cos\theta_B(-\lambda'\cos^2\theta_s\cos\phi_B+\lambda\cos\phi_B\sin^2\theta_s+\tilde\lambda\sin2\theta_s\sin\phi_B)}{m_e\Delta_\text{HL}}\\
        -\frac{6m_\parallel\mu_B\cos\theta_B(\tilde\lambda\cos\phi_B\sin2\theta_s+(\lambda\cos^2\theta_s-\lambda'\sin^2\theta_s)\sin\phi_B)}{m_e\Delta_\text{HL}}\\
        0
    \end{pmatrix},\\
    \boldsymbol{\xi}&=\begin{pmatrix}
        \frac{6m_\parallel\mu_B\cos\theta_B((\lambda+\lambda')\cos\phi_B\sin2\theta_s+2\tilde\lambda\cos2\theta_s\sin\phi_B)}{m_e\Delta_\text{HL}}\\
        \frac{6m_\parallel\mu_B\cos\theta_B((\lambda+\lambda')\sin\phi_B\sin2\theta_s-2\tilde\lambda\cos2\theta_s\cos\phi_B)}{m_e\Delta_\text{HL}}\\
        0
    \end{pmatrix},
\end{align}
\end{subequations}
\end{widetext}
with the magnetic field $\boldsymbol B=B(\cos\theta_B\cos\phi_B,\cos\theta_B\sin\phi_B,\sin\theta_B)$ whereas $H_z^{(0)}$ is straightforwardly computed using $\tfrac{1}{2}\mu_B \sigma\cdot \hat g \boldsymbol B$ and neglecting terms with squared momenta. For Figs.~\ref{fig:temp_uniform} and~\ref{fig:temp_lateral} we used $\theta_s=20^\circ,\theta_B=0^\circ,\phi_B=50^\circ$, whilst in Fig.~\ref{fig:spatial_plots} we used $\theta_s=0^\circ,\theta_B=0^\circ,\phi_B=60^\circ$. For all three figures we further used parameters provided in Tab.~\ref{tab:appendix_parameters_for_g_tensor}.

For Fig.~\ref{fig:temp_uniform} we note that we can recast the Hamiltonian Eq.~\eqref{eq:Effective_spin_Hamiltonian_breathing_protocol} in the case $\omega_\alpha=\omega_{\alpha,0}(1+\gamma \sin\omega_d t)$ to one of the form
\begin{widetext}
\begin{subequations}
\begin{align}
    \tilde H_\text{eff}=&\frac{\hbar\boldsymbol \omega_q\cdot\boldsymbol\sigma}{2}+\hbar\boldsymbol \Omega\cdot\boldsymbol\sigma\sin\omega_d t\\
    \hbar\boldsymbol\omega_q=&\left(3qB_x+\frac{6\tilde\kappa b_v B_x(\varepsilon_{yy}-\varepsilon_{xx})}{\Delta_\text{HL}},-3qB_y+\frac{6\tilde\kappa b_v B_y(\varepsilon_{yy}-\varepsilon_{xx})}{\Delta_\text{HL}},B_z\left(\frac{27q}{2}-2\gamma_h+6\kappa\right)\right)^T+\frac{\hbar}{4}(\omega_{x,0}\boldsymbol\nu+\omega_{y,0}\boldsymbol\mu)\\
    \hbar\boldsymbol \Omega=&\frac{\hbar}{16}\gamma \mu_B(\omega_{x,0}\boldsymbol\nu+\omega_{y,0}\boldsymbol\mu)
\end{align}
\end{subequations}
\end{widetext}
In this way, we define the angle illustrated in Fig.~\ref{fig:temp_uniform}(c) between the driving term and the constant Larmor vector as
\begin{align}\label{eq:angle_uniform_driving}
    \varphi =\arccos \frac{\boldsymbol \omega_q\cdot\boldsymbol \Omega}{|\boldsymbol \omega_q||\boldsymbol \Omega|}.
\end{align}
\section{Unitary transformation of position and momentum operators}\label{app:Unitary_transformations}
Here, we summarize the unitary transformations used in the main text. In which way the transformations act on a specific operator can be evidently derived using the Baker-Campbell-Hausdorff formula in the form
\begin{align}
    e^{B}Ae^{-B}=\sum_{k=0}^\infty\frac{[B,A]_k}{k!}
\end{align}
where $[B,A]_k=[B,[B,A]_{k-1}]$ with $[B,A]_0=A$ denotes the left-nested commutator. For instance with the 2-D rotation $R_z(\theta)=e^{-i(xp_y-yp_x)/\hbar}$ of the $(x,y)$-plane used in App.~\ref{app:Explicit formula of spin-orbit coefficients} as well as the Baker-Campbell-Hausdorff formula one can derive~\eqref{eq:2D_coordinate_rotation_on_position_and_momenta}.

The translation operator $T=e^{-ix_0p_x/\hbar}$ along the $x$-coordinate shifts position as
\begin{align}
    T^\dagger xT=x+x_0.
\end{align}
For time-dependent positions $x_0(t)$ it contributes to the transformed Hamiltonian via a dynamical term $i\hbar\dot T^\dagger T=-\dot x_0(t)p_x$. The operators $p_x,y$ and $p_y$ are invariant under $T$. The squeezing operator $R_x(\alpha)=\exp\left(-\frac{i}{2\hbar}\Lambda\{x,p_x\}\right)$ transforms position and momentum as
\begin{subequations}
\begin{align}
R_x^\dagger xR_x=&e^\Lambda x\\
R_x^\dagger p_x R_x=&e^{-\Lambda}p_x.
\end{align}
\end{subequations}
In the main text we choose, e.g., $\Lambda=\ln\frac{\ell_x(t)}{\ell_{x,0}}=\ln\sqrt{\frac{\omega_{x,0}}{\omega_x(t)}}$ keeping track of the lengthening and squeezing of the dot along the $x$-direction. Due to the time-dependence of $\Lambda$ the dynamical term from the squeezing operator is non-trivial and in the aforementioned case $i\hbar\dot R_xR_x=\frac{\dot\omega_x(t)}{4\omega_x(t)}\{x,p_x\}$. In the same way, one can define the squeezing operator $R_y$ for the $y$-coordinate. Again, the operators $x,p_x$ and $y,p_y$ are invariant under transformations $R_y$ and $R_x$, respectively.

For the spatial breathing shuttle protocol we used $R=\exp\left(-\frac{i}{2\hbar}\Lambda(x+vt)\{y,p_y\}\right)$ with $\Lambda(x+vt)=\ln\sqrt{\frac{\omega_{y,0}}{\omega_y(x+vt)}}$. While this transformation acts trivially on $x$, it shifts the momentum along the $x$ direction as
\begin{align}
    R^\dagger p_xR=p_x-\tfrac{1}{2}\Lambda'(x+vt) \{y,p_y\}
\end{align}
using the commutation relation $[p_x,R]=-i\hbar \partial_x R$ and therefore
\begin{align}
\begin{aligned}
    R^\dagger p_x^2R=&p_x^2+\left(\frac{i\hbar}{2}\Lambda''(x+vt)-\Lambda'(x+vt)p_x\right)\{y,p_y\}\\&+\frac{1}{4}\Lambda'^2(x+vt)\{y,p_y\}^2.
\end{aligned}
\end{align}
For the $y$-orbital operators $R$ acts according to
\begin{align}
    R^\dagger yR&=\sqrt{\frac{\omega_{y,0}}{\omega_{y}(x+vt)}}y,\\
    R^\dagger p_yR&=\sqrt{\frac{\omega_{y}(x+vt)}{\omega_{y,0}}}p_y.
\end{align}

After the transformations used in~\ref{sec:spatial_effective_spin_Hamiltonian}, the Hamiltonian becomes
\begin{widetext}
\begin{align}
\begin{aligned}
    H=&\left[p_x^2+\left(\frac{i\hbar}{2}\Lambda''(x+vt)-\Lambda'(x+vt)p_x\right)\{y,p_y\}+\frac{1}{4}\Lambda'^2(x+vt)\{y,p_y\}^2\right](\mathbbm 1_2+B\boldsymbol\nu\boldsymbol\sigma)/2m\\
    &+\frac{\omega_y(x+vt)}{\omega_{y,0}}\frac{p_y^2}{2m}(\mathbbm 1_2+B\boldsymbol\mu\boldsymbol\sigma)+\frac{m\omega_x^2}{2}x^2+\frac{\omega_{y,0}}{\omega_y(x+vt)}\frac{m\omega_{y}^2(x+vt)}{2}y^2+\frac{v\omega_y'(x+vt)}{4\omega_y(x+vt)}\{y,p_y\}\\&-v\left[p_x+\frac{\omega_y'(x+vt)}{4\omega_y(x+vt)}\{y,p_y\}\right].
\end{aligned}
\end{align}
\end{widetext}
For the projection into the orbital ground state we note that $\{y,p_y\}=i\hbar[(a^\dagger)^2-a^2]$ excites or relaxes the $y$-orbital states two-fold. Within a low order perturbation theory, we ignore terms $\sim\{y,p_y\}$. Furthermore, $\bra{0_y}\{y,p_y\}^2\ket{0_y}=2\hbar^2$. As such, a projection to the effective low-energy $2\times2$ Hamiltonian yields the one provided in Eq.~\eqref{eq:effective_Hamiltonian_spatial_breathing}.

\bibliography{library}% Produces the bibliography via BibTeX.

%apsrev4-2.bst 2019-01-14 (MD) hand-edited version of apsrev4-1.bst
%Control: key (0)
%Control: author (8) initials jnrlst
%Control: editor formatted (1) identically to author
%Control: production of article title (0) allowed
%Control: page (0) single
%Control: year (1) truncated
%Control: production of eprint (0) enabled
\providecommand{\noopsort}[1]{}\providecommand{\singleletter}[1]{#1}%
\begin{thebibliography}{62}%
\makeatletter
\providecommand \@ifxundefined [1]{%
 \@ifx{#1\undefined}
}%
\providecommand \@ifnum [1]{%
 \ifnum #1\expandafter \@firstoftwo
 \else \expandafter \@secondoftwo
 \fi
}%
\providecommand \@ifx [1]{%
 \ifx #1\expandafter \@firstoftwo
 \else \expandafter \@secondoftwo
 \fi
}%
\providecommand \natexlab [1]{#1}%
\providecommand \enquote  [1]{``#1''}%
\providecommand \bibnamefont  [1]{#1}%
\providecommand \bibfnamefont [1]{#1}%
\providecommand \citenamefont [1]{#1}%
\providecommand \href@noop [0]{\@secondoftwo}%
\providecommand \href [0]{\begingroup \@sanitize@url \@href}%
\providecommand \@href[1]{\@@startlink{#1}\@@href}%
\providecommand \@@href[1]{\endgroup#1\@@endlink}%
\providecommand \@sanitize@url [0]{\catcode `\\12\catcode `\$12\catcode `\&12\catcode `\#12\catcode `\^12\catcode `\_12\catcode `\%12\relax}%
\providecommand \@@startlink[1]{}%
\providecommand \@@endlink[0]{}%
\providecommand \url  [0]{\begingroup\@sanitize@url \@url }%
\providecommand \@url [1]{\endgroup\@href {#1}{\urlprefix }}%
\providecommand \urlprefix  [0]{URL }%
\providecommand \Eprint [0]{\href }%
\providecommand \doibase [0]{https://doi.org/}%
\providecommand \selectlanguage [0]{\@gobble}%
\providecommand \bibinfo  [0]{\@secondoftwo}%
\providecommand \bibfield  [0]{\@secondoftwo}%
\providecommand \translation [1]{[#1]}%
\providecommand \BibitemOpen [0]{}%
\providecommand \bibitemStop [0]{}%
\providecommand \bibitemNoStop [0]{.\EOS\space}%
\providecommand \EOS [0]{\spacefactor3000\relax}%
\providecommand \BibitemShut  [1]{\csname bibitem#1\endcsname}%
\let\auto@bib@innerbib\@empty
%</preamble>
\bibitem [{\citenamefont {Burkard}\ \emph {et~al.}(2023)\citenamefont {Burkard}, \citenamefont {Ladd}, \citenamefont {Pan}, \citenamefont {Nichol},\ and\ \citenamefont {Petta}}]{Burkard2023-ex}%
  \BibitemOpen
  \bibfield  {author} {\bibinfo {author} {\bibfnamefont {G.}~\bibnamefont {Burkard}}, \bibinfo {author} {\bibfnamefont {T.~D.}\ \bibnamefont {Ladd}}, \bibinfo {author} {\bibfnamefont {A.}~\bibnamefont {Pan}}, \bibinfo {author} {\bibfnamefont {J.~M.}\ \bibnamefont {Nichol}},\ and\ \bibinfo {author} {\bibfnamefont {J.~R.}\ \bibnamefont {Petta}},\ }\bibfield  {title} {\bibinfo {title} {Semiconductor spin qubits},\ }\href {https://doi.org/10.1103/RevModPhys.95.025003} {\bibfield  {journal} {\bibinfo  {journal} {Rev. Mod. Phys.}\ }\textbf {\bibinfo {volume} {95}},\ \bibinfo {pages} {025003} (\bibinfo {year} {2023})}\BibitemShut {NoStop}%
\bibitem [{\citenamefont {Zwerver}\ \emph {et~al.}(2022)\citenamefont {Zwerver}, \citenamefont {Kr{\"a}henmann}, \citenamefont {Watson}, \citenamefont {Lampert}, \citenamefont {George}, \citenamefont {Pillarisetty}, \citenamefont {Bojarski}, \citenamefont {Amin}, \citenamefont {Amitonov}, \citenamefont {Boter}, \citenamefont {Caudillo}, \citenamefont {{Corras-Serrano}}, \citenamefont {Dehollain}, \citenamefont {Droulers}, \citenamefont {Henry}, \citenamefont {Kotlyar}, \citenamefont {Lodari}, \citenamefont {L{\"u}thi}, \citenamefont {Michalak}, \citenamefont {Mueller}, \citenamefont {Neyens}, \citenamefont {Roberts}, \citenamefont {Samkharadze}, \citenamefont {Zheng}, \citenamefont {Zietz}, \citenamefont {Scappucci}, \citenamefont {Veldhorst}, \citenamefont {Vandersypen},\ and\ \citenamefont {Clarke}}]{zwerverQubitsMadeAdvanced2022}%
  \BibitemOpen
  \bibfield  {author} {\bibinfo {author} {\bibfnamefont {A.~M.~J.}\ \bibnamefont {Zwerver}}, \bibinfo {author} {\bibfnamefont {T.}~\bibnamefont {Kr{\"a}henmann}}, \bibinfo {author} {\bibfnamefont {T.~F.}\ \bibnamefont {Watson}}, \bibinfo {author} {\bibfnamefont {L.}~\bibnamefont {Lampert}}, \bibinfo {author} {\bibfnamefont {H.~C.}\ \bibnamefont {George}}, \bibinfo {author} {\bibfnamefont {R.}~\bibnamefont {Pillarisetty}}, \bibinfo {author} {\bibfnamefont {S.~A.}\ \bibnamefont {Bojarski}}, \bibinfo {author} {\bibfnamefont {P.}~\bibnamefont {Amin}}, \bibinfo {author} {\bibfnamefont {S.~V.}\ \bibnamefont {Amitonov}}, \bibinfo {author} {\bibfnamefont {J.~M.}\ \bibnamefont {Boter}}, \bibinfo {author} {\bibfnamefont {R.}~\bibnamefont {Caudillo}}, \bibinfo {author} {\bibfnamefont {D.}~\bibnamefont {{Corras-Serrano}}}, \bibinfo {author} {\bibfnamefont {J.~P.}\ \bibnamefont {Dehollain}}, \bibinfo {author} {\bibfnamefont {G.}~\bibnamefont {Droulers}}, \bibinfo {author} {\bibfnamefont {E.~M.}\ \bibnamefont {Henry}},
  \bibinfo {author} {\bibfnamefont {R.}~\bibnamefont {Kotlyar}}, \bibinfo {author} {\bibfnamefont {M.}~\bibnamefont {Lodari}}, \bibinfo {author} {\bibfnamefont {F.}~\bibnamefont {L{\"u}thi}}, \bibinfo {author} {\bibfnamefont {D.~J.}\ \bibnamefont {Michalak}}, \bibinfo {author} {\bibfnamefont {B.~K.}\ \bibnamefont {Mueller}}, \bibinfo {author} {\bibfnamefont {S.}~\bibnamefont {Neyens}}, \bibinfo {author} {\bibfnamefont {J.}~\bibnamefont {Roberts}}, \bibinfo {author} {\bibfnamefont {N.}~\bibnamefont {Samkharadze}}, \bibinfo {author} {\bibfnamefont {G.}~\bibnamefont {Zheng}}, \bibinfo {author} {\bibfnamefont {O.~K.}\ \bibnamefont {Zietz}}, \bibinfo {author} {\bibfnamefont {G.}~\bibnamefont {Scappucci}}, \bibinfo {author} {\bibfnamefont {M.}~\bibnamefont {Veldhorst}}, \bibinfo {author} {\bibfnamefont {L.~M.~K.}\ \bibnamefont {Vandersypen}},\ and\ \bibinfo {author} {\bibfnamefont {J.~S.}\ \bibnamefont {Clarke}},\ }\bibfield  {title} {\bibinfo {title} {Qubits made by advanced semiconductor manufacturing},\ }\href
  {https://doi.org/10.1038/s41928-022-00727-9} {\bibfield  {journal} {\bibinfo  {journal} {Nature Electronics}\ }\textbf {\bibinfo {volume} {5}},\ \bibinfo {pages} {184} (\bibinfo {year} {2022})}\BibitemShut {NoStop}%
\bibitem [{\citenamefont {Steinacker}\ \emph {et~al.}(2025)\citenamefont {Steinacker}, \citenamefont {Dumoulin~Stuyck}, \citenamefont {Lim}, \citenamefont {Tanttu}, \citenamefont {Feng}, \citenamefont {Serrano}, \citenamefont {Nickl}, \citenamefont {Candido}, \citenamefont {Cifuentes}, \citenamefont {Vahapoglu}, \citenamefont {Bartee}, \citenamefont {Hudson}, \citenamefont {Chan}, \citenamefont {Kubicek}, \citenamefont {Jussot}, \citenamefont {Canvel}, \citenamefont {Beyne}, \citenamefont {Shimura}, \citenamefont {Loo}, \citenamefont {Godfrin}, \citenamefont {Raes}, \citenamefont {Baudot}, \citenamefont {Wan}, \citenamefont {Laucht}, \citenamefont {Yang}, \citenamefont {Saraiva}, \citenamefont {Escott}, \citenamefont {De~Greve},\ and\ \citenamefont {Dzurak}}]{steinackerIndustrycompatibleSiliconSpinqubit2025}%
  \BibitemOpen
  \bibfield  {author} {\bibinfo {author} {\bibfnamefont {P.}~\bibnamefont {Steinacker}}, \bibinfo {author} {\bibfnamefont {N.}~\bibnamefont {Dumoulin~Stuyck}}, \bibinfo {author} {\bibfnamefont {W.~H.}\ \bibnamefont {Lim}}, \bibinfo {author} {\bibfnamefont {T.}~\bibnamefont {Tanttu}}, \bibinfo {author} {\bibfnamefont {M.}~\bibnamefont {Feng}}, \bibinfo {author} {\bibfnamefont {S.}~\bibnamefont {Serrano}}, \bibinfo {author} {\bibfnamefont {A.}~\bibnamefont {Nickl}}, \bibinfo {author} {\bibfnamefont {M.}~\bibnamefont {Candido}}, \bibinfo {author} {\bibfnamefont {J.~D.}\ \bibnamefont {Cifuentes}}, \bibinfo {author} {\bibfnamefont {E.}~\bibnamefont {Vahapoglu}}, \bibinfo {author} {\bibfnamefont {S.~K.}\ \bibnamefont {Bartee}}, \bibinfo {author} {\bibfnamefont {F.~E.}\ \bibnamefont {Hudson}}, \bibinfo {author} {\bibfnamefont {K.~W.}\ \bibnamefont {Chan}}, \bibinfo {author} {\bibfnamefont {S.}~\bibnamefont {Kubicek}}, \bibinfo {author} {\bibfnamefont {J.}~\bibnamefont {Jussot}}, \bibinfo {author} {\bibfnamefont
  {Y.}~\bibnamefont {Canvel}}, \bibinfo {author} {\bibfnamefont {S.}~\bibnamefont {Beyne}}, \bibinfo {author} {\bibfnamefont {Y.}~\bibnamefont {Shimura}}, \bibinfo {author} {\bibfnamefont {R.}~\bibnamefont {Loo}}, \bibinfo {author} {\bibfnamefont {C.}~\bibnamefont {Godfrin}}, \bibinfo {author} {\bibfnamefont {B.}~\bibnamefont {Raes}}, \bibinfo {author} {\bibfnamefont {S.}~\bibnamefont {Baudot}}, \bibinfo {author} {\bibfnamefont {D.}~\bibnamefont {Wan}}, \bibinfo {author} {\bibfnamefont {A.}~\bibnamefont {Laucht}}, \bibinfo {author} {\bibfnamefont {C.~H.}\ \bibnamefont {Yang}}, \bibinfo {author} {\bibfnamefont {A.}~\bibnamefont {Saraiva}}, \bibinfo {author} {\bibfnamefont {C.~C.}\ \bibnamefont {Escott}}, \bibinfo {author} {\bibfnamefont {K.}~\bibnamefont {De~Greve}},\ and\ \bibinfo {author} {\bibfnamefont {A.~S.}\ \bibnamefont {Dzurak}},\ }\bibfield  {title} {\bibinfo {title} {Industry-compatible silicon spin-qubit unit cells exceeding 99\% fidelity},\ }\href {https://doi.org/10.1038/s41586-025-09531-9}
  {\bibfield  {journal} {\bibinfo  {journal} {Nature}\ }\textbf {\bibinfo {volume} {646}},\ \bibinfo {pages} {81} (\bibinfo {year} {2025})}\BibitemShut {NoStop}%
\bibitem [{\citenamefont {George}\ \emph {et~al.}(2025)\citenamefont {George}, \citenamefont {M{\k a}dzik}, \citenamefont {Henry}, \citenamefont {Wagner}, \citenamefont {Islam}, \citenamefont {Borjans}, \citenamefont {Connors}, \citenamefont {Corrigan}, \citenamefont {Curry}, \citenamefont {Harper}, \citenamefont {Keith}, \citenamefont {Lampert}, \citenamefont {Luthi}, \citenamefont {Mohiyaddin}, \citenamefont {Murcia}, \citenamefont {Nair}, \citenamefont {Nahm}, \citenamefont {Nethwewala}, \citenamefont {Neyens}, \citenamefont {Patra}, \citenamefont {Raharjo}, \citenamefont {Rogan}, \citenamefont {Savytskyy}, \citenamefont {Watson}, \citenamefont {Ziegler}, \citenamefont {Zietz}, \citenamefont {Pellerano}, \citenamefont {Pillarisetty}, \citenamefont {Bishop}, \citenamefont {Bojarski}, \citenamefont {Roberts},\ and\ \citenamefont {Clarke}}]{george12SpinQubitArraysFabricated2025}%
  \BibitemOpen
  \bibfield  {author} {\bibinfo {author} {\bibfnamefont {H.~C.}\ \bibnamefont {George}}, \bibinfo {author} {\bibfnamefont {M.~T.}\ \bibnamefont {M{\k a}dzik}}, \bibinfo {author} {\bibfnamefont {E.~M.}\ \bibnamefont {Henry}}, \bibinfo {author} {\bibfnamefont {A.~J.}\ \bibnamefont {Wagner}}, \bibinfo {author} {\bibfnamefont {M.~M.}\ \bibnamefont {Islam}}, \bibinfo {author} {\bibfnamefont {F.}~\bibnamefont {Borjans}}, \bibinfo {author} {\bibfnamefont {E.~J.}\ \bibnamefont {Connors}}, \bibinfo {author} {\bibfnamefont {J.}~\bibnamefont {Corrigan}}, \bibinfo {author} {\bibfnamefont {M.}~\bibnamefont {Curry}}, \bibinfo {author} {\bibfnamefont {M.~K.}\ \bibnamefont {Harper}}, \bibinfo {author} {\bibfnamefont {D.}~\bibnamefont {Keith}}, \bibinfo {author} {\bibfnamefont {L.}~\bibnamefont {Lampert}}, \bibinfo {author} {\bibfnamefont {F.}~\bibnamefont {Luthi}}, \bibinfo {author} {\bibfnamefont {F.~A.}\ \bibnamefont {Mohiyaddin}}, \bibinfo {author} {\bibfnamefont {S.}~\bibnamefont {Murcia}}, \bibinfo {author}
  {\bibfnamefont {R.}~\bibnamefont {Nair}}, \bibinfo {author} {\bibfnamefont {R.}~\bibnamefont {Nahm}}, \bibinfo {author} {\bibfnamefont {A.}~\bibnamefont {Nethwewala}}, \bibinfo {author} {\bibfnamefont {S.}~\bibnamefont {Neyens}}, \bibinfo {author} {\bibfnamefont {B.}~\bibnamefont {Patra}}, \bibinfo {author} {\bibfnamefont {R.~D.}\ \bibnamefont {Raharjo}}, \bibinfo {author} {\bibfnamefont {C.}~\bibnamefont {Rogan}}, \bibinfo {author} {\bibfnamefont {R.}~\bibnamefont {Savytskyy}}, \bibinfo {author} {\bibfnamefont {T.~F.}\ \bibnamefont {Watson}}, \bibinfo {author} {\bibfnamefont {J.}~\bibnamefont {Ziegler}}, \bibinfo {author} {\bibfnamefont {O.~K.}\ \bibnamefont {Zietz}}, \bibinfo {author} {\bibfnamefont {S.}~\bibnamefont {Pellerano}}, \bibinfo {author} {\bibfnamefont {R.}~\bibnamefont {Pillarisetty}}, \bibinfo {author} {\bibfnamefont {N.~C.}\ \bibnamefont {Bishop}}, \bibinfo {author} {\bibfnamefont {S.~A.}\ \bibnamefont {Bojarski}}, \bibinfo {author} {\bibfnamefont {J.}~\bibnamefont {Roberts}},\ and\ \bibinfo
  {author} {\bibfnamefont {J.~S.}\ \bibnamefont {Clarke}},\ }\bibfield  {title} {\bibinfo {title} {12-{{Spin-Qubit Arrays Fabricated}} on a 300 mm {{Semiconductor Manufacturing Line}}},\ }\href {https://doi.org/10.1021/acs.nanolett.4c05205} {\bibfield  {journal} {\bibinfo  {journal} {Nano Letters}\ }\textbf {\bibinfo {volume} {25}},\ \bibinfo {pages} {793} (\bibinfo {year} {2025})}\BibitemShut {NoStop}%
\bibitem [{\citenamefont {Nickl}\ \emph {et~al.}(2025)\citenamefont {Nickl}, \citenamefont {Stuyck}, \citenamefont {Steinacker}, \citenamefont {Cifuentes}, \citenamefont {Serrano}, \citenamefont {Feng}, \citenamefont {Vahapoglu}, \citenamefont {Hudson}, \citenamefont {Chan}, \citenamefont {Kubicek}, \citenamefont {Jussot}, \citenamefont {Canvel}, \citenamefont {Beyne}, \citenamefont {Shimura}, \citenamefont {Loo}, \citenamefont {Godfrin}, \citenamefont {Raes}, \citenamefont {Baudot}, \citenamefont {Wan}, \citenamefont {Laucht}, \citenamefont {Yang}, \citenamefont {Lim}, \citenamefont {Saraiva}, \citenamefont {Escott}, \citenamefont {Greve}, \citenamefont {Dzurak},\ and\ \citenamefont {Tanttu}}]{nicklEightQubitOperation3002025}%
  \BibitemOpen
  \bibfield  {author} {\bibinfo {author} {\bibfnamefont {A.}~\bibnamefont {Nickl}}, \bibinfo {author} {\bibfnamefont {N.~D.}\ \bibnamefont {Stuyck}}, \bibinfo {author} {\bibfnamefont {P.}~\bibnamefont {Steinacker}}, \bibinfo {author} {\bibfnamefont {J.~D.}\ \bibnamefont {Cifuentes}}, \bibinfo {author} {\bibfnamefont {S.}~\bibnamefont {Serrano}}, \bibinfo {author} {\bibfnamefont {M.}~\bibnamefont {Feng}}, \bibinfo {author} {\bibfnamefont {E.}~\bibnamefont {Vahapoglu}}, \bibinfo {author} {\bibfnamefont {F.~E.}\ \bibnamefont {Hudson}}, \bibinfo {author} {\bibfnamefont {K.~W.}\ \bibnamefont {Chan}}, \bibinfo {author} {\bibfnamefont {S.}~\bibnamefont {Kubicek}}, \bibinfo {author} {\bibfnamefont {J.}~\bibnamefont {Jussot}}, \bibinfo {author} {\bibfnamefont {Y.}~\bibnamefont {Canvel}}, \bibinfo {author} {\bibfnamefont {S.}~\bibnamefont {Beyne}}, \bibinfo {author} {\bibfnamefont {Y.}~\bibnamefont {Shimura}}, \bibinfo {author} {\bibfnamefont {R.}~\bibnamefont {Loo}}, \bibinfo {author} {\bibfnamefont {C.}~\bibnamefont
  {Godfrin}}, \bibinfo {author} {\bibfnamefont {B.}~\bibnamefont {Raes}}, \bibinfo {author} {\bibfnamefont {S.}~\bibnamefont {Baudot}}, \bibinfo {author} {\bibfnamefont {D.}~\bibnamefont {Wan}}, \bibinfo {author} {\bibfnamefont {A.}~\bibnamefont {Laucht}}, \bibinfo {author} {\bibfnamefont {C.-H.}\ \bibnamefont {Yang}}, \bibinfo {author} {\bibfnamefont {W.~H.}\ \bibnamefont {Lim}}, \bibinfo {author} {\bibfnamefont {A.}~\bibnamefont {Saraiva}}, \bibinfo {author} {\bibfnamefont {C.~C.}\ \bibnamefont {Escott}}, \bibinfo {author} {\bibfnamefont {K.~D.}\ \bibnamefont {Greve}}, \bibinfo {author} {\bibfnamefont {A.~S.}\ \bibnamefont {Dzurak}},\ and\ \bibinfo {author} {\bibfnamefont {T.}~\bibnamefont {Tanttu}},\ }\href {https://doi.org/10.48550/arXiv.2512.10174} {\bibinfo {title} {Eight-{{Qubit Operation}} of a 300 mm {{SiMOS Foundry-Fabricated Device}}}} (\bibinfo {year} {2025}),\ \Eprint {https://arxiv.org/abs/2512.10174} {2512.10174} \BibitemShut {NoStop}%
\bibitem [{\citenamefont {Vorreiter}\ \emph {et~al.}(2025)\citenamefont {Vorreiter}, \citenamefont {Huang}, \citenamefont {Liles}, \citenamefont {Hillier}, \citenamefont {Li}, \citenamefont {Raes}, \citenamefont {Kubicek}, \citenamefont {Jussot}, \citenamefont {Beyne}, \citenamefont {Godfrin}, \citenamefont {Sharma}, \citenamefont {Wan}, \citenamefont {Stuyck}, \citenamefont {Gilbert}, \citenamefont {Yang}, \citenamefont {Dzurak}, \citenamefont {Greve},\ and\ \citenamefont {Hamilton}}]{vorreiterPrecisionHighspeedQuantum2025}%
  \BibitemOpen
  \bibfield  {author} {\bibinfo {author} {\bibfnamefont {I.}~\bibnamefont {Vorreiter}}, \bibinfo {author} {\bibfnamefont {J.~Y.}\ \bibnamefont {Huang}}, \bibinfo {author} {\bibfnamefont {S.~D.}\ \bibnamefont {Liles}}, \bibinfo {author} {\bibfnamefont {J.}~\bibnamefont {Hillier}}, \bibinfo {author} {\bibfnamefont {R.}~\bibnamefont {Li}}, \bibinfo {author} {\bibfnamefont {B.}~\bibnamefont {Raes}}, \bibinfo {author} {\bibfnamefont {S.}~\bibnamefont {Kubicek}}, \bibinfo {author} {\bibfnamefont {J.}~\bibnamefont {Jussot}}, \bibinfo {author} {\bibfnamefont {S.}~\bibnamefont {Beyne}}, \bibinfo {author} {\bibfnamefont {C.}~\bibnamefont {Godfrin}}, \bibinfo {author} {\bibfnamefont {S.}~\bibnamefont {Sharma}}, \bibinfo {author} {\bibfnamefont {D.}~\bibnamefont {Wan}}, \bibinfo {author} {\bibfnamefont {N.~D.}\ \bibnamefont {Stuyck}}, \bibinfo {author} {\bibfnamefont {W.}~\bibnamefont {Gilbert}}, \bibinfo {author} {\bibfnamefont {C.~H.}\ \bibnamefont {Yang}}, \bibinfo {author} {\bibfnamefont {A.~S.}\ \bibnamefont
  {Dzurak}}, \bibinfo {author} {\bibfnamefont {K.~D.}\ \bibnamefont {Greve}},\ and\ \bibinfo {author} {\bibfnamefont {A.~R.}\ \bibnamefont {Hamilton}},\ }\href {https://doi.org/10.48550/arXiv.2508.00446} {\bibinfo {title} {Precision high-speed quantum logic with holes on a natural silicon foundry platform}} (\bibinfo {year} {2025}),\ \Eprint {https://arxiv.org/abs/2508.00446} {2508.00446 [cond-mat]} \BibitemShut {NoStop}%
\bibitem [{\citenamefont {Taylor}\ \emph {et~al.}(2005)\citenamefont {Taylor}, \citenamefont {Engel}, \citenamefont {D\"{u}r}, \citenamefont {Yacoby}, \citenamefont {Marcus}, \citenamefont {Zoller},\ and\ \citenamefont {Lukin}}]{Taylor2005}%
  \BibitemOpen
  \bibfield  {author} {\bibinfo {author} {\bibfnamefont {J.~M.}\ \bibnamefont {Taylor}}, \bibinfo {author} {\bibfnamefont {H.-A.}\ \bibnamefont {Engel}}, \bibinfo {author} {\bibfnamefont {W.}~\bibnamefont {D\"{u}r}}, \bibinfo {author} {\bibfnamefont {A.}~\bibnamefont {Yacoby}}, \bibinfo {author} {\bibfnamefont {C.~M.}\ \bibnamefont {Marcus}}, \bibinfo {author} {\bibfnamefont {P.}~\bibnamefont {Zoller}},\ and\ \bibinfo {author} {\bibfnamefont {M.~D.}\ \bibnamefont {Lukin}},\ }\bibfield  {title} {\bibinfo {title} {Fault-tolerant architecture for quantum computation using electrically controlled semiconductor spins},\ }\href {https://doi.org/10.1038/nphys174} {\bibfield  {journal} {\bibinfo  {journal} {Nature Physics}\ }\textbf {\bibinfo {volume} {1}},\ \bibinfo {pages} {177} (\bibinfo {year} {2005})}\BibitemShut {NoStop}%
\bibitem [{\citenamefont {Vandersypen}\ \emph {et~al.}(2017)\citenamefont {Vandersypen}, \citenamefont {Bluhm}, \citenamefont {Clarke}, \citenamefont {Dzurak}, \citenamefont {Ishihara}, \citenamefont {Morello}, \citenamefont {Reilly}, \citenamefont {Schreiber},\ and\ \citenamefont {Veldhorst}}]{Vandersypen2017}%
  \BibitemOpen
  \bibfield  {author} {\bibinfo {author} {\bibfnamefont {L.~M.~K.}\ \bibnamefont {Vandersypen}}, \bibinfo {author} {\bibfnamefont {H.}~\bibnamefont {Bluhm}}, \bibinfo {author} {\bibfnamefont {J.~S.}\ \bibnamefont {Clarke}}, \bibinfo {author} {\bibfnamefont {A.~S.}\ \bibnamefont {Dzurak}}, \bibinfo {author} {\bibfnamefont {R.}~\bibnamefont {Ishihara}}, \bibinfo {author} {\bibfnamefont {A.}~\bibnamefont {Morello}}, \bibinfo {author} {\bibfnamefont {D.~J.}\ \bibnamefont {Reilly}}, \bibinfo {author} {\bibfnamefont {L.~R.}\ \bibnamefont {Schreiber}},\ and\ \bibinfo {author} {\bibfnamefont {M.}~\bibnamefont {Veldhorst}},\ }\bibfield  {title} {\bibinfo {title} {Interfacing spin qubits in quantum dots and donors---hot, dense, and coherent},\ }\href {https://doi.org/10.1038/s41534-017-0038-y} {\bibfield  {journal} {\bibinfo  {journal} {npj Quantum Information}\ }\textbf {\bibinfo {volume} {3}},\ \bibinfo {pages} {34} (\bibinfo {year} {2017})}\BibitemShut {NoStop}%
\bibitem [{\citenamefont {K{\"u}nne}\ \emph {et~al.}(2024)\citenamefont {K{\"u}nne}, \citenamefont {Willmes}, \citenamefont {Oberl{\"a}nder}, \citenamefont {Gorjaew}, \citenamefont {Teske}, \citenamefont {Bhardwaj}, \citenamefont {Beer}, \citenamefont {Kammerloher}, \citenamefont {Otten}, \citenamefont {Seidler}, \citenamefont {Xue}, \citenamefont {Schreiber},\ and\ \citenamefont {Bluhm}}]{Knne2024}%
  \BibitemOpen
  \bibfield  {author} {\bibinfo {author} {\bibfnamefont {M.}~\bibnamefont {K{\"u}nne}}, \bibinfo {author} {\bibfnamefont {A.}~\bibnamefont {Willmes}}, \bibinfo {author} {\bibfnamefont {M.}~\bibnamefont {Oberl{\"a}nder}}, \bibinfo {author} {\bibfnamefont {C.}~\bibnamefont {Gorjaew}}, \bibinfo {author} {\bibfnamefont {J.~D.}\ \bibnamefont {Teske}}, \bibinfo {author} {\bibfnamefont {H.}~\bibnamefont {Bhardwaj}}, \bibinfo {author} {\bibfnamefont {M.}~\bibnamefont {Beer}}, \bibinfo {author} {\bibfnamefont {E.}~\bibnamefont {Kammerloher}}, \bibinfo {author} {\bibfnamefont {R.}~\bibnamefont {Otten}}, \bibinfo {author} {\bibfnamefont {I.}~\bibnamefont {Seidler}}, \bibinfo {author} {\bibfnamefont {R.}~\bibnamefont {Xue}}, \bibinfo {author} {\bibfnamefont {L.~R.}\ \bibnamefont {Schreiber}},\ and\ \bibinfo {author} {\bibfnamefont {H.}~\bibnamefont {Bluhm}},\ }\bibfield  {title} {\bibinfo {title} {The {{SpinBus}} architecture for scaling spin qubits with electron shuttling},\ }\href
  {https://doi.org/10.1038/s41467-024-49182-4} {\bibfield  {journal} {\bibinfo  {journal} {Nature Communications}\ }\textbf {\bibinfo {volume} {15}},\ \bibinfo {pages} {4977} (\bibinfo {year} {2024})}\BibitemShut {NoStop}%
\bibitem [{\citenamefont {Siegel}\ \emph {et~al.}(2026)\citenamefont {Siegel}, \citenamefont {Cai}, \citenamefont {Jnane}, \citenamefont {Koczor}, \citenamefont {Pexton}, \citenamefont {Strikis},\ and\ \citenamefont {Benjamin}}]{494s-jd8h}%
  \BibitemOpen
  \bibfield  {author} {\bibinfo {author} {\bibfnamefont {A.}~\bibnamefont {Siegel}}, \bibinfo {author} {\bibfnamefont {Z.}~\bibnamefont {Cai}}, \bibinfo {author} {\bibfnamefont {H.}~\bibnamefont {Jnane}}, \bibinfo {author} {\bibfnamefont {B.}~\bibnamefont {Koczor}}, \bibinfo {author} {\bibfnamefont {S.}~\bibnamefont {Pexton}}, \bibinfo {author} {\bibfnamefont {A.}~\bibnamefont {Strikis}},\ and\ \bibinfo {author} {\bibfnamefont {S.}~\bibnamefont {Benjamin}},\ }\bibfield  {title} {\bibinfo {title} {Quantum snakes on a plane: Mobile, low-dimensional logical qubits on a 2d surface},\ }\href {https://doi.org/10.1103/494s-jd8h} {\bibfield  {journal} {\bibinfo  {journal} {PRX Quantum}\ }\textbf {\bibinfo {volume} {7}},\ \bibinfo {pages} {010339} (\bibinfo {year} {2026})}\BibitemShut {NoStop}%
\bibitem [{\citenamefont {Langrock}\ \emph {et~al.}(2023)\citenamefont {Langrock}, \citenamefont {Krzywda}, \citenamefont {Focke}, \citenamefont {Seidler}, \citenamefont {Schreiber},\ and\ \citenamefont {Cywi{\'n}ski}}]{langrockBlueprintScalableSpin2023}%
  \BibitemOpen
  \bibfield  {author} {\bibinfo {author} {\bibfnamefont {V.}~\bibnamefont {Langrock}}, \bibinfo {author} {\bibfnamefont {J.~A.}\ \bibnamefont {Krzywda}}, \bibinfo {author} {\bibfnamefont {N.}~\bibnamefont {Focke}}, \bibinfo {author} {\bibfnamefont {I.}~\bibnamefont {Seidler}}, \bibinfo {author} {\bibfnamefont {L.~R.}\ \bibnamefont {Schreiber}},\ and\ \bibinfo {author} {\bibfnamefont {{\L}.}~\bibnamefont {Cywi{\'n}ski}},\ }\bibfield  {title} {\bibinfo {title} {Blueprint of a {{Scalable Spin Qubit Shuttle Device}} for {{Coherent Mid-Range Qubit Transfer}} in {{Disordered Si}}/{{SiGe}}/{{SiO}}},\ }\href {https://doi.org/10.1103/PRXQuantum.4.020305} {\bibfield  {journal} {\bibinfo  {journal} {PRX Quantum}\ }\textbf {\bibinfo {volume} {4}},\ \bibinfo {pages} {020305} (\bibinfo {year} {2023})}\BibitemShut {NoStop}%
\bibitem [{\citenamefont {Seidler}\ \emph {et~al.}(2022)\citenamefont {Seidler}, \citenamefont {Struck}, \citenamefont {Xue}, \citenamefont {Focke}, \citenamefont {Trellenkamp}, \citenamefont {Bluhm},\ and\ \citenamefont {Schreiber}}]{Seidler2022}%
  \BibitemOpen
  \bibfield  {author} {\bibinfo {author} {\bibfnamefont {I.}~\bibnamefont {Seidler}}, \bibinfo {author} {\bibfnamefont {T.}~\bibnamefont {Struck}}, \bibinfo {author} {\bibfnamefont {R.}~\bibnamefont {Xue}}, \bibinfo {author} {\bibfnamefont {N.}~\bibnamefont {Focke}}, \bibinfo {author} {\bibfnamefont {S.}~\bibnamefont {Trellenkamp}}, \bibinfo {author} {\bibfnamefont {H.}~\bibnamefont {Bluhm}},\ and\ \bibinfo {author} {\bibfnamefont {L.~R.}\ \bibnamefont {Schreiber}},\ }\bibfield  {title} {\bibinfo {title} {Conveyor-mode single-electron shuttling in {{Si}}/{{SiGe}} for a scalable quantum computing architecture},\ }\href {https://doi.org/10.1038/s41534-022-00615-2} {\bibfield  {journal} {\bibinfo  {journal} {npj Quantum Information}\ }\textbf {\bibinfo {volume} {8}},\ \bibinfo {pages} {100} (\bibinfo {year} {2022})}\BibitemShut {NoStop}%
\bibitem [{\citenamefont {Struck}\ \emph {et~al.}(2024)\citenamefont {Struck}, \citenamefont {Volmer}, \citenamefont {Visser}, \citenamefont {Offermann}, \citenamefont {Xue}, \citenamefont {Tu}, \citenamefont {Trellenkamp}, \citenamefont {Cywi{\'n}ski}, \citenamefont {Bluhm},\ and\ \citenamefont {Schreiber}}]{Struck2024}%
  \BibitemOpen
  \bibfield  {author} {\bibinfo {author} {\bibfnamefont {T.}~\bibnamefont {Struck}}, \bibinfo {author} {\bibfnamefont {M.}~\bibnamefont {Volmer}}, \bibinfo {author} {\bibfnamefont {L.}~\bibnamefont {Visser}}, \bibinfo {author} {\bibfnamefont {T.}~\bibnamefont {Offermann}}, \bibinfo {author} {\bibfnamefont {R.}~\bibnamefont {Xue}}, \bibinfo {author} {\bibfnamefont {J.-S.}\ \bibnamefont {Tu}}, \bibinfo {author} {\bibfnamefont {S.}~\bibnamefont {Trellenkamp}}, \bibinfo {author} {\bibfnamefont {{\L}.}~\bibnamefont {Cywi{\'n}ski}}, \bibinfo {author} {\bibfnamefont {H.}~\bibnamefont {Bluhm}},\ and\ \bibinfo {author} {\bibfnamefont {L.~R.}\ \bibnamefont {Schreiber}},\ }\bibfield  {title} {\bibinfo {title} {Spin-{{EPR-pair}} separation by conveyor-mode single electron shuttling in {{Si}}/{{SiGe}}},\ }\href {https://doi.org/10.1038/s41467-024-45583-7} {\bibfield  {journal} {\bibinfo  {journal} {Nature Communications}\ }\textbf {\bibinfo {volume} {15}},\ \bibinfo {pages} {1325} (\bibinfo {year} {2024})}\BibitemShut
  {NoStop}%
\bibitem [{\citenamefont {De~Smet}\ \emph {et~al.}(2025)\citenamefont {De~Smet}, \citenamefont {Matsumoto}, \citenamefont {Zwerver}, \citenamefont {Tryputen}, \citenamefont {de~Snoo}, \citenamefont {Amitonov}, \citenamefont {Katiraee-Far}, \citenamefont {Sammak}, \citenamefont {Samkharadze}, \citenamefont {G\"{u}l}, \citenamefont {Wasserman}, \citenamefont {Greplov{\'a}}, \citenamefont {Rimbach-Russ}, \citenamefont {Scappucci},\ and\ \citenamefont {Vandersypen}}]{DeSmet2025}%
  \BibitemOpen
  \bibfield  {author} {\bibinfo {author} {\bibfnamefont {M.}~\bibnamefont {De~Smet}}, \bibinfo {author} {\bibfnamefont {Y.}~\bibnamefont {Matsumoto}}, \bibinfo {author} {\bibfnamefont {A.-M.~J.}\ \bibnamefont {Zwerver}}, \bibinfo {author} {\bibfnamefont {L.}~\bibnamefont {Tryputen}}, \bibinfo {author} {\bibfnamefont {S.~L.}\ \bibnamefont {de~Snoo}}, \bibinfo {author} {\bibfnamefont {S.~V.}\ \bibnamefont {Amitonov}}, \bibinfo {author} {\bibfnamefont {S.~R.}\ \bibnamefont {Katiraee-Far}}, \bibinfo {author} {\bibfnamefont {A.}~\bibnamefont {Sammak}}, \bibinfo {author} {\bibfnamefont {N.}~\bibnamefont {Samkharadze}}, \bibinfo {author} {\bibfnamefont {O.}~\bibnamefont {G\"{u}l}}, \bibinfo {author} {\bibfnamefont {R.~N.~M.}\ \bibnamefont {Wasserman}}, \bibinfo {author} {\bibfnamefont {E.}~\bibnamefont {Greplov{\'a}}}, \bibinfo {author} {\bibfnamefont {M.}~\bibnamefont {Rimbach-Russ}}, \bibinfo {author} {\bibfnamefont {G.}~\bibnamefont {Scappucci}},\ and\ \bibinfo {author} {\bibfnamefont {L.~M.~K.}\ \bibnamefont
  {Vandersypen}},\ }\bibfield  {title} {\bibinfo {title} {High-fidelity single-spin shuttling in silicon},\ }\href {https://doi.org/10.1038/s41565-025-01920-5} {\bibfield  {journal} {\bibinfo  {journal} {Nature Nanotechnology}\ }\textbf {\bibinfo {volume} {20}},\ \bibinfo {pages} {866} (\bibinfo {year} {2025})}\BibitemShut {NoStop}%
\bibitem [{\citenamefont {Xue}\ \emph {et~al.}(2024)\citenamefont {Xue}, \citenamefont {Beer}, \citenamefont {Seidler}, \citenamefont {Humpohl}, \citenamefont {Tu}, \citenamefont {Trellenkamp}, \citenamefont {Struck}, \citenamefont {Bluhm},\ and\ \citenamefont {Schreiber}}]{xueSiSiGeQuBus2024}%
  \BibitemOpen
  \bibfield  {author} {\bibinfo {author} {\bibfnamefont {R.}~\bibnamefont {Xue}}, \bibinfo {author} {\bibfnamefont {M.}~\bibnamefont {Beer}}, \bibinfo {author} {\bibfnamefont {I.}~\bibnamefont {Seidler}}, \bibinfo {author} {\bibfnamefont {S.}~\bibnamefont {Humpohl}}, \bibinfo {author} {\bibfnamefont {J.-S.}\ \bibnamefont {Tu}}, \bibinfo {author} {\bibfnamefont {S.}~\bibnamefont {Trellenkamp}}, \bibinfo {author} {\bibfnamefont {T.}~\bibnamefont {Struck}}, \bibinfo {author} {\bibfnamefont {H.}~\bibnamefont {Bluhm}},\ and\ \bibinfo {author} {\bibfnamefont {L.~R.}\ \bibnamefont {Schreiber}},\ }\bibfield  {title} {\bibinfo {title} {Si/{{SiGe QuBus}} for single electron information-processing devices with memory and micron-scale connectivity function},\ }\href {https://doi.org/10.1038/s41467-024-46519-x} {\bibfield  {journal} {\bibinfo  {journal} {Nature Communications}\ }\textbf {\bibinfo {volume} {15}},\ \bibinfo {pages} {2296} (\bibinfo {year} {2024})}\BibitemShut {NoStop}%
\bibitem [{\citenamefont {Ademi}\ \emph {et~al.}(2025)\citenamefont {Ademi}, \citenamefont {Bassi}, \citenamefont {Yu}, \citenamefont {de~Snoo}, \citenamefont {Oosterhout}, \citenamefont {Sammak}, \citenamefont {Vandersypen}, \citenamefont {Scappucci}, \citenamefont {D{\'e}prez},\ and\ \citenamefont {Veldhorst}}]{ademi2025distributingentanglementdistantsemiconductor}%
  \BibitemOpen
  \bibfield  {author} {\bibinfo {author} {\bibfnamefont {Z.}~\bibnamefont {Ademi}}, \bibinfo {author} {\bibfnamefont {M.}~\bibnamefont {Bassi}}, \bibinfo {author} {\bibfnamefont {C.~X.}\ \bibnamefont {Yu}}, \bibinfo {author} {\bibfnamefont {S.~L.}\ \bibnamefont {de~Snoo}}, \bibinfo {author} {\bibfnamefont {S.~D.}\ \bibnamefont {Oosterhout}}, \bibinfo {author} {\bibfnamefont {A.}~\bibnamefont {Sammak}}, \bibinfo {author} {\bibfnamefont {L.~M.~K.}\ \bibnamefont {Vandersypen}}, \bibinfo {author} {\bibfnamefont {G.}~\bibnamefont {Scappucci}}, \bibinfo {author} {\bibfnamefont {C.}~\bibnamefont {D{\'e}prez}},\ and\ \bibinfo {author} {\bibfnamefont {M.}~\bibnamefont {Veldhorst}},\ }\href {https://arxiv.org/abs/2510.26860} {\bibinfo {title} {Distributing entanglement between distant semiconductor qubit registers using a shared-control shuttling link}} (\bibinfo {year} {2025}),\ \Eprint {https://arxiv.org/abs/2510.26860} {arXiv:2510.26860} \BibitemShut {NoStop}%
\bibitem [{\citenamefont {Krzywda}\ \emph {et~al.}(2026)\citenamefont {Krzywda}, \citenamefont {Matsumoto}, \citenamefont {Smet}, \citenamefont {Tryputen}, \citenamefont {de~Snoo}, \citenamefont {Amitonov}, \citenamefont {van Nieuwenburg}, \citenamefont {Scappucci},\ and\ \citenamefont {Vandersypen}}]{krzywda2026coherenceprotectionmobilespin}%
  \BibitemOpen
  \bibfield  {author} {\bibinfo {author} {\bibfnamefont {J.~A.}\ \bibnamefont {Krzywda}}, \bibinfo {author} {\bibfnamefont {Y.}~\bibnamefont {Matsumoto}}, \bibinfo {author} {\bibfnamefont {M.~D.}\ \bibnamefont {Smet}}, \bibinfo {author} {\bibfnamefont {L.}~\bibnamefont {Tryputen}}, \bibinfo {author} {\bibfnamefont {S.~L.}\ \bibnamefont {de~Snoo}}, \bibinfo {author} {\bibfnamefont {S.~V.}\ \bibnamefont {Amitonov}}, \bibinfo {author} {\bibfnamefont {E.}~\bibnamefont {van Nieuwenburg}}, \bibinfo {author} {\bibfnamefont {G.}~\bibnamefont {Scappucci}},\ and\ \bibinfo {author} {\bibfnamefont {L.~M.~K.}\ \bibnamefont {Vandersypen}},\ }\href {https://arxiv.org/abs/2602.09179} {\bibinfo {title} {Coherence protection for mobile spin qubits in silicon}} (\bibinfo {year} {2026}),\ \Eprint {https://arxiv.org/abs/2602.09179} {arXiv:2602.09179} \BibitemShut {NoStop}%
\bibitem [{\citenamefont {Matsumoto}\ \emph {et~al.}(2025)\citenamefont {Matsumoto}, \citenamefont {Smet}, \citenamefont {Tryputen}, \citenamefont {de~Snoo}, \citenamefont {Amitonov}, \citenamefont {Sammak}, \citenamefont {{Rimbach-Russ}}, \citenamefont {Scappucci},\ and\ \citenamefont {Vandersypen}}]{matsumoto2025twoqubitlogicteleportationmobile}%
  \BibitemOpen
  \bibfield  {author} {\bibinfo {author} {\bibfnamefont {Y.}~\bibnamefont {Matsumoto}}, \bibinfo {author} {\bibfnamefont {M.~D.}\ \bibnamefont {Smet}}, \bibinfo {author} {\bibfnamefont {L.}~\bibnamefont {Tryputen}}, \bibinfo {author} {\bibfnamefont {S.~L.}\ \bibnamefont {de~Snoo}}, \bibinfo {author} {\bibfnamefont {S.~V.}\ \bibnamefont {Amitonov}}, \bibinfo {author} {\bibfnamefont {A.}~\bibnamefont {Sammak}}, \bibinfo {author} {\bibfnamefont {M.}~\bibnamefont {{Rimbach-Russ}}}, \bibinfo {author} {\bibfnamefont {G.}~\bibnamefont {Scappucci}},\ and\ \bibinfo {author} {\bibfnamefont {L.~M.~K.}\ \bibnamefont {Vandersypen}},\ }\href {https://doi.org/10.48550/arXiv.2503.15434} {\bibinfo {title} {Two-qubit logic and teleportation with mobile spin qubits in silicon}} (\bibinfo {year} {2025}),\ \Eprint {https://arxiv.org/abs/2503.15434} {2503.15434} \BibitemShut {NoStop}%
\bibitem [{\citenamefont {Chekhovich}\ \emph {et~al.}(2013)\citenamefont {Chekhovich}, \citenamefont {Makhonin}, \citenamefont {Tartakovskii}, \citenamefont {Yacoby}, \citenamefont {Bluhm}, \citenamefont {Nowack},\ and\ \citenamefont {Vandersypen}}]{Chekhovich2013}%
  \BibitemOpen
  \bibfield  {author} {\bibinfo {author} {\bibfnamefont {E.~A.}\ \bibnamefont {Chekhovich}}, \bibinfo {author} {\bibfnamefont {M.~N.}\ \bibnamefont {Makhonin}}, \bibinfo {author} {\bibfnamefont {A.~I.}\ \bibnamefont {Tartakovskii}}, \bibinfo {author} {\bibfnamefont {A.}~\bibnamefont {Yacoby}}, \bibinfo {author} {\bibfnamefont {H.}~\bibnamefont {Bluhm}}, \bibinfo {author} {\bibfnamefont {K.~C.}\ \bibnamefont {Nowack}},\ and\ \bibinfo {author} {\bibfnamefont {L.~M.~K.}\ \bibnamefont {Vandersypen}},\ }\bibfield  {title} {\bibinfo {title} {Nuclear spin effects in semiconductor quantum dots},\ }\href {https://doi.org/10.1038/nmat3652} {\bibfield  {journal} {\bibinfo  {journal} {Nature Materials}\ }\textbf {\bibinfo {volume} {12}},\ \bibinfo {pages} {494} (\bibinfo {year} {2013})}\BibitemShut {NoStop}%
\bibitem [{\citenamefont {Schriefl}\ \emph {et~al.}(2006)\citenamefont {Schriefl}, \citenamefont {Makhlin}, \citenamefont {Shnirman},\ and\ \citenamefont {Sch\"{o}n}}]{Schriefl2006}%
  \BibitemOpen
  \bibfield  {author} {\bibinfo {author} {\bibfnamefont {J.}~\bibnamefont {Schriefl}}, \bibinfo {author} {\bibfnamefont {Y.}~\bibnamefont {Makhlin}}, \bibinfo {author} {\bibfnamefont {A.}~\bibnamefont {Shnirman}},\ and\ \bibinfo {author} {\bibfnamefont {G.}~\bibnamefont {Sch\"{o}n}},\ }\bibfield  {title} {\bibinfo {title} {Decoherence from ensembles of two-level fluctuators},\ }\href {https://doi.org/10.1088/1367-2630/8/1/001} {\bibfield  {journal} {\bibinfo  {journal} {New Journal of Physics}\ }\textbf {\bibinfo {volume} {8}},\ \bibinfo {pages} {1} (\bibinfo {year} {2006})}\BibitemShut {NoStop}%
\bibitem [{\citenamefont {Struck}\ \emph {et~al.}(2020)\citenamefont {Struck}, \citenamefont {Hollmann}, \citenamefont {Schauer}, \citenamefont {Fedorets}, \citenamefont {Schmidbauer}, \citenamefont {Sawano}, \citenamefont {Riemann}, \citenamefont {Abrosimov}, \citenamefont {Cywi{\'n}ski}, \citenamefont {Bougeard},\ and\ \citenamefont {Schreiber}}]{struckLowfrequencySpinQubit2020}%
  \BibitemOpen
  \bibfield  {author} {\bibinfo {author} {\bibfnamefont {T.}~\bibnamefont {Struck}}, \bibinfo {author} {\bibfnamefont {A.}~\bibnamefont {Hollmann}}, \bibinfo {author} {\bibfnamefont {F.}~\bibnamefont {Schauer}}, \bibinfo {author} {\bibfnamefont {O.}~\bibnamefont {Fedorets}}, \bibinfo {author} {\bibfnamefont {A.}~\bibnamefont {Schmidbauer}}, \bibinfo {author} {\bibfnamefont {K.}~\bibnamefont {Sawano}}, \bibinfo {author} {\bibfnamefont {H.}~\bibnamefont {Riemann}}, \bibinfo {author} {\bibfnamefont {N.~V.}\ \bibnamefont {Abrosimov}}, \bibinfo {author} {\bibfnamefont {{\L}.}~\bibnamefont {Cywi{\'n}ski}}, \bibinfo {author} {\bibfnamefont {D.}~\bibnamefont {Bougeard}},\ and\ \bibinfo {author} {\bibfnamefont {L.~R.}\ \bibnamefont {Schreiber}},\ }\bibfield  {title} {\bibinfo {title} {Low-frequency spin qubit energy splitting noise in highly purified {{28Si}}/{{SiGe}}},\ }\href {https://doi.org/10.1038/s41534-020-0276-2} {\bibfield  {journal} {\bibinfo  {journal} {npj Quantum Information}\ }\textbf {\bibinfo {volume}
  {6}},\ \bibinfo {pages} {40} (\bibinfo {year} {2020})}\BibitemShut {NoStop}%
\bibitem [{\citenamefont {Paquelet~Wuetz}\ \emph {et~al.}(2023)\citenamefont {Paquelet~Wuetz}, \citenamefont {Degli~Esposti}, \citenamefont {Zwerver}, \citenamefont {Amitonov}, \citenamefont {Botifoll}, \citenamefont {Arbiol}, \citenamefont {Vandersypen}, \citenamefont {Russ},\ and\ \citenamefont {Scappucci}}]{paqueletwuetzReducingChargeNoise2023}%
  \BibitemOpen
  \bibfield  {author} {\bibinfo {author} {\bibfnamefont {B.}~\bibnamefont {Paquelet~Wuetz}}, \bibinfo {author} {\bibfnamefont {D.}~\bibnamefont {Degli~Esposti}}, \bibinfo {author} {\bibfnamefont {A.-M.~J.}\ \bibnamefont {Zwerver}}, \bibinfo {author} {\bibfnamefont {S.~V.}\ \bibnamefont {Amitonov}}, \bibinfo {author} {\bibfnamefont {M.}~\bibnamefont {Botifoll}}, \bibinfo {author} {\bibfnamefont {J.}~\bibnamefont {Arbiol}}, \bibinfo {author} {\bibfnamefont {L.~M.~K.}\ \bibnamefont {Vandersypen}}, \bibinfo {author} {\bibfnamefont {M.}~\bibnamefont {Russ}},\ and\ \bibinfo {author} {\bibfnamefont {G.}~\bibnamefont {Scappucci}},\ }\bibfield  {title} {\bibinfo {title} {Reducing charge noise in quantum dots by using thin silicon quantum wells},\ }\href {https://doi.org/10.1038/s41467-023-36951-w} {\bibfield  {journal} {\bibinfo  {journal} {Nature Communications}\ }\textbf {\bibinfo {volume} {14}},\ \bibinfo {pages} {1385} (\bibinfo {year} {2023})}\BibitemShut {NoStop}%
\bibitem [{\citenamefont {K{\c e}pa}\ \emph {et~al.}(2023)\citenamefont {K{\c e}pa}, \citenamefont {Focke}, \citenamefont {Cywi{\'n}ski},\ and\ \citenamefont {Krzywda}}]{Kpa2023}%
  \BibitemOpen
  \bibfield  {author} {\bibinfo {author} {\bibfnamefont {M.}~\bibnamefont {K{\c e}pa}}, \bibinfo {author} {\bibfnamefont {N.}~\bibnamefont {Focke}}, \bibinfo {author} {\bibfnamefont {{\L}.}~\bibnamefont {Cywi{\'n}ski}},\ and\ \bibinfo {author} {\bibfnamefont {J.~A.}\ \bibnamefont {Krzywda}},\ }\bibfield  {title} {\bibinfo {title} {Simulation of 1 / f charge noise affecting a quantum dot in a {{Si}}/{{SiGe}} structure},\ }\href {https://doi.org/10.1063/5.0151029} {\bibfield  {journal} {\bibinfo  {journal} {Applied Physics Letters}\ }\textbf {\bibinfo {volume} {123}},\ \bibinfo {pages} {034005} (\bibinfo {year} {2023})}\BibitemShut {NoStop}%
\bibitem [{\citenamefont {Shehata}\ \emph {et~al.}(2023)\citenamefont {Shehata}, \citenamefont {Simion}, \citenamefont {Li}, \citenamefont {Mohiyaddin}, \citenamefont {Wan}, \citenamefont {Mongillo}, \citenamefont {Govoreanu}, \citenamefont {Radu}, \citenamefont {De~Greve},\ and\ \citenamefont {Van~Dorpe}}]{shehataModelingSemiconductorSpin2023}%
  \BibitemOpen
  \bibfield  {author} {\bibinfo {author} {\bibfnamefont {M.~M. E.~K.}\ \bibnamefont {Shehata}}, \bibinfo {author} {\bibfnamefont {G.}~\bibnamefont {Simion}}, \bibinfo {author} {\bibfnamefont {R.}~\bibnamefont {Li}}, \bibinfo {author} {\bibfnamefont {F.~A.}\ \bibnamefont {Mohiyaddin}}, \bibinfo {author} {\bibfnamefont {D.}~\bibnamefont {Wan}}, \bibinfo {author} {\bibfnamefont {M.}~\bibnamefont {Mongillo}}, \bibinfo {author} {\bibfnamefont {B.}~\bibnamefont {Govoreanu}}, \bibinfo {author} {\bibfnamefont {I.}~\bibnamefont {Radu}}, \bibinfo {author} {\bibfnamefont {K.}~\bibnamefont {De~Greve}},\ and\ \bibinfo {author} {\bibfnamefont {P.}~\bibnamefont {Van~Dorpe}},\ }\bibfield  {title} {\bibinfo {title} {Modeling semiconductor spin qubits and their charge noise environment for quantum gate fidelity estimation},\ }\href {https://doi.org/10.1103/PhysRevB.108.045305} {\bibfield  {journal} {\bibinfo  {journal} {Physical Review B}\ }\textbf {\bibinfo {volume} {108}},\ \bibinfo {pages} {045305} (\bibinfo {year}
  {2023})}\BibitemShut {NoStop}%
\bibitem [{\citenamefont {Wang}\ \emph {et~al.}(2025)\citenamefont {Wang}, \citenamefont {Gholizadeh}, \citenamefont {Hu}, \citenamefont {Das~Sarma},\ and\ \citenamefont {Culcer}}]{wangDephasingPlanarGe2025}%
  \BibitemOpen
  \bibfield  {author} {\bibinfo {author} {\bibfnamefont {Z.}~\bibnamefont {Wang}}, \bibinfo {author} {\bibfnamefont {S.}~\bibnamefont {Gholizadeh}}, \bibinfo {author} {\bibfnamefont {X.}~\bibnamefont {Hu}}, \bibinfo {author} {\bibfnamefont {S.}~\bibnamefont {Das~Sarma}},\ and\ \bibinfo {author} {\bibfnamefont {D.}~\bibnamefont {Culcer}},\ }\bibfield  {title} {\bibinfo {title} {Dephasing of planar {{Ge}} hole spin qubits due to $1/f$ charge noise},\ }\href {https://doi.org/10.1103/PhysRevB.111.155403} {\bibfield  {journal} {\bibinfo  {journal} {Physical Review B}\ }\textbf {\bibinfo {volume} {111}},\ \bibinfo {pages} {155403} (\bibinfo {year} {2025})}\BibitemShut {NoStop}%
\bibitem [{\citenamefont {Wang}\ \emph {et~al.}(2024{\natexlab{a}})\citenamefont {Wang}, \citenamefont {Ercan}, \citenamefont {Gyure}, \citenamefont {Scappucci}, \citenamefont {Veldhorst},\ and\ \citenamefont {{Rimbach-Russ}}}]{wangModelingPlanarGermanium2024}%
  \BibitemOpen
  \bibfield  {author} {\bibinfo {author} {\bibfnamefont {C.-A.}\ \bibnamefont {Wang}}, \bibinfo {author} {\bibfnamefont {H.~E.}\ \bibnamefont {Ercan}}, \bibinfo {author} {\bibfnamefont {M.~F.}\ \bibnamefont {Gyure}}, \bibinfo {author} {\bibfnamefont {G.}~\bibnamefont {Scappucci}}, \bibinfo {author} {\bibfnamefont {M.}~\bibnamefont {Veldhorst}},\ and\ \bibinfo {author} {\bibfnamefont {M.}~\bibnamefont {{Rimbach-Russ}}},\ }\bibfield  {title} {\bibinfo {title} {Modeling of planar germanium hole qubits in electric and magnetic fields},\ }\href {https://doi.org/10.1038/s41534-024-00897-8} {\bibfield  {journal} {\bibinfo  {journal} {npj Quantum Information}\ }\textbf {\bibinfo {volume} {10}},\ \bibinfo {pages} {1} (\bibinfo {year} {2024}{\natexlab{a}})}\BibitemShut {NoStop}%
\bibitem [{\citenamefont {Wang}\ \emph {et~al.}(2022)\citenamefont {Wang}, \citenamefont {Xu}, \citenamefont {Gao}, \citenamefont {Liu}, \citenamefont {Ma}, \citenamefont {Zhang}, \citenamefont {Wang}, \citenamefont {Cao}, \citenamefont {Wang}, \citenamefont {Zhang}, \citenamefont {Culcer}, \citenamefont {Hu}, \citenamefont {Jiang}, \citenamefont {Li}, \citenamefont {Guo},\ and\ \citenamefont {Guo}}]{wangUltrafastCoherentControl2022}%
  \BibitemOpen
  \bibfield  {author} {\bibinfo {author} {\bibfnamefont {K.}~\bibnamefont {Wang}}, \bibinfo {author} {\bibfnamefont {G.}~\bibnamefont {Xu}}, \bibinfo {author} {\bibfnamefont {F.}~\bibnamefont {Gao}}, \bibinfo {author} {\bibfnamefont {H.}~\bibnamefont {Liu}}, \bibinfo {author} {\bibfnamefont {R.-L.}\ \bibnamefont {Ma}}, \bibinfo {author} {\bibfnamefont {X.}~\bibnamefont {Zhang}}, \bibinfo {author} {\bibfnamefont {Z.}~\bibnamefont {Wang}}, \bibinfo {author} {\bibfnamefont {G.}~\bibnamefont {Cao}}, \bibinfo {author} {\bibfnamefont {T.}~\bibnamefont {Wang}}, \bibinfo {author} {\bibfnamefont {J.-J.}\ \bibnamefont {Zhang}}, \bibinfo {author} {\bibfnamefont {D.}~\bibnamefont {Culcer}}, \bibinfo {author} {\bibfnamefont {X.}~\bibnamefont {Hu}}, \bibinfo {author} {\bibfnamefont {H.-W.}\ \bibnamefont {Jiang}}, \bibinfo {author} {\bibfnamefont {H.-O.}\ \bibnamefont {Li}}, \bibinfo {author} {\bibfnamefont {G.-C.}\ \bibnamefont {Guo}},\ and\ \bibinfo {author} {\bibfnamefont {G.-P.}\ \bibnamefont {Guo}},\ }\bibfield
  {title} {\bibinfo {title} {Ultrafast coherent control of a hole spin qubit in a germanium quantum dot},\ }\href {https://doi.org/10.1038/s41467-021-27880-7} {\bibfield  {journal} {\bibinfo  {journal} {Nature Communications}\ }\textbf {\bibinfo {volume} {13}},\ \bibinfo {pages} {206} (\bibinfo {year} {2022})}\BibitemShut {NoStop}%
\bibitem [{\citenamefont {Bosco}\ \emph {et~al.}(2024)\citenamefont {Bosco}, \citenamefont {Zou},\ and\ \citenamefont {Loss}}]{Bosco2024}%
  \BibitemOpen
  \bibfield  {author} {\bibinfo {author} {\bibfnamefont {S.}~\bibnamefont {Bosco}}, \bibinfo {author} {\bibfnamefont {J.}~\bibnamefont {Zou}},\ and\ \bibinfo {author} {\bibfnamefont {D.}~\bibnamefont {Loss}},\ }\bibfield  {title} {\bibinfo {title} {High-fidelity spin qubit shuttling via large spin-orbit interactions},\ }\href {https://doi.org/10.1103/PRXQuantum.5.020353} {\bibfield  {journal} {\bibinfo  {journal} {PRX Quantum}\ }\textbf {\bibinfo {volume} {5}},\ \bibinfo {pages} {020353} (\bibinfo {year} {2024})}\BibitemShut {NoStop}%
\bibitem [{\citenamefont {Cywi\ifmmode~\acute{n}\else \'{n}\fi{}ski}\ \emph {et~al.}(2008)\citenamefont {Cywi\ifmmode~\acute{n}\else \'{n}\fi{}ski}, \citenamefont {Lutchyn}, \citenamefont {Nave},\ and\ \citenamefont {Das~Sarma}}]{Cywinski2008-aj}%
  \BibitemOpen
  \bibfield  {author} {\bibinfo {author} {\bibfnamefont {L.}~\bibnamefont {Cywi\ifmmode~\acute{n}\else \'{n}\fi{}ski}}, \bibinfo {author} {\bibfnamefont {R.~M.}\ \bibnamefont {Lutchyn}}, \bibinfo {author} {\bibfnamefont {C.~P.}\ \bibnamefont {Nave}},\ and\ \bibinfo {author} {\bibfnamefont {S.}~\bibnamefont {Das~Sarma}},\ }\bibfield  {title} {\bibinfo {title} {How to enhance dephasing time in superconducting qubits},\ }\href {https://doi.org/10.1103/PhysRevB.77.174509} {\bibfield  {journal} {\bibinfo  {journal} {Phys. Rev. B}\ }\textbf {\bibinfo {volume} {77}},\ \bibinfo {pages} {174509} (\bibinfo {year} {2008})}\BibitemShut {NoStop}%
\bibitem [{\citenamefont {Laucht}\ \emph {et~al.}(2016)\citenamefont {Laucht}, \citenamefont {Kalra}, \citenamefont {Simmons}, \citenamefont {Dehollain}, \citenamefont {Muhonen}, \citenamefont {Mohiyaddin}, \citenamefont {Freer}, \citenamefont {Hudson}, \citenamefont {Itoh}, \citenamefont {Jamieson}, \citenamefont {McCallum}, \citenamefont {Dzurak},\ and\ \citenamefont {Morello}}]{Laucht2016}%
  \BibitemOpen
  \bibfield  {author} {\bibinfo {author} {\bibfnamefont {A.}~\bibnamefont {Laucht}}, \bibinfo {author} {\bibfnamefont {R.}~\bibnamefont {Kalra}}, \bibinfo {author} {\bibfnamefont {S.}~\bibnamefont {Simmons}}, \bibinfo {author} {\bibfnamefont {J.~P.}\ \bibnamefont {Dehollain}}, \bibinfo {author} {\bibfnamefont {J.~T.}\ \bibnamefont {Muhonen}}, \bibinfo {author} {\bibfnamefont {F.~A.}\ \bibnamefont {Mohiyaddin}}, \bibinfo {author} {\bibfnamefont {S.}~\bibnamefont {Freer}}, \bibinfo {author} {\bibfnamefont {F.~E.}\ \bibnamefont {Hudson}}, \bibinfo {author} {\bibfnamefont {K.~M.}\ \bibnamefont {Itoh}}, \bibinfo {author} {\bibfnamefont {D.~N.}\ \bibnamefont {Jamieson}}, \bibinfo {author} {\bibfnamefont {J.~C.}\ \bibnamefont {McCallum}}, \bibinfo {author} {\bibfnamefont {A.~S.}\ \bibnamefont {Dzurak}},\ and\ \bibinfo {author} {\bibfnamefont {A.}~\bibnamefont {Morello}},\ }\bibfield  {title} {\bibinfo {title} {A dressed spin qubit in silicon},\ }\href {https://doi.org/10.1038/nnano.2016.178} {\bibfield  {journal}
  {\bibinfo  {journal} {Nature Nanotechnology}\ }\textbf {\bibinfo {volume} {12}},\ \bibinfo {pages} {61} (\bibinfo {year} {2016})}\BibitemShut {NoStop}%
\bibitem [{\citenamefont {Tsoukalas}\ \emph {et~al.}(2026)\citenamefont {Tsoukalas}, \citenamefont {{von L{\"u}pke}}, \citenamefont {Orekhov}, \citenamefont {Het{\'e}nyi}, \citenamefont {Seidler}, \citenamefont {Sommer}, \citenamefont {Kelly}, \citenamefont {Massai}, \citenamefont {Aldeghi}, \citenamefont {{Pita-Vidal}}, \citenamefont {Hendrickx}, \citenamefont {Bedell}, \citenamefont {Paredes}, \citenamefont {Schupp}, \citenamefont {Mergenthaler}, \citenamefont {Salis}, \citenamefont {Fuhrer},\ and\ \citenamefont {{Harvey-Collard}}}]{tsoukalas2025dressedsinglettripletqubitgermanium}%
  \BibitemOpen
  \bibfield  {author} {\bibinfo {author} {\bibfnamefont {K.}~\bibnamefont {Tsoukalas}}, \bibinfo {author} {\bibfnamefont {U.}~\bibnamefont {{von L{\"u}pke}}}, \bibinfo {author} {\bibfnamefont {A.}~\bibnamefont {Orekhov}}, \bibinfo {author} {\bibfnamefont {B.}~\bibnamefont {Het{\'e}nyi}}, \bibinfo {author} {\bibfnamefont {I.}~\bibnamefont {Seidler}}, \bibinfo {author} {\bibfnamefont {L.}~\bibnamefont {Sommer}}, \bibinfo {author} {\bibfnamefont {E.~G.}\ \bibnamefont {Kelly}}, \bibinfo {author} {\bibfnamefont {L.}~\bibnamefont {Massai}}, \bibinfo {author} {\bibfnamefont {M.}~\bibnamefont {Aldeghi}}, \bibinfo {author} {\bibfnamefont {M.}~\bibnamefont {{Pita-Vidal}}}, \bibinfo {author} {\bibfnamefont {N.~W.}\ \bibnamefont {Hendrickx}}, \bibinfo {author} {\bibfnamefont {S.~W.}\ \bibnamefont {Bedell}}, \bibinfo {author} {\bibfnamefont {S.}~\bibnamefont {Paredes}}, \bibinfo {author} {\bibfnamefont {F.~J.}\ \bibnamefont {Schupp}}, \bibinfo {author} {\bibfnamefont {M.}~\bibnamefont {Mergenthaler}}, \bibinfo {author}
  {\bibfnamefont {G.}~\bibnamefont {Salis}}, \bibinfo {author} {\bibfnamefont {A.}~\bibnamefont {Fuhrer}},\ and\ \bibinfo {author} {\bibfnamefont {P.}~\bibnamefont {{Harvey-Collard}}},\ }\bibfield  {title} {\bibinfo {title} {A dressed singlet-triplet qubit in germanium},\ }\href {https://doi.org/10.1038/s41467-025-65569-3} {\bibfield  {journal} {\bibinfo  {journal} {Nature Communications}\ }\textbf {\bibinfo {volume} {17}},\ \bibinfo {pages} {699} (\bibinfo {year} {2026})}\BibitemShut {NoStop}%
\bibitem [{\citenamefont {Scappucci}\ \emph {et~al.}(2020)\citenamefont {Scappucci}, \citenamefont {Kloeffel}, \citenamefont {Zwanenburg}, \citenamefont {Loss}, \citenamefont {Myronov}, \citenamefont {Zhang}, \citenamefont {De~Franceschi}, \citenamefont {Katsaros},\ and\ \citenamefont {Veldhorst}}]{Scappucci2020}%
  \BibitemOpen
  \bibfield  {author} {\bibinfo {author} {\bibfnamefont {G.}~\bibnamefont {Scappucci}}, \bibinfo {author} {\bibfnamefont {C.}~\bibnamefont {Kloeffel}}, \bibinfo {author} {\bibfnamefont {F.~A.}\ \bibnamefont {Zwanenburg}}, \bibinfo {author} {\bibfnamefont {D.}~\bibnamefont {Loss}}, \bibinfo {author} {\bibfnamefont {M.}~\bibnamefont {Myronov}}, \bibinfo {author} {\bibfnamefont {J.-J.}\ \bibnamefont {Zhang}}, \bibinfo {author} {\bibfnamefont {S.}~\bibnamefont {De~Franceschi}}, \bibinfo {author} {\bibfnamefont {G.}~\bibnamefont {Katsaros}},\ and\ \bibinfo {author} {\bibfnamefont {M.}~\bibnamefont {Veldhorst}},\ }\bibfield  {title} {\bibinfo {title} {The germanium quantum information route},\ }\href {https://doi.org/10.1038/s41578-020-00262-z} {\bibfield  {journal} {\bibinfo  {journal} {Nature Reviews Materials}\ }\textbf {\bibinfo {volume} {6}},\ \bibinfo {pages} {926} (\bibinfo {year} {2020})}\BibitemShut {NoStop}%
\bibitem [{\citenamefont {Jirovec}\ \emph {et~al.}(2021)\citenamefont {Jirovec}, \citenamefont {Hofmann}, \citenamefont {Ballabio}, \citenamefont {Mutter}, \citenamefont {Tavani}, \citenamefont {Botifoll}, \citenamefont {Crippa}, \citenamefont {Kukucka}, \citenamefont {Sagi}, \citenamefont {Martins}, \citenamefont {{Saez-Mollejo}}, \citenamefont {Prieto}, \citenamefont {Borovkov}, \citenamefont {Arbiol}, \citenamefont {Chrastina}, \citenamefont {Isella},\ and\ \citenamefont {Katsaros}}]{jirovecSinglettripletHoleSpin2021}%
  \BibitemOpen
  \bibfield  {author} {\bibinfo {author} {\bibfnamefont {D.}~\bibnamefont {Jirovec}}, \bibinfo {author} {\bibfnamefont {A.}~\bibnamefont {Hofmann}}, \bibinfo {author} {\bibfnamefont {A.}~\bibnamefont {Ballabio}}, \bibinfo {author} {\bibfnamefont {P.~M.}\ \bibnamefont {Mutter}}, \bibinfo {author} {\bibfnamefont {G.}~\bibnamefont {Tavani}}, \bibinfo {author} {\bibfnamefont {M.}~\bibnamefont {Botifoll}}, \bibinfo {author} {\bibfnamefont {A.}~\bibnamefont {Crippa}}, \bibinfo {author} {\bibfnamefont {J.}~\bibnamefont {Kukucka}}, \bibinfo {author} {\bibfnamefont {O.}~\bibnamefont {Sagi}}, \bibinfo {author} {\bibfnamefont {F.}~\bibnamefont {Martins}}, \bibinfo {author} {\bibfnamefont {J.}~\bibnamefont {{Saez-Mollejo}}}, \bibinfo {author} {\bibfnamefont {I.}~\bibnamefont {Prieto}}, \bibinfo {author} {\bibfnamefont {M.}~\bibnamefont {Borovkov}}, \bibinfo {author} {\bibfnamefont {J.}~\bibnamefont {Arbiol}}, \bibinfo {author} {\bibfnamefont {D.}~\bibnamefont {Chrastina}}, \bibinfo {author} {\bibfnamefont
  {G.}~\bibnamefont {Isella}},\ and\ \bibinfo {author} {\bibfnamefont {G.}~\bibnamefont {Katsaros}},\ }\bibfield  {title} {\bibinfo {title} {A singlet-triplet hole spin qubit in planar {{Ge}}},\ }\href {https://doi.org/10.1038/s41563-021-01022-2} {\bibfield  {journal} {\bibinfo  {journal} {Nature Materials}\ }\textbf {\bibinfo {volume} {20}},\ \bibinfo {pages} {1106} (\bibinfo {year} {2021})}\BibitemShut {NoStop}%
\bibitem [{\citenamefont {Hendrickx}\ \emph {et~al.}(2021)\citenamefont {Hendrickx}, \citenamefont {Lawrie}, \citenamefont {Russ}, \citenamefont {{van Riggelen}}, \citenamefont {{de Snoo}}, \citenamefont {Schouten}, \citenamefont {Sammak}, \citenamefont {Scappucci},\ and\ \citenamefont {Veldhorst}}]{hendrickxFourqubitGermaniumQuantum2021}%
  \BibitemOpen
  \bibfield  {author} {\bibinfo {author} {\bibfnamefont {N.~W.}\ \bibnamefont {Hendrickx}}, \bibinfo {author} {\bibfnamefont {W.~I.~L.}\ \bibnamefont {Lawrie}}, \bibinfo {author} {\bibfnamefont {M.}~\bibnamefont {Russ}}, \bibinfo {author} {\bibfnamefont {F.}~\bibnamefont {{van Riggelen}}}, \bibinfo {author} {\bibfnamefont {S.~L.}\ \bibnamefont {{de Snoo}}}, \bibinfo {author} {\bibfnamefont {R.~N.}\ \bibnamefont {Schouten}}, \bibinfo {author} {\bibfnamefont {A.}~\bibnamefont {Sammak}}, \bibinfo {author} {\bibfnamefont {G.}~\bibnamefont {Scappucci}},\ and\ \bibinfo {author} {\bibfnamefont {M.}~\bibnamefont {Veldhorst}},\ }\bibfield  {title} {\bibinfo {title} {A four-qubit germanium quantum processor},\ }\href {https://doi.org/10.1038/s41586-021-03332-6} {\bibfield  {journal} {\bibinfo  {journal} {Nature}\ }\textbf {\bibinfo {volume} {591}},\ \bibinfo {pages} {580} (\bibinfo {year} {2021})}\BibitemShut {NoStop}%
\bibitem [{\citenamefont {Hendrickx}\ \emph {et~al.}(2024)\citenamefont {Hendrickx}, \citenamefont {Massai}, \citenamefont {Mergenthaler}, \citenamefont {Schupp}, \citenamefont {Paredes}, \citenamefont {Bedell}, \citenamefont {Salis},\ and\ \citenamefont {Fuhrer}}]{hendrickxSweetspotOperationGermanium2024}%
  \BibitemOpen
  \bibfield  {author} {\bibinfo {author} {\bibfnamefont {N.~W.}\ \bibnamefont {Hendrickx}}, \bibinfo {author} {\bibfnamefont {L.}~\bibnamefont {Massai}}, \bibinfo {author} {\bibfnamefont {M.}~\bibnamefont {Mergenthaler}}, \bibinfo {author} {\bibfnamefont {F.~J.}\ \bibnamefont {Schupp}}, \bibinfo {author} {\bibfnamefont {S.}~\bibnamefont {Paredes}}, \bibinfo {author} {\bibfnamefont {S.~W.}\ \bibnamefont {Bedell}}, \bibinfo {author} {\bibfnamefont {G.}~\bibnamefont {Salis}},\ and\ \bibinfo {author} {\bibfnamefont {A.}~\bibnamefont {Fuhrer}},\ }\bibfield  {title} {\bibinfo {title} {Sweet-spot operation of a germanium hole spin qubit with highly anisotropic noise sensitivity},\ }\href {https://doi.org/10.1038/s41563-024-01857-5} {\bibfield  {journal} {\bibinfo  {journal} {Nature Materials}\ }\textbf {\bibinfo {volume} {23}},\ \bibinfo {pages} {920} (\bibinfo {year} {2024})}\BibitemShut {NoStop}%
\bibitem [{\citenamefont {Wang}\ \emph {et~al.}(2024{\natexlab{b}})\citenamefont {Wang}, \citenamefont {John}, \citenamefont {Tidjani}, \citenamefont {Yu}, \citenamefont {Ivlev}, \citenamefont {D{\'e}prez}, \citenamefont {{van Riggelen-Doelman}}, \citenamefont {Woods}, \citenamefont {Hendrickx}, \citenamefont {Lawrie}, \citenamefont {Stehouwer}, \citenamefont {Oosterhout}, \citenamefont {Sammak}, \citenamefont {Friesen}, \citenamefont {Scappucci}, \citenamefont {{de Snoo}}, \citenamefont {{Rimbach-Russ}}, \citenamefont {Borsoi},\ and\ \citenamefont {Veldhorst}}]{wangOperatingSemiconductorQuantum2024}%
  \BibitemOpen
  \bibfield  {author} {\bibinfo {author} {\bibfnamefont {C.-A.}\ \bibnamefont {Wang}}, \bibinfo {author} {\bibfnamefont {V.}~\bibnamefont {John}}, \bibinfo {author} {\bibfnamefont {H.}~\bibnamefont {Tidjani}}, \bibinfo {author} {\bibfnamefont {C.~X.}\ \bibnamefont {Yu}}, \bibinfo {author} {\bibfnamefont {A.~S.}\ \bibnamefont {Ivlev}}, \bibinfo {author} {\bibfnamefont {C.}~\bibnamefont {D{\'e}prez}}, \bibinfo {author} {\bibfnamefont {F.}~\bibnamefont {{van Riggelen-Doelman}}}, \bibinfo {author} {\bibfnamefont {B.~D.}\ \bibnamefont {Woods}}, \bibinfo {author} {\bibfnamefont {N.~W.}\ \bibnamefont {Hendrickx}}, \bibinfo {author} {\bibfnamefont {W.~I.~L.}\ \bibnamefont {Lawrie}}, \bibinfo {author} {\bibfnamefont {L.~E.~A.}\ \bibnamefont {Stehouwer}}, \bibinfo {author} {\bibfnamefont {S.~D.}\ \bibnamefont {Oosterhout}}, \bibinfo {author} {\bibfnamefont {A.}~\bibnamefont {Sammak}}, \bibinfo {author} {\bibfnamefont {M.}~\bibnamefont {Friesen}}, \bibinfo {author} {\bibfnamefont {G.}~\bibnamefont {Scappucci}}, \bibinfo
  {author} {\bibfnamefont {S.~L.}\ \bibnamefont {{de Snoo}}}, \bibinfo {author} {\bibfnamefont {M.}~\bibnamefont {{Rimbach-Russ}}}, \bibinfo {author} {\bibfnamefont {F.}~\bibnamefont {Borsoi}},\ and\ \bibinfo {author} {\bibfnamefont {M.}~\bibnamefont {Veldhorst}},\ }\bibfield  {title} {\bibinfo {title} {Operating semiconductor quantum processors with hopping spins},\ }\href {https://doi.org/10.1126/science.ado5915} {\bibfield  {journal} {\bibinfo  {journal} {Science}\ }\textbf {\bibinfo {volume} {385}},\ \bibinfo {pages} {447} (\bibinfo {year} {2024}{\natexlab{b}})}\BibitemShut {NoStop}%
\bibitem [{\citenamefont {{Saez-Mollejo}}\ \emph {et~al.}(2025)\citenamefont {{Saez-Mollejo}}, \citenamefont {Jirovec}, \citenamefont {Schell}, \citenamefont {Kukucka}, \citenamefont {Calcaterra}, \citenamefont {Chrastina}, \citenamefont {Isella}, \citenamefont {{Rimbach-Russ}}, \citenamefont {Bosco},\ and\ \citenamefont {Katsaros}}]{saez-mollejoExchangeAnisotropiesMicrowavedriven2025}%
  \BibitemOpen
  \bibfield  {author} {\bibinfo {author} {\bibfnamefont {J.}~\bibnamefont {{Saez-Mollejo}}}, \bibinfo {author} {\bibfnamefont {D.}~\bibnamefont {Jirovec}}, \bibinfo {author} {\bibfnamefont {Y.}~\bibnamefont {Schell}}, \bibinfo {author} {\bibfnamefont {J.}~\bibnamefont {Kukucka}}, \bibinfo {author} {\bibfnamefont {S.}~\bibnamefont {Calcaterra}}, \bibinfo {author} {\bibfnamefont {D.}~\bibnamefont {Chrastina}}, \bibinfo {author} {\bibfnamefont {G.}~\bibnamefont {Isella}}, \bibinfo {author} {\bibfnamefont {M.}~\bibnamefont {{Rimbach-Russ}}}, \bibinfo {author} {\bibfnamefont {S.}~\bibnamefont {Bosco}},\ and\ \bibinfo {author} {\bibfnamefont {G.}~\bibnamefont {Katsaros}},\ }\bibfield  {title} {\bibinfo {title} {Exchange anisotropies in microwave-driven singlet-triplet qubits},\ }\href {https://doi.org/10.1038/s41467-025-58969-y} {\bibfield  {journal} {\bibinfo  {journal} {Nature Communications}\ }\textbf {\bibinfo {volume} {16}},\ \bibinfo {pages} {3862} (\bibinfo {year} {2025})}\BibitemShut {NoStop}%
\bibitem [{\citenamefont {John}\ \emph {et~al.}(2025)\citenamefont {John}, \citenamefont {Yu}, \citenamefont {{van Straaten}}, \citenamefont {{Rodr{\'i}guez-Mena}}, \citenamefont {Rodr{\'i}guez}, \citenamefont {Oosterhout}, \citenamefont {Stehouwer}, \citenamefont {Scappucci}, \citenamefont {{Rimbach-Russ}}, \citenamefont {Bosco}, \citenamefont {Borsoi}, \citenamefont {Niquet},\ and\ \citenamefont {Veldhorst}}]{johnRobustLocalisedControl2025}%
  \BibitemOpen
  \bibfield  {author} {\bibinfo {author} {\bibfnamefont {V.}~\bibnamefont {John}}, \bibinfo {author} {\bibfnamefont {C.~X.}\ \bibnamefont {Yu}}, \bibinfo {author} {\bibfnamefont {B.}~\bibnamefont {{van Straaten}}}, \bibinfo {author} {\bibfnamefont {E.~A.}\ \bibnamefont {{Rodr{\'i}guez-Mena}}}, \bibinfo {author} {\bibfnamefont {M.}~\bibnamefont {Rodr{\'i}guez}}, \bibinfo {author} {\bibfnamefont {S.~D.}\ \bibnamefont {Oosterhout}}, \bibinfo {author} {\bibfnamefont {L.~E.~A.}\ \bibnamefont {Stehouwer}}, \bibinfo {author} {\bibfnamefont {G.}~\bibnamefont {Scappucci}}, \bibinfo {author} {\bibfnamefont {M.}~\bibnamefont {{Rimbach-Russ}}}, \bibinfo {author} {\bibfnamefont {S.}~\bibnamefont {Bosco}}, \bibinfo {author} {\bibfnamefont {F.}~\bibnamefont {Borsoi}}, \bibinfo {author} {\bibfnamefont {Y.-M.}\ \bibnamefont {Niquet}},\ and\ \bibinfo {author} {\bibfnamefont {M.}~\bibnamefont {Veldhorst}},\ }\bibfield  {title} {\bibinfo {title} {Robust and localised control of a 10-spin qubit array in germanium},\ }\href
  {https://doi.org/10.1038/s41467-025-65577-3} {\bibfield  {journal} {\bibinfo  {journal} {Nature Communications}\ }\textbf {\bibinfo {volume} {16}},\ \bibinfo {pages} {10560} (\bibinfo {year} {2025})}\BibitemShut {NoStop}%
\bibitem [{\citenamefont {Ivlev}\ \emph {et~al.}(2025)\citenamefont {Ivlev}, \citenamefont {Crielaard}, \citenamefont {Meyer}, \citenamefont {Lawrie}, \citenamefont {Hendrickx}, \citenamefont {Sammak}, \citenamefont {Matsumoto}, \citenamefont {Vandersypen}, \citenamefont {Scappucci}, \citenamefont {D{\'e}prez},\ and\ \citenamefont {Veldhorst}}]{ivlevOperatingSemiconductorQubits2025}%
  \BibitemOpen
  \bibfield  {author} {\bibinfo {author} {\bibfnamefont {A.~S.}\ \bibnamefont {Ivlev}}, \bibinfo {author} {\bibfnamefont {D.~R.}\ \bibnamefont {Crielaard}}, \bibinfo {author} {\bibfnamefont {M.}~\bibnamefont {Meyer}}, \bibinfo {author} {\bibfnamefont {W.~I.~L.}\ \bibnamefont {Lawrie}}, \bibinfo {author} {\bibfnamefont {N.~W.}\ \bibnamefont {Hendrickx}}, \bibinfo {author} {\bibfnamefont {A.}~\bibnamefont {Sammak}}, \bibinfo {author} {\bibfnamefont {Y.}~\bibnamefont {Matsumoto}}, \bibinfo {author} {\bibfnamefont {L.~M.~K.}\ \bibnamefont {Vandersypen}}, \bibinfo {author} {\bibfnamefont {G.}~\bibnamefont {Scappucci}}, \bibinfo {author} {\bibfnamefont {C.}~\bibnamefont {D{\'e}prez}},\ and\ \bibinfo {author} {\bibfnamefont {M.}~\bibnamefont {Veldhorst}},\ }\bibfield  {title} {\bibinfo {title} {Operating {{Semiconductor Qubits}} without {{Individual Barrier Gates}}},\ }\href {https://doi.org/10.1103/xhq3-4jxz} {\bibfield  {journal} {\bibinfo  {journal} {Physical Review X}\ }\textbf {\bibinfo {volume} {15}},\
  \bibinfo {pages} {031042} (\bibinfo {year} {2025})}\BibitemShut {NoStop}%
\bibitem [{\citenamefont {Bosco}\ \emph {et~al.}(2021{\natexlab{a}})\citenamefont {Bosco}, \citenamefont {Benito}, \citenamefont {Adelsberger},\ and\ \citenamefont {Loss}}]{Bosco_2021}%
  \BibitemOpen
  \bibfield  {author} {\bibinfo {author} {\bibfnamefont {S.}~\bibnamefont {Bosco}}, \bibinfo {author} {\bibfnamefont {M.}~\bibnamefont {Benito}}, \bibinfo {author} {\bibfnamefont {C.}~\bibnamefont {Adelsberger}},\ and\ \bibinfo {author} {\bibfnamefont {D.}~\bibnamefont {Loss}},\ }\bibfield  {title} {\bibinfo {title} {Squeezed hole spin qubits in ge quantum dots with ultrafast gates at low power},\ }\href {https://doi.org/10.1103/PhysRevB.104.115425} {\bibfield  {journal} {\bibinfo  {journal} {Phys. Rev. B}\ }\textbf {\bibinfo {volume} {104}},\ \bibinfo {pages} {115425} (\bibinfo {year} {2021}{\natexlab{a}})}\BibitemShut {NoStop}%
\bibitem [{\citenamefont {Martinez}\ \emph {et~al.}(2022)\citenamefont {Martinez}, \citenamefont {Abadillo-Uriel}, \citenamefont {Rodr\'{\i}guez-Mena},\ and\ \citenamefont {Niquet}}]{Martinez_2022}%
  \BibitemOpen
  \bibfield  {author} {\bibinfo {author} {\bibfnamefont {B.}~\bibnamefont {Martinez}}, \bibinfo {author} {\bibfnamefont {J.~C.}\ \bibnamefont {Abadillo-Uriel}}, \bibinfo {author} {\bibfnamefont {E.~A.}\ \bibnamefont {Rodr\'{\i}guez-Mena}},\ and\ \bibinfo {author} {\bibfnamefont {Y.-M.}\ \bibnamefont {Niquet}},\ }\bibfield  {title} {\bibinfo {title} {Hole spin manipulation in inhomogeneous and nonseparable electric fields},\ }\href {https://doi.org/10.1103/PhysRevB.106.235426} {\bibfield  {journal} {\bibinfo  {journal} {Phys. Rev. B}\ }\textbf {\bibinfo {volume} {106}},\ \bibinfo {pages} {235426} (\bibinfo {year} {2022})}\BibitemShut {NoStop}%
\bibitem [{\citenamefont {Sarkar}\ \emph {et~al.}(2023)\citenamefont {Sarkar}, \citenamefont {Wang}, \citenamefont {Rendell}, \citenamefont {Hendrickx}, \citenamefont {Veldhorst}, \citenamefont {Scappucci}, \citenamefont {Khalifa}, \citenamefont {Salfi}, \citenamefont {Saraiva}, \citenamefont {Dzurak}, \citenamefont {Hamilton},\ and\ \citenamefont {Culcer}}]{Sarkar_2023}%
  \BibitemOpen
  \bibfield  {author} {\bibinfo {author} {\bibfnamefont {A.}~\bibnamefont {Sarkar}}, \bibinfo {author} {\bibfnamefont {Z.}~\bibnamefont {Wang}}, \bibinfo {author} {\bibfnamefont {M.}~\bibnamefont {Rendell}}, \bibinfo {author} {\bibfnamefont {N.~W.}\ \bibnamefont {Hendrickx}}, \bibinfo {author} {\bibfnamefont {M.}~\bibnamefont {Veldhorst}}, \bibinfo {author} {\bibfnamefont {G.}~\bibnamefont {Scappucci}}, \bibinfo {author} {\bibfnamefont {M.}~\bibnamefont {Khalifa}}, \bibinfo {author} {\bibfnamefont {J.}~\bibnamefont {Salfi}}, \bibinfo {author} {\bibfnamefont {A.}~\bibnamefont {Saraiva}}, \bibinfo {author} {\bibfnamefont {A.~S.}\ \bibnamefont {Dzurak}}, \bibinfo {author} {\bibfnamefont {A.~R.}\ \bibnamefont {Hamilton}},\ and\ \bibinfo {author} {\bibfnamefont {D.}~\bibnamefont {Culcer}},\ }\bibfield  {title} {\bibinfo {title} {Electrical operation of planar ge hole spin qubits in an in-plane magnetic field},\ }\href {https://doi.org/10.1103/PhysRevB.108.245301} {\bibfield  {journal} {\bibinfo  {journal} {Phys.
  Rev. B}\ }\textbf {\bibinfo {volume} {108}},\ \bibinfo {pages} {245301} (\bibinfo {year} {2023})}\BibitemShut {NoStop}%
\bibitem [{\citenamefont {Liles}\ \emph {et~al.}(2021)\citenamefont {Liles}, \citenamefont {Martins}, \citenamefont {Miserev}, \citenamefont {Kiselev}, \citenamefont {Thorvaldson}, \citenamefont {Rendell}, \citenamefont {Jin}, \citenamefont {Hudson}, \citenamefont {Veldhorst}, \citenamefont {Itoh}, \citenamefont {Sushkov}, \citenamefont {Ladd}, \citenamefont {Dzurak},\ and\ \citenamefont {Hamilton}}]{Liles_2021}%
  \BibitemOpen
  \bibfield  {author} {\bibinfo {author} {\bibfnamefont {S.~D.}\ \bibnamefont {Liles}}, \bibinfo {author} {\bibfnamefont {F.}~\bibnamefont {Martins}}, \bibinfo {author} {\bibfnamefont {D.~S.}\ \bibnamefont {Miserev}}, \bibinfo {author} {\bibfnamefont {A.~A.}\ \bibnamefont {Kiselev}}, \bibinfo {author} {\bibfnamefont {I.~D.}\ \bibnamefont {Thorvaldson}}, \bibinfo {author} {\bibfnamefont {M.~J.}\ \bibnamefont {Rendell}}, \bibinfo {author} {\bibfnamefont {I.~K.}\ \bibnamefont {Jin}}, \bibinfo {author} {\bibfnamefont {F.~E.}\ \bibnamefont {Hudson}}, \bibinfo {author} {\bibfnamefont {M.}~\bibnamefont {Veldhorst}}, \bibinfo {author} {\bibfnamefont {K.~M.}\ \bibnamefont {Itoh}}, \bibinfo {author} {\bibfnamefont {O.~P.}\ \bibnamefont {Sushkov}}, \bibinfo {author} {\bibfnamefont {T.~D.}\ \bibnamefont {Ladd}}, \bibinfo {author} {\bibfnamefont {A.~S.}\ \bibnamefont {Dzurak}},\ and\ \bibinfo {author} {\bibfnamefont {A.~R.}\ \bibnamefont {Hamilton}},\ }\bibfield  {title} {\bibinfo {title} {Electrical control of the g
  tensor of the first hole in a silicon mos quantum dot},\ }\href {https://doi.org/10.1103/physrevb.104.235303} {\bibfield  {journal} {\bibinfo  {journal} {Physical Review B}\ }\textbf {\bibinfo {volume} {104}},\ \bibinfo {pages} {235303} (\bibinfo {year} {2021})}\BibitemShut {NoStop}%
\bibitem [{\citenamefont {Corley-Wiciak}\ \emph {et~al.}(2023)\citenamefont {Corley-Wiciak}, \citenamefont {Richter}, \citenamefont {Zoellner}, \citenamefont {Zaitsev}, \citenamefont {Manganelli}, \citenamefont {Zatterin}, \citenamefont {Sch{\"u}lli}, \citenamefont {Corley-Wiciak}, \citenamefont {Katzer}, \citenamefont {Reichmann}, \citenamefont {Klesse}, \citenamefont {Hendrickx}, \citenamefont {Sammak}, \citenamefont {Veldhorst}, \citenamefont {Scappucci}, \citenamefont {Virgilio},\ and\ \citenamefont {Capellini}}]{strainmapping}%
  \BibitemOpen
  \bibfield  {author} {\bibinfo {author} {\bibfnamefont {C.}~\bibnamefont {Corley-Wiciak}}, \bibinfo {author} {\bibfnamefont {C.}~\bibnamefont {Richter}}, \bibinfo {author} {\bibfnamefont {M.~H.}\ \bibnamefont {Zoellner}}, \bibinfo {author} {\bibfnamefont {I.}~\bibnamefont {Zaitsev}}, \bibinfo {author} {\bibfnamefont {C.~L.}\ \bibnamefont {Manganelli}}, \bibinfo {author} {\bibfnamefont {E.}~\bibnamefont {Zatterin}}, \bibinfo {author} {\bibfnamefont {T.~U.}\ \bibnamefont {Sch{\"u}lli}}, \bibinfo {author} {\bibfnamefont {A.~A.}\ \bibnamefont {Corley-Wiciak}}, \bibinfo {author} {\bibfnamefont {J.}~\bibnamefont {Katzer}}, \bibinfo {author} {\bibfnamefont {F.}~\bibnamefont {Reichmann}}, \bibinfo {author} {\bibfnamefont {W.~M.}\ \bibnamefont {Klesse}}, \bibinfo {author} {\bibfnamefont {N.~W.}\ \bibnamefont {Hendrickx}}, \bibinfo {author} {\bibfnamefont {A.}~\bibnamefont {Sammak}}, \bibinfo {author} {\bibfnamefont {M.}~\bibnamefont {Veldhorst}}, \bibinfo {author} {\bibfnamefont {G.}~\bibnamefont {Scappucci}},
  \bibinfo {author} {\bibfnamefont {M.}~\bibnamefont {Virgilio}},\ and\ \bibinfo {author} {\bibfnamefont {G.}~\bibnamefont {Capellini}},\ }\bibfield  {title} {\bibinfo {title} {Nanoscale mapping of the 3d strain tensor in a germanium quantum well hosting a functional spin qubit device},\ }\href {https://doi.org/10.1021/acsami.2c17395} {\bibfield  {journal} {\bibinfo  {journal} {ACS Applied Materials \& Interfaces}\ }\textbf {\bibinfo {volume} {15}},\ \bibinfo {pages} {3119} (\bibinfo {year} {2023})}\BibitemShut {NoStop}%
\bibitem [{\citenamefont {Green}\ \emph {et~al.}(2013)\citenamefont {Green}, \citenamefont {Sastrawan}, \citenamefont {Uys},\ and\ \citenamefont {Biercuk}}]{Green_2013}%
  \BibitemOpen
  \bibfield  {author} {\bibinfo {author} {\bibfnamefont {T.~J.}\ \bibnamefont {Green}}, \bibinfo {author} {\bibfnamefont {J.}~\bibnamefont {Sastrawan}}, \bibinfo {author} {\bibfnamefont {H.}~\bibnamefont {Uys}},\ and\ \bibinfo {author} {\bibfnamefont {M.~J.}\ \bibnamefont {Biercuk}},\ }\bibfield  {title} {\bibinfo {title} {Arbitrary quantum control of qubits in the presence of universal noise},\ }\href {https://doi.org/10.1088/1367-2630/15/9/095004} {\bibfield  {journal} {\bibinfo  {journal} {New Journal of Physics}\ }\textbf {\bibinfo {volume} {15}},\ \bibinfo {pages} {095004} (\bibinfo {year} {2013})}\BibitemShut {NoStop}%
\bibitem [{\citenamefont {Yan}\ \emph {et~al.}(2015)\citenamefont {Yan}, \citenamefont {L{\"u}},\ and\ \citenamefont {Zheng}}]{Yan2015-pm}%
  \BibitemOpen
  \bibfield  {author} {\bibinfo {author} {\bibfnamefont {Y.}~\bibnamefont {Yan}}, \bibinfo {author} {\bibfnamefont {Z.}~\bibnamefont {L{\"u}}},\ and\ \bibinfo {author} {\bibfnamefont {H.}~\bibnamefont {Zheng}},\ }\bibfield  {title} {\bibinfo {title} {{Bloch-Siegert} shift of the rabi model},\ }\href@noop {} {\bibfield  {journal} {\bibinfo  {journal} {Phys. Rev. A}\ }\textbf {\bibinfo {volume} {91}} (\bibinfo {year} {2015})}\BibitemShut {NoStop}%
\bibitem [{\citenamefont {Wu}\ and\ \citenamefont {Yang}(2007)}]{PhysRevLett.98.013601}%
  \BibitemOpen
  \bibfield  {author} {\bibinfo {author} {\bibfnamefont {Y.}~\bibnamefont {Wu}}\ and\ \bibinfo {author} {\bibfnamefont {X.}~\bibnamefont {Yang}},\ }\bibfield  {title} {\bibinfo {title} {Strong-coupling theory of periodically driven two-level systems},\ }\href {https://doi.org/10.1103/PhysRevLett.98.013601} {\bibfield  {journal} {\bibinfo  {journal} {Phys. Rev. Lett.}\ }\textbf {\bibinfo {volume} {98}},\ \bibinfo {pages} {013601} (\bibinfo {year} {2007})}\BibitemShut {NoStop}%
\bibitem [{\citenamefont {Zeuch}\ \emph {et~al.}(2020)\citenamefont {Zeuch}, \citenamefont {Hassler}, \citenamefont {Slim},\ and\ \citenamefont {DiVincenzo}}]{Zeuch2020}%
  \BibitemOpen
  \bibfield  {author} {\bibinfo {author} {\bibfnamefont {D.}~\bibnamefont {Zeuch}}, \bibinfo {author} {\bibfnamefont {F.}~\bibnamefont {Hassler}}, \bibinfo {author} {\bibfnamefont {J.~J.}\ \bibnamefont {Slim}},\ and\ \bibinfo {author} {\bibfnamefont {D.~P.}\ \bibnamefont {DiVincenzo}},\ }\bibfield  {title} {\bibinfo {title} {Exact rotating wave approximation},\ }\href {https://doi.org/10.1016/j.aop.2020.168327} {\bibfield  {journal} {\bibinfo  {journal} {Annals of Physics}\ }\textbf {\bibinfo {volume} {423}},\ \bibinfo {pages} {168327} (\bibinfo {year} {2020})}\BibitemShut {NoStop}%
\bibitem [{\citenamefont {Rimbach-Russ}\ \emph {et~al.}(2023)\citenamefont {Rimbach-Russ}, \citenamefont {Philips}, \citenamefont {Xue},\ and\ \citenamefont {Vandersypen}}]{RimbachRuss2023}%
  \BibitemOpen
  \bibfield  {author} {\bibinfo {author} {\bibfnamefont {M.}~\bibnamefont {Rimbach-Russ}}, \bibinfo {author} {\bibfnamefont {S.~G.~J.}\ \bibnamefont {Philips}}, \bibinfo {author} {\bibfnamefont {X.}~\bibnamefont {Xue}},\ and\ \bibinfo {author} {\bibfnamefont {L.~M.~K.}\ \bibnamefont {Vandersypen}},\ }\bibfield  {title} {\bibinfo {title} {Simple framework for systematic high-fidelity gate operations},\ }\href {https://doi.org/10.1088/2058-9565/acf786} {\bibfield  {journal} {\bibinfo  {journal} {Quantum Science and Technology}\ }\textbf {\bibinfo {volume} {8}},\ \bibinfo {pages} {045025} (\bibinfo {year} {2023})}\BibitemShut {NoStop}%
\bibitem [{\citenamefont {Kloeffel}\ and\ \citenamefont {Loss}(2013)}]{kloeffelProspectsSpinBasedQuantum2013}%
  \BibitemOpen
  \bibfield  {author} {\bibinfo {author} {\bibfnamefont {C.}~\bibnamefont {Kloeffel}}\ and\ \bibinfo {author} {\bibfnamefont {D.}~\bibnamefont {Loss}},\ }\bibfield  {title} {\bibinfo {title} {Prospects for {{Spin-Based Quantum Computing}} in {{Quantum Dots}}},\ }\href {https://doi.org/10.1146/annurev-conmatphys-030212-184248} {\bibfield  {journal} {\bibinfo  {journal} {Annu. Rev. Condens. Matter Phys.}\ }\textbf {\bibinfo {volume} {4}},\ \bibinfo {pages} {51} (\bibinfo {year} {2013})}\BibitemShut {NoStop}%
\bibitem [{\citenamefont {Malkoc}\ \emph {et~al.}(2022)\citenamefont {Malkoc}, \citenamefont {Stano},\ and\ \citenamefont {Loss}}]{PhysRevLett.129.247701}%
  \BibitemOpen
  \bibfield  {author} {\bibinfo {author} {\bibfnamefont {O.}~\bibnamefont {Malkoc}}, \bibinfo {author} {\bibfnamefont {P.}~\bibnamefont {Stano}},\ and\ \bibinfo {author} {\bibfnamefont {D.}~\bibnamefont {Loss}},\ }\bibfield  {title} {\bibinfo {title} {Charge-noise-induced dephasing in silicon hole-spin qubits},\ }\href {https://doi.org/10.1103/PhysRevLett.129.247701} {\bibfield  {journal} {\bibinfo  {journal} {Phys. Rev. Lett.}\ }\textbf {\bibinfo {volume} {129}},\ \bibinfo {pages} {247701} (\bibinfo {year} {2022})}\BibitemShut {NoStop}%
\bibitem [{\citenamefont {Froning}\ \emph {et~al.}(2021)\citenamefont {Froning}, \citenamefont {Ran\ifmmode \check{c}\else \v{c}\fi{}i\ifmmode~\acute{c}\else \'{c}\fi{}}, \citenamefont {Het\'enyi}, \citenamefont {Bosco}, \citenamefont {Rehmann}, \citenamefont {Li}, \citenamefont {Bakkers}, \citenamefont {Zwanenburg}, \citenamefont {Loss}, \citenamefont {Zumb\"uhl},\ and\ \citenamefont {Braakman}}]{PhysRevResearch.3.013081}%
  \BibitemOpen
  \bibfield  {author} {\bibinfo {author} {\bibfnamefont {F.~N.~M.}\ \bibnamefont {Froning}}, \bibinfo {author} {\bibfnamefont {M.~J.}\ \bibnamefont {Ran\ifmmode \check{c}\else \v{c}\fi{}i\ifmmode~\acute{c}\else \'{c}\fi{}}}, \bibinfo {author} {\bibfnamefont {B.}~\bibnamefont {Het\'enyi}}, \bibinfo {author} {\bibfnamefont {S.}~\bibnamefont {Bosco}}, \bibinfo {author} {\bibfnamefont {M.~K.}\ \bibnamefont {Rehmann}}, \bibinfo {author} {\bibfnamefont {A.}~\bibnamefont {Li}}, \bibinfo {author} {\bibfnamefont {E.~P. A.~M.}\ \bibnamefont {Bakkers}}, \bibinfo {author} {\bibfnamefont {F.~A.}\ \bibnamefont {Zwanenburg}}, \bibinfo {author} {\bibfnamefont {D.}~\bibnamefont {Loss}}, \bibinfo {author} {\bibfnamefont {D.~M.}\ \bibnamefont {Zumb\"uhl}},\ and\ \bibinfo {author} {\bibfnamefont {F.~R.}\ \bibnamefont {Braakman}},\ }\bibfield  {title} {\bibinfo {title} {Strong spin-orbit interaction and $g$-factor renormalization of hole spins in ge/si nanowire quantum dots},\ }\href
  {https://doi.org/10.1103/PhysRevResearch.3.013081} {\bibfield  {journal} {\bibinfo  {journal} {Phys. Rev. Res.}\ }\textbf {\bibinfo {volume} {3}},\ \bibinfo {pages} {013081} (\bibinfo {year} {2021})}\BibitemShut {NoStop}%
\bibitem [{\citenamefont {Bosco}\ \emph {et~al.}(2021{\natexlab{b}})\citenamefont {Bosco}, \citenamefont {Het\'enyi},\ and\ \citenamefont {Loss}}]{PRXQuantum.2.010348}%
  \BibitemOpen
  \bibfield  {author} {\bibinfo {author} {\bibfnamefont {S.}~\bibnamefont {Bosco}}, \bibinfo {author} {\bibfnamefont {B.}~\bibnamefont {Het\'enyi}},\ and\ \bibinfo {author} {\bibfnamefont {D.}~\bibnamefont {Loss}},\ }\bibfield  {title} {\bibinfo {title} {Hole spin qubits in $\mathrm{Si}$ finfets with fully tunable spin-orbit coupling and sweet spots for charge noise},\ }\href {https://doi.org/10.1103/PRXQuantum.2.010348} {\bibfield  {journal} {\bibinfo  {journal} {PRX Quantum}\ }\textbf {\bibinfo {volume} {2}},\ \bibinfo {pages} {010348} (\bibinfo {year} {2021}{\natexlab{b}})}\BibitemShut {NoStop}%
\bibitem [{\citenamefont {Bassi}\ \emph {et~al.}(2025)\citenamefont {Bassi}, \citenamefont {Rodr{\'i}guez-Mena}, \citenamefont {Brun}, \citenamefont {Zihlmann}, \citenamefont {Nguyen}, \citenamefont {Champain}, \citenamefont {Abadillo-Uriel}, \citenamefont {Bertrand}, \citenamefont {Niebojewski}, \citenamefont {Maurand}, \citenamefont {Niquet}, \citenamefont {Jehl}, \citenamefont {De~Franceschi},\ and\ \citenamefont {Schmitt}}]{Bassi2025}%
  \BibitemOpen
  \bibfield  {author} {\bibinfo {author} {\bibfnamefont {M.}~\bibnamefont {Bassi}}, \bibinfo {author} {\bibfnamefont {E.~A.}\ \bibnamefont {Rodr{\'i}guez-Mena}}, \bibinfo {author} {\bibfnamefont {B.}~\bibnamefont {Brun}}, \bibinfo {author} {\bibfnamefont {S.}~\bibnamefont {Zihlmann}}, \bibinfo {author} {\bibfnamefont {T.}~\bibnamefont {Nguyen}}, \bibinfo {author} {\bibfnamefont {V.}~\bibnamefont {Champain}}, \bibinfo {author} {\bibfnamefont {J.~C.}\ \bibnamefont {Abadillo-Uriel}}, \bibinfo {author} {\bibfnamefont {B.}~\bibnamefont {Bertrand}}, \bibinfo {author} {\bibfnamefont {H.}~\bibnamefont {Niebojewski}}, \bibinfo {author} {\bibfnamefont {R.}~\bibnamefont {Maurand}}, \bibinfo {author} {\bibfnamefont {Y.-M.}\ \bibnamefont {Niquet}}, \bibinfo {author} {\bibfnamefont {X.}~\bibnamefont {Jehl}}, \bibinfo {author} {\bibfnamefont {S.}~\bibnamefont {De~Franceschi}},\ and\ \bibinfo {author} {\bibfnamefont {V.}~\bibnamefont {Schmitt}},\ }\bibfield  {title} {\bibinfo {title} {Optimal operation of hole spin qubits},\
  }\href {https://doi.org/10.1038/s41567-025-03106-1} {\bibfield  {journal} {\bibinfo  {journal} {Nature Physics}\ }\textbf {\bibinfo {volume} {22}},\ \bibinfo {pages} {75} (\bibinfo {year} {2025})}\BibitemShut {NoStop}%
\bibitem [{\citenamefont {Yu}\ \emph {et~al.}(2023)\citenamefont {Yu}, \citenamefont {Zihlmann}, \citenamefont {Abadillo-Uriel}, \citenamefont {Michal}, \citenamefont {Rambal}, \citenamefont {Niebojewski}, \citenamefont {Bedecarrats}, \citenamefont {Vinet}, \citenamefont {Dumur}, \citenamefont {Filippone}, \citenamefont {Bertrand}, \citenamefont {De~Franceschi}, \citenamefont {Niquet},\ and\ \citenamefont {Maurand}}]{Yu2023}%
  \BibitemOpen
  \bibfield  {author} {\bibinfo {author} {\bibfnamefont {C.~X.}\ \bibnamefont {Yu}}, \bibinfo {author} {\bibfnamefont {S.}~\bibnamefont {Zihlmann}}, \bibinfo {author} {\bibfnamefont {J.~C.}\ \bibnamefont {Abadillo-Uriel}}, \bibinfo {author} {\bibfnamefont {V.~P.}\ \bibnamefont {Michal}}, \bibinfo {author} {\bibfnamefont {N.}~\bibnamefont {Rambal}}, \bibinfo {author} {\bibfnamefont {H.}~\bibnamefont {Niebojewski}}, \bibinfo {author} {\bibfnamefont {T.}~\bibnamefont {Bedecarrats}}, \bibinfo {author} {\bibfnamefont {M.}~\bibnamefont {Vinet}}, \bibinfo {author} {\bibfnamefont {{\'E}.}~\bibnamefont {Dumur}}, \bibinfo {author} {\bibfnamefont {M.}~\bibnamefont {Filippone}}, \bibinfo {author} {\bibfnamefont {B.}~\bibnamefont {Bertrand}}, \bibinfo {author} {\bibfnamefont {S.}~\bibnamefont {De~Franceschi}}, \bibinfo {author} {\bibfnamefont {Y.-M.}\ \bibnamefont {Niquet}},\ and\ \bibinfo {author} {\bibfnamefont {R.}~\bibnamefont {Maurand}},\ }\bibfield  {title} {\bibinfo {title} {Strong coupling between a photon and a
  hole spin in silicon},\ }\href {https://doi.org/10.1038/s41565-023-01332-3} {\bibfield  {journal} {\bibinfo  {journal} {Nature Nanotechnology}\ }\textbf {\bibinfo {volume} {18}},\ \bibinfo {pages} {741} (\bibinfo {year} {2023})}\BibitemShut {NoStop}%
\bibitem [{\citenamefont {Adelsberger}\ \emph {et~al.}(2022{\natexlab{a}})\citenamefont {Adelsberger}, \citenamefont {Benito}, \citenamefont {Bosco}, \citenamefont {Klinovaja},\ and\ \citenamefont {Loss}}]{spindependentmass}%
  \BibitemOpen
  \bibfield  {author} {\bibinfo {author} {\bibfnamefont {C.}~\bibnamefont {Adelsberger}}, \bibinfo {author} {\bibfnamefont {M.}~\bibnamefont {Benito}}, \bibinfo {author} {\bibfnamefont {S.}~\bibnamefont {Bosco}}, \bibinfo {author} {\bibfnamefont {J.}~\bibnamefont {Klinovaja}},\ and\ \bibinfo {author} {\bibfnamefont {D.}~\bibnamefont {Loss}},\ }\bibfield  {title} {\bibinfo {title} {Hole-spin qubits in ge nanowire quantum dots: Interplay of orbital magnetic field, strain, and growth direction},\ }\href {https://doi.org/10.1103/PhysRevB.105.075308} {\bibfield  {journal} {\bibinfo  {journal} {Phys. Rev. B}\ }\textbf {\bibinfo {volume} {105}},\ \bibinfo {pages} {075308} (\bibinfo {year} {2022}{\natexlab{a}})}\BibitemShut {NoStop}%
\bibitem [{\citenamefont {Adelsberger}\ \emph {et~al.}(2022{\natexlab{b}})\citenamefont {Adelsberger}, \citenamefont {Bosco}, \citenamefont {Klinovaja},\ and\ \citenamefont {Loss}}]{PhysRevB.106.235408}%
  \BibitemOpen
  \bibfield  {author} {\bibinfo {author} {\bibfnamefont {C.}~\bibnamefont {Adelsberger}}, \bibinfo {author} {\bibfnamefont {S.}~\bibnamefont {Bosco}}, \bibinfo {author} {\bibfnamefont {J.}~\bibnamefont {Klinovaja}},\ and\ \bibinfo {author} {\bibfnamefont {D.}~\bibnamefont {Loss}},\ }\bibfield  {title} {\bibinfo {title} {Enhanced orbital magnetic field effects in ge hole nanowires},\ }\href {https://doi.org/10.1103/PhysRevB.106.235408} {\bibfield  {journal} {\bibinfo  {journal} {Phys. Rev. B}\ }\textbf {\bibinfo {volume} {106}},\ \bibinfo {pages} {235408} (\bibinfo {year} {2022}{\natexlab{b}})}\BibitemShut {NoStop}%
\bibitem [{\citenamefont {Terrazos}\ \emph {et~al.}(2021)\citenamefont {Terrazos}, \citenamefont {Marcellina}, \citenamefont {Wang}, \citenamefont {Coppersmith}, \citenamefont {Friesen}, \citenamefont {Hamilton}, \citenamefont {Hu}, \citenamefont {Koiller}, \citenamefont {Saraiva}, \citenamefont {Culcer},\ and\ \citenamefont {Capaz}}]{Terrazos_2021}%
  \BibitemOpen
  \bibfield  {author} {\bibinfo {author} {\bibfnamefont {L.~A.}\ \bibnamefont {Terrazos}}, \bibinfo {author} {\bibfnamefont {E.}~\bibnamefont {Marcellina}}, \bibinfo {author} {\bibfnamefont {Z.}~\bibnamefont {Wang}}, \bibinfo {author} {\bibfnamefont {S.~N.}\ \bibnamefont {Coppersmith}}, \bibinfo {author} {\bibfnamefont {M.}~\bibnamefont {Friesen}}, \bibinfo {author} {\bibfnamefont {A.~R.}\ \bibnamefont {Hamilton}}, \bibinfo {author} {\bibfnamefont {X.}~\bibnamefont {Hu}}, \bibinfo {author} {\bibfnamefont {B.}~\bibnamefont {Koiller}}, \bibinfo {author} {\bibfnamefont {A.~L.}\ \bibnamefont {Saraiva}}, \bibinfo {author} {\bibfnamefont {D.}~\bibnamefont {Culcer}},\ and\ \bibinfo {author} {\bibfnamefont {R.~B.}\ \bibnamefont {Capaz}},\ }\bibfield  {title} {\bibinfo {title} {Theory of hole-spin qubits in strained germanium quantum dots},\ }\href {https://doi.org/10.1103/PhysRevB.103.125201} {\bibfield  {journal} {\bibinfo  {journal} {Phys. Rev. B}\ }\textbf {\bibinfo {volume} {103}},\ \bibinfo {pages} {125201}
  (\bibinfo {year} {2021})}\BibitemShut {NoStop}%
\bibitem [{\citenamefont {Bosco}\ and\ \citenamefont {Loss}(2021)}]{Bosco2021}%
  \BibitemOpen
  \bibfield  {author} {\bibinfo {author} {\bibfnamefont {S.}~\bibnamefont {Bosco}}\ and\ \bibinfo {author} {\bibfnamefont {D.}~\bibnamefont {Loss}},\ }\bibfield  {title} {\bibinfo {title} {Fully tunable hyperfine interactions of hole spin qubits in si and ge quantum dots},\ }\href {https://doi.org/10.1103/physrevlett.127.190501} {\bibfield  {journal} {\bibinfo  {journal} {Physical Review Letters}\ }\textbf {\bibinfo {volume} {127}},\ \bibinfo {pages} {190501} (\bibinfo {year} {2021})}\BibitemShut {NoStop}%
\bibitem [{\citenamefont {Fischer}\ \emph {et~al.}(2008)\citenamefont {Fischer}, \citenamefont {Coish}, \citenamefont {Bulaev},\ and\ \citenamefont {Loss}}]{Fischer2008}%
  \BibitemOpen
  \bibfield  {author} {\bibinfo {author} {\bibfnamefont {J.}~\bibnamefont {Fischer}}, \bibinfo {author} {\bibfnamefont {W.~A.}\ \bibnamefont {Coish}}, \bibinfo {author} {\bibfnamefont {D.~V.}\ \bibnamefont {Bulaev}},\ and\ \bibinfo {author} {\bibfnamefont {D.}~\bibnamefont {Loss}},\ }\bibfield  {title} {\bibinfo {title} {Spin decoherence of a heavy hole coupled to nuclear spins in a quantum dot},\ }\href {https://doi.org/10.1103/PhysRevB.78.155329} {\bibfield  {journal} {\bibinfo  {journal} {Phys. Rev. B}\ }\textbf {\bibinfo {volume} {78}},\ \bibinfo {pages} {155329} (\bibinfo {year} {2008})}\BibitemShut {NoStop}%
\bibitem [{\citenamefont {Abadillo-Uriel}\ \emph {et~al.}(2023)\citenamefont {Abadillo-Uriel}, \citenamefont {Rodr\'{\i}guez-Mena}, \citenamefont {Martinez},\ and\ \citenamefont {Niquet}}]{urielstrainholedriving}%
  \BibitemOpen
  \bibfield  {author} {\bibinfo {author} {\bibfnamefont {J.~C.}\ \bibnamefont {Abadillo-Uriel}}, \bibinfo {author} {\bibfnamefont {E.~A.}\ \bibnamefont {Rodr\'{\i}guez-Mena}}, \bibinfo {author} {\bibfnamefont {B.}~\bibnamefont {Martinez}},\ and\ \bibinfo {author} {\bibfnamefont {Y.-M.}\ \bibnamefont {Niquet}},\ }\bibfield  {title} {\bibinfo {title} {Hole-spin driving by strain-induced spin-orbit interactions},\ }\href {https://doi.org/10.1103/PhysRevLett.131.097002} {\bibfield  {journal} {\bibinfo  {journal} {Phys. Rev. Lett.}\ }\textbf {\bibinfo {volume} {131}},\ \bibinfo {pages} {097002} (\bibinfo {year} {2023})}\BibitemShut {NoStop}%
\bibitem [{\citenamefont {Rimbach-Russ}\ \emph {et~al.}(2025)\citenamefont {Rimbach-Russ}, \citenamefont {John}, \citenamefont {van Straaten},\ and\ \citenamefont {Bosco}}]{mvtj-zhrl}%
  \BibitemOpen
  \bibfield  {author} {\bibinfo {author} {\bibfnamefont {M.}~\bibnamefont {Rimbach-Russ}}, \bibinfo {author} {\bibfnamefont {V.}~\bibnamefont {John}}, \bibinfo {author} {\bibfnamefont {B.}~\bibnamefont {van Straaten}},\ and\ \bibinfo {author} {\bibfnamefont {S.}~\bibnamefont {Bosco}},\ }\bibfield  {title} {\bibinfo {title} {Gapless single-spin qubit},\ }\href {https://doi.org/10.1103/mvtj-zhrl} {\bibfield  {journal} {\bibinfo  {journal} {Phys. Rev. Lett.}\ }\textbf {\bibinfo {volume} {135}},\ \bibinfo {pages} {197001} (\bibinfo {year} {2025})}\BibitemShut {NoStop}%
\end{thebibliography}%

\end{document}